\documentclass[jog]{svjour}
\usepackage{times}
\usepackage{color}
\usepackage{graphics}
\usepackage{amssymb}
\usepackage{graphicx}
\usepackage{lineno}
\newcommand\arcdeg{\mbox{$^\circ$}}%
\newcommand\arcmin{\mbox{$^\prime$}}%
\newcommand\fs{\mbox{$.\!\!^{\mathrm s}$}}%
\newcommand\farcs{\mbox{$.\!\!^{\prime\prime}$}}%
\begin{document}
\title{The impacts of source structure on geodetic parameters demonstrated by the radio source 3C371}
\author{Ming H. Xu\inst{1,2}, Robert Heinkelmann\inst{2}, James M. Anderson\inst{2}, Julian Mora-Diaz\inst{2}, Maria Karbon\inst{2}, Harald Schuh\inst{2,3}, and Guang L. Wang\inst{1} 
}                     
\offprints{}          
\institute{Key Laboratory of Planetary Sciences, Shanghai Astronomical Observatory, Chinese Academy of Sciences, 200030 Shanghai, China \and ~Helmholtz Centre Potsdam, GFZ German Research Centre for Geosciences, Telegrafenberg 14473 Potsdam, Germany \and 
~Institute of Geodesy and Geoinformation Science, Technische Universit\"{a}t Berlin, Stra${\beta}$e des 17. Juni 135, 10623, Berlin, Germany}
\date{Received: date / Revised version: date}
%
\abstract{
Closure quantities measured by very long baseline interferometry (VLBI) observations are independent of instrumental and propagation instabilities 
and antenna gain factors, but are sensitive to source structure. 
A new method is proposed to calculate a structure index based on the median values of closure 
quantities rather than the brightness distribution of a source. The results are comparable to 
structure indices based on imaging observations at other epochs and demonstrate the flexibility of
deriving structure indices from exactly the same observations as used for geodetic analysis and without
imaging analysis.
A three-component model for the structure of source 3C371 is developed by model-fitting closure phases. 
It provides a
real case of tracing how the structure effect identified by closure phases in the same observations as 
the delay observables 
affects the geodetic analysis, and investigating
which geodetic parameters are corrupted to what extent by the structure effect. Using the resulting
structure correction based on the three-component model of source 3C371, two solutions, with and without correcting the
structure effect, are made. With corrections, the overall rms of this source 
is reduced by 1 ps, and the impacts of the structure effect introduced by this single source are 
up to 1.4 mm on station positions
and up to 4.4 microarcseconds on Earth orientation parameters.
This study is considered as a starting point for handling the source structure effect on geodetic VLBI 
from geodetic sessions themselves. 
\keywords{source structure effect, structure index, VLBI, CONT14, quasars: individual (3C371)}} 
\maketitle
\section{Introduction}
\label{intro}
{It is well known from astrophysical studies of imaging that structures of 
geodetic radio sources are generally asymmetric, time-dependent, and frequency-dependent 
\citep[e.g.,][]{cha90a, ojh04, ojh05, pin07, lis09, cha10, fom11, lis13}.
The effects of source structures in geodetic very long baseline interferometry 
(VLBI) have been studied for decades \citep[e.g.,][]{cam88, cha88, tan88, ulv88, cha90b,  fey96, tor07, sha15, pla16}, 
and, for instance, by studying a series of ten
Research and Development VLBI (RDV) sessions, \citet{sov02} concluded that the structure effects contributed 8 ps -- 30 ps WRMS 
residual delay, and were the three major contributors along with the instrumental and tropospheric delays 
in geodetic VLBI. However, this effect is still ignored as noise in routine
geodetic VLBI data analysis so far.

In order to reach the future goals of VGOS \citep{pet09}, including 1~mm position accuracy, delay errors from individual sources should also be below approximately 1~mm~/~c~$\sim 3$~ps, implying that for a typical baseline length of 8000~km for geodetic VLBI observations, the astrometric positions of sources must be accurate to about 25~microarcseconds ($\mu$as). The motions and brightness fluctuations of the radio components of the set of regularly observed geodetic quasars are not known well, but worst-case limits can be estimated from studies of other active galactic nuclei source samples. Using the largest high-cadence study of flat-spectrum radio quasars  \citep[MOJAVE;][]{lis09}, the most common jet speed is 200~$\mu$as~yr$^{-1}$, with a maximum jet speed of 2500~$\mu$as~yr$^{-1}$, while \citet{fom11} demonstrated that the jet speed of the geodetic source $0556+238$ is about 100~$\mu$as~yr$^{-1}$.  Such motions would require source structure to typically be redetermined 8 times per year. Alternatively, brightness fluctuations, such as flares in the jet core regions, can also affect the effective astrometric position of sources.  Numerous long-term (many years) single-dish monitoring programs show that the rise times to double the total flux densities of sources can be as small as a few months or even a few weeks \citep[see, for example,][]{all85, lis09, fuh14, par14, max16}.  For a simple structure model of two approximately-equal-brightness components separated by 500~$\mu$as with one component undergoing a flare, an astrometric centroid shift of 25~$\mu$as results from just a 10~\% change in the total brightness.

Clearly, source structure changes must be monitored on timescales far shorter than a year in order to meet the future VGOS goals, and we cannot expect that source structure can be handled merely by selecting sources in VLBI observations based on structure indices that were in many cases obtained from single-epoch observations and separated in time from the geodetic observations by more than a decade.  Instead, geodetic VLBI should be able to determine source structure properties from the same observations that are being used to determine geodetic information.  Therefore, we propose a method that uses geodetic observations to derive structure indices and study the impacts of structure effects on those same observations. We use an individual source in this paper as a demonstration case; in practice all sources with suitable observations can be analyzed using this method.

\section{Closure quantities}
\label{sec2}
It is difficult to use group delay, fringe phase, and observed amplitude to study structures 
of radio sources 
owing to instabilities introduced by the atmosphere, independent local oscillators, 
and varying antenna gains at each site. However, VLBI observations are 
baseline dependent, and for an interferometer array with more than two stations there are  
redundancies allowing the formation of closure quantities that are independent of atmospheric effects, clock fluctuations,
and any station-based errors.

\subsection{Closure delay}
\label{sec2.1}
We defined closure delay as the sum of the delays around a 
closed triangle of baselines. 
Closure delay is a direct and important criterion of how much the source structure 
affects delay observables. It can also be used to determine the measurement noise in geodetic VLBI observables
and thus indicate the precision level of delay observables. For a detailed discussion about closure delay, please refer to \citet{xu16}.
\subsection{Closure phase}
\label{sec2.2}
Closure phase has been used by the astrophysical community to make images of radio sources for decades \citep{rog74,pea84,pea88}.
It is well know that phase delays in VLBI observations are more accurate than group delays, but due to the unresolved
ambiguity issue, phase in fact has not been used in geodetic VLBI. Closure phase actually circumvents the ambiguity issue.

\subsection{Closure amplitude}
\label{sec2.3}
Amplitude is generally not calibrated in geodetic VLBI observations, but closure amplitude, 
independent of the gain of each individual station, is a good observable for the study of source structure.
With four stations, $a$, $b$, $c$, and $d$, it is possible to form combinations of amplitudes that are independent of the antenna\rq{}s gain factors by using
    \begin{equation}
     \label{eq_closure3}
    A_{\mbox{\scriptsize abcd}}=\frac{A_{\mbox{\scriptsize ab}}A_{\mbox{\scriptsize cd}}}{A_{\mbox{\scriptsize ac}}A_{\mbox{\scriptsize bd}}},
    \end{equation}
where, for instance, $A_{\mbox{\scriptsize ab}}$ is the observed amplitude on baseline $ab$. These combinations are called closure amplitude \citep{rea80}.
If all six interferometer baselines formed by the 
four stations are correlated in one scan, three closure amplitudes with different values can be obtained, for example 
$A_{\mbox{\scriptsize abcd}}$, $A_{\mbox{\scriptsize abdc}}$, and $A_{\mbox{\scriptsize adcb}}$, only two of which are independent. For a comprehensive discusion about closure phase and closure amplitude, please refer
to \citet{pea84} and \citet{tho07} and the references therein.

\subsection{Calculations and data}
\label{sec2.3}
The interpretation of closure quantities is very challenging, as, unlike the visibility,
the sky brightness distribution cannot be obtained from them by a simple Fourier transform relationship.
Moreover, knowledge of both the absolute strength and the absolute position of the source is lost in these 
closure quantities. However, closure quantities have the advantage of showing 
the magnitudes of source structure effects without the need for calibration or imaging. Source structure effects in geodetic VLBI data analysis
have merely been considered in terms of a structure index, so that point-like sources can be selected and extended sources 
can be avoided in scheduling of VLBI observations. However, for the 1 mm accuracy goal of VGOS, such an approach is unlikely to be sufficient as there are not enough sources with low structure indices ($<$3) to cover the sky uniformly.  For most geodetic sources,
the structure indices are single epoch and thus might not represent the magnitudes of structure effects in observations after several years.
Therefore, it should be more effective and illuminating to determine the magnitudes of structure effects based on 
closure quantities from geodetic observations themselves.

To demonstrate this, we use the data from CONT14\footnote{http://ivscc.gsfc.nasa.gov/program/cont14/}
observations \citep{not15} at X band.
CONT14, as a campaign of continuous VLBI observations conducted by the International VLBI Service for 
Geodesy and Astrometry (IVS) over 15 days with 17 globally distributed stations, 
was intended to
acquire state-of-the-art VLBI data with the highest accuracy that the then existing VLBI
system was capable of. Since only 71 radio sources were observed in this campaign, one can expect that
most of these radio sources have enough observations with good $uv$ coverages to 
get meaningful statistical information from closure quantities.

The specific equations that were used for the calculations of closure quantities from geodetic VLBI observations 
are shown in Appendix A. Closure quantities were calculated based on these equations.
Next, the median and the rms values of the magnitudes of closure delays, closure phases, 
and closure amplitude logarithms were determined
for each source. Due to the sensitivity of rms value to outliers with large magnitudes, 
rms values were derived in an iterative way: 
closure quantities with magnitudes 5 times larger than the rms were identified as outliers
until no outliers remained. For most radio sources, only a few percent
of the closure quantities were excluded in this procedure.
The statistics for the 65 radio sources that have more than 30 closure relations in CONT14 observations
 are presented in Table \ref{tab1}.

\section{Structure index}

Structure index \citep{fey1997,fey2000} plays an important role in geodetic/astrometric VLBI as an indicator 
of the magnitude of the structure effect for each source: (1) extended sources can be avoided 
and compact sources can be selected in VLBI observations; (2) one of the criteria for selecting 
defining sources in ICRF2
is that they had structure indices smaller than 3 \citep{fey2015}; (3) sources with high astrometric quality
can be selected as candidate sources for aligning optical and radio catalogs \citep[e.g.][]{bou08, bou11, le16}.
Structure index can be calculated from closure delays based on observations if the thresholds for closure delays 
$\tau^{\mbox{\scriptsize th}}_{\mbox{\scriptsize closure}}$
are related to the thresholds for calculated structure corrections in delay observables $\tau^{\mbox{\scriptsize th}}_{\mbox{\scriptsize delay}}$ 
that were used by \citet{fey1997} to calculate structure index. The relation
between these two thresholds is,  
    \begin{equation}
     \label{eq_threshold}
 \tau^{\mbox{\scriptsize th}}_{\mbox{\scriptsize closure}} = \sqrt{3}\sqrt{\tau_{\mbox{\scriptsize noise}}^{2}+{(\tau^{\mbox{\scriptsize th}}_{\mbox{\scriptsize delay}})}^2},
    \end{equation}
where $\tau_{\mbox{\scriptsize noise}}$ is the median value of measurement noises. 
The rms of closure delays for 0727-115, which has minimal source structure and shows no closure structure with baseline orientation \citep[see,][]{xu16}, was 8~ps, which suggests that the measurement noise of individual group delays is below 4.8~ps  (8~ps$/\sqrt{3}$).
According to the relative relationship between rms and median values, the median 
value of the measurement noise was set to be 2.8 ps. Furthermore, note that changing this 
value to around 5 ps does not change classifying the structure indices between 2 and 3 or 3 and 4 --- it 
only affects distinguishing a structure index of 1 from 2.
In order to facilitate comparison to the Bordeaux VLBI image database\footnote{http://www.obs.u-bordeaux1.fr/BVID/} (BVID), which unfortunately only contains integer structure indices near in time to the CONT14 sessions, we define an integer closure delay structure index as follows. Using the same thresholds of structure delays for classifying structure groups in \citet{fey1997}, and assuming $\tau_{\mbox{\scriptsize noise}}$ = 2.8 ps, the structure index (SI) can also be calculated from the median value of closure delays by,  

\begin{equation}
     \label{eq_index}
 \mbox{\emph{SI}}=\left\{
\begin{array}{l}
1,\quad\mbox{if}\; \phantom{0}0\phantom{.0}~\mbox{ps} \;\leq \;\left|\tau_{\mbox{\scriptsize closure}}\right|^{\mbox{\scriptsize med}} \; < \; \phantom{0}7.1~\mbox{ps},\\
2,\quad\mbox{if}\; \phantom{0}7.1~\mbox{ps}           \;\leq \;\left|\tau_{\mbox{\scriptsize closure}}\right|^{\mbox{\scriptsize med}} \; < \; 18.0~\mbox{ps}, \\
3,\quad\mbox{if}\;  18.0~\mbox{ps}                    \;\leq \;\left|\tau_{\mbox{\scriptsize closure}}\right|^{\mbox{\scriptsize med}} \; < \; 52.2~\mbox{ps},\\
4,\quad\mbox{if}\;  52.2~\mbox{ps}                    \;\leq \;\left|\tau_{\mbox{\scriptsize closure}}\right|^{\mbox{\scriptsize med}} \; < \; \infty. 
\end{array} \right. 
    \end{equation}

We further define a continuous closure structure index that closely approximates equation \ref{eq_index} of an integer structure index as:
    \begin{equation}
     \label{eq_ind2}
 \mbox{\emph{SI}} \equiv  \ln \frac{\left|\tau_{\mbox{\scriptsize closure}}\right|^{\mbox{\scriptsize med}}}{1~\mbox{ps}}.
    \end{equation}
This equation is for closure delays, and a complete set of equations for 
the three kinds of closure quantities are
presented in Appendix B.
\begin{figure}
\resizebox{0.48\textwidth}{!}{%
  \includegraphics{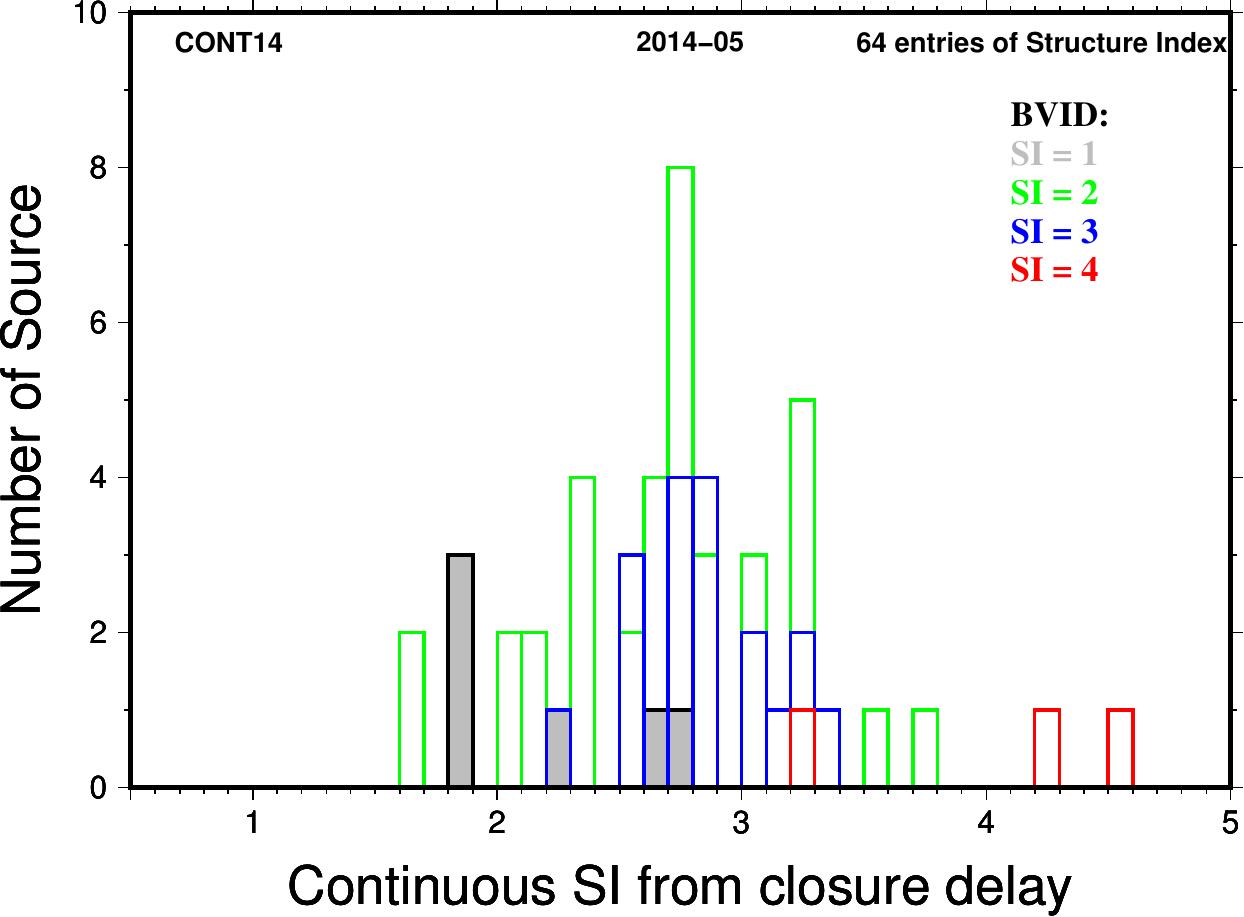}
}
\caption{
The distribution of continuous structure indices of 64 radio sources. 
Each color represents one group of structure index in the BVID and each group is shown independently.}
\label{si65}       
\end{figure}

Integer structure indices for the 65 radio sources were derived according to equation \ref{eq_index} 
and shown in column 11
of Table \ref{tab1} labeled as closure delay (CD1), while the continuous structure indices 
calculated from equation \ref{eq_ind2} based on closure delays are shown in the last column referred to as CD2.
For comparison, the most recent 
structure indices at X band from the BVID, which unfortunately are integer,
are presented in column 10. 
The BVID structure indices 
are in the time range of April 2004 to July 2013 with an average of 2011,
about three years earlier than CONT14 observations.
There is one source, 0637$-$752, that is not found in the BVID 
and it was identified to have structure index
of 3 at X~band from closure delays. Structure indices 
of 14 radio sources were found to have increased, either from 1 to 2 
(3 sources) or 2 to 3 (11 sources); those of 13 radio sources
have decreased, from 3 to 2 (10 sources), 2 to 1 (2 sources), 
or 4 to 3 (1 source); while the remaining 37 radio sources have the same structure indices. 
The most frequent changes between our identified structure indices and those from the BVID are 
the transitions between structure indices of 2 and 3.
The distribution of continuous structure indices are presented in Fig. \ref{si65}.
The rms of the differences between the continuous structure indices from closure delays and those
from the BVID is 0.61. 

%
%
\begin{table*}
\caption{
Statistics of closure quantities and structure indices for 65 radio sources in CONT14 observations. Column~2 is 
the total number of triangles and
column~7 is the total number of quadrangles 
for each source formed based on these observations.}
\label{tab1}       
\begin{tabular}{lrrrrrrccccc}
\hline\noalign{\smallskip}
SOURCE &$N_{\mbox{\scriptsize tri}}$& $\left|\tau_{\mbox{\scriptsize closure}}\right|^{\mbox{\scriptsize med}}$ & $\tau_{\mbox{\scriptsize closure}}^{\mbox{\scriptsize rms}}$& $\left|\phi_{\mbox{\scriptsize closure}}\right|^{\mbox{\scriptsize med}}$ & $\phi_{\mbox{\scriptsize closure}}^{\mbox{\scriptsize rms}}$ &$N_{\mbox{\scriptsize qua}}$ & $\left|\ln{A}_{\mbox{\scriptsize closure}}\right|^{\mbox{\scriptsize med}}$ & $(\ln{A}_{\mbox{\scriptsize closure}})^{\mbox{\scriptsize rms}}$& \multicolumn{3}{c}{structure index} \\
&&(ps)&(ps)&(deg.)&(deg.)& && &BVID& CD1 & CD2 \\
\noalign{\smallskip}\hline\noalign{\smallskip}
 0014$+$813 &   15141 &       42.7&    64.9 &       18.8&   65.9 &   69062 &     0.548 &  0.961  & 2   & 3 &     3.8 \\
 0016$+$731 &   23307 &        5.7&    10.0 &        3.3&    6.1 &  114970 &     0.097 &  0.164  & 2   & 1 &     1.7 \\
 0017$+$200 &    3380 &       15.3&    33.1 &        6.8&   16.2 &   12585 &     0.209 &  0.360  & 2   & 2 &     2.7 \\
 0059$+$581 &   33755 &        6.5&    12.8 &        3.4&    7.7 &  167895 &     0.143 &  0.240  & 1   & 1 &     1.9 \\
 0106$+$013 &    3279 &       16.3&    36.5 &        9.4&   19.0 &    9668 &     0.251 &  0.591  & 1   & 2 &     2.8 \\
 0119$+$115 &     980 &       17.9&    31.1 &        5.1&    8.6 &    1555 &     0.206 &  0.469  & 2   & 2 &     2.9 \\
 0212$+$735 &   18631 &       28.2&    69.5 &       25.1&   54.7 &   92255 &     0.578 &  1.079  & 4   & 3 &     3.3 \\
 0229$+$131 &    3136 &       31.7&    83.8 &       17.2&   58.9 &   10937 &     0.418 &  0.806  & 2   & 3 &     3.5 \\
 0322$+$222 &     640 &      23.7 &    47.5 &      13.7 &   25.0 &    2075 &     0.262 &  0.377  & 2   & 3 &     3.2 \\
 0332$-$403 &    2676 &       19.2&    31.4 &        9.4&   17.4 &    4361 &     0.172 &  0.317  & 2   & 3 &     3.0 \\
 0420$-$014 &   10479 &       11.0&    20.0 &        7.1&   13.4 &   40717 &     0.186 &  0.374  & 2   & 2 &     2.4 \\
 0454$-$234 &    5681 &       13.1&    21.3 &        4.0&    7.0 &   13608 &     0.115 &  0.168  & 2   & 2 &     2.6 \\
 0528$+$134 &     620 &      19.1 &    34.1 &      10.0 &   17.9 &     951 &     0.171 &  0.239  & 3   & 3 &     2.9 \\
 0529$+$483 &   15027 &       24.0&    44.2 &        9.4&   15.9 &   65080 &     0.190 &  0.317  & 2   & 3 &     3.2 \\
 0537$-$441 &    3780 &       13.6&    21.7 &        6.6&   11.0 &    6412 &     0.141 &  0.230  & 2   & 2 &     2.6 \\
 0607$-$157 &    7432 &       26.6&    87.2 &       10.8&   47.0 &   19547 &     0.646 &  1.102  & 3   & 3 &     3.3 \\
 0637$-$752 &    1174 &       31.5&    86.3 &       15.8&   58.4 &    1230 &     0.342 &  1.031  & -   & 3 &     3.4 \\
 0642$+$449 &   22154 &       21.1&   119.0 &       11.4&   48.4 &   98559 &     0.462 &  0.824  & 2   & 3 &     3.0 \\
 0657$+$172 &   10946 &       16.4&    30.4 &        8.6&   18.4 &   51460 &     0.187 &  0.315  & 2   & 2 &     2.8 \\
 0716$+$714 &   26767 &        9.3&    16.2 &        3.7&    6.2 &  132691 &     0.077 &  0.130  & 1   & 2 &     2.2 \\
 0727$-$115 &   11224 &        6.3&     8.2 &        1.8&    2.9 &   35874 &     0.121 &  0.221  & 1   & 1 &     1.8 \\
 0738$+$313 &   13994 &      100.6&   232.3 &       36.4&   75.1 &   67137 &     0.749 &  1.153  & 4   & 4 &     4.6 \\
 0748$+$126 &   16018 &       12.7&    21.8 &        8.1&   13.2 &   72260 &     0.169 &  0.284  & 3   & 2 &     2.5 \\
 0749$+$540 &    5339 &       17.2&    30.0 &       11.1&   20.7 &   20256 &     0.212 &  0.380  & 3   & 2 &     2.8 \\
 0814$+$425 &    8470 &       15.2&    26.1 &        8.1&   12.1 &   36226 &     0.126 &  0.189  & 2   & 2 &     2.7 \\
 0827$+$243 &    3045 &       13.7&    25.0 &        4.9&    8.1 &   10953 &     0.131 &  0.266  & 1   & 2 &     2.6 \\
 0919$-$260 &     434 &      25.8 &    45.7 &      11.2 &   15.8 &     489 &     0.497 &  0.553  & 3   & 3 &     3.3 \\
 0955$+$476 &    7032 &       11.1&    18.6 &        4.1&    7.3 &   27993 &     0.085 &  0.142  & 2   & 2 &     2.4 \\
 1044$+$719 &    2303 &        5.4&     9.6 &        3.1&    6.5 &    8551 &     0.102 &  0.152  & 2   & 1 &     1.7 \\
 1053$+$815 &     862 &       11.5&    19.0 &        5.4&    8.4 &    2297 &     0.088 &  0.133  & 2   & 2 &     2.4 \\
 1104$-$445 &     298 &      17.8 &    27.7 &      11.4 &   17.9 &     198 &     0.248 &  0.365  & 3   & 2 &     2.9 \\
 1124$-$186 &    5214 &       17.5&    28.6 &        7.6&   13.2 &   13743 &     0.184 &  0.363  & 2   & 2 &     2.9 \\
 1156$+$295 &    1754 &       20.3&    40.5 &        8.7&   23.5 &    6471 &     0.179 &  0.265  & 3   & 3 &     3.0 \\
 1308$+$326 &     316 &      10.8 &    16.9 &       4.8 &    8.3 &     574 &     0.136 &  0.188  & 2   & 2 &     2.4 \\
 1406$-$076 &     282 &      16.0 &    27.4 &       7.4 &   12.9 &     645 &     0.451 &  0.234  & 3   & 2 &     2.8 \\
 1417$+$385 &    2433 &       14.0&    25.1 &        3.8&    7.2 &    9204 &     0.083 &  0.129  & 2   & 2 &     2.6 \\
 1424$-$418 &    5482 &        7.0&    11.0 &        2.6&    5.4 &   10295 &     0.115 &  0.177  & 1   & 1 &     1.9 \\
 1448$+$762 &    1486 &       12.6&    25.3 &        6.1&   11.8 &    4897 &     0.291 &  0.584  & 3   & 2 &     2.5 \\
 1519$-$273 &    3341 &       16.6&    28.3 &        6.1&   10.6 &    6716 &     0.124 &  0.185  & 2   & 2 &     2.8 \\
 1611$+$343 &   15235 &       12.8&    21.0 &        6.5&   11.2 &   69563 &     0.176 &  0.296  & 3   & 2 &     2.5 \\
 1639$-$062 &    5593 &       19.3&    32.8 &        6.2&   11.2 &   16341 &     0.152 &  0.273  & 2   & 3 &     3.0 \\
 1739$+$522 &   31157 &       29.6&    57.8 &       14.0&   33.2 &  150648 &     0.339 &  0.584  & 3   & 3 &     3.4 \\
 1741$-$038 &    6184 &        7.4&    13.3 &        4.9&    8.4 &   18738 &     0.134 &  0.255  & 2   & 2 &     2.0 \\
 1751$+$288 &   10501 &        7.9&    14.8 &        3.9&    8.0 &   47261 &     0.096 &  0.164  & 2   & 2 &     2.1 \\
 1806$+$456 &    3458 &       11.7&    24.9 &        4.2&    8.7 &   12480 &     0.138 &  0.278  & 2   & 2 &     2.5 \\
 1846$+$322 &   14942 &       28.0&    63.2 &        7.4&   33.0 &   66839 &     0.309 &  0.551  & 2   & 3 &     3.3 \\
 1921$-$293 &    1225 &        7.7&    12.8 &        2.1&    3.7 &    1686 &     0.178 &  0.509  & 2   & 2 &     2.0 \\
 1954$-$388 &    1496 &       17.7&    28.2 &        9.9&   14.9 &    2214 &     0.160 &  0.295  & 3   & 2 &     2.9 \\
 2000$+$472 &   13718 &       17.2&    33.8 &        4.9&   10.3 &   61370 &     0.231 &  0.413  & 3   & 2 &     2.8 \\
 2052$-$474 &     369 &      25.2 &    40.6 &       6.9 &   10.2 &     286 &     0.261 &  0.350  & 2   & 3 &     3.2 \\
 2059$+$034 &     556 &      23.9 &    33.8 &       8.5 &   19.9 &    2008 &     0.357 &  0.450  & 2   & 3 &     3.2 \\
 2113$+$293 &     281 &      21.6 &    31.8 &       7.8 &   12.0 &     667 &     0.256 &  0.351  & 3   & 3 &     3.1 \\
 2121$+$053 &     168 &      18.3 &    27.2 &      14.3 &   20.1 &     352 &     0.254 &  0.347  & 3   & 3 &     2.9 \\
 2145$+$067 &    4631 &       15.9&    30.0 &       13.5&   39.0 &   14173 &     0.304 &  0.687  & 2   & 2 &     2.8 \\
 2201$+$315 &      84 &      14.2 &    22.5 &       5.1 &    7.4 &      86 &     0.239 &  0.324  & 2   & 2 &     2.7 \\
 2209$+$236 &    4018 &       18.8&    46.2 &        7.3&   25.2 &   16516 &     0.321 &  0.574  & 2   & 3 &     2.9 \\
 2214$+$350 &    3377 &       16.5&    29.7 &        5.6&    9.7 &   12589 &     0.111 &  0.170  & 2   & 2 &     2.8\\ 
 2227$-$088 &    1732 &       12.6&    23.5 &        7.0&   11.4 &    5597 &     0.145 &  0.267  & 2   & 2 &     2.5\\ 
 2234$+$282 &     293 &       9.8 &    15.6 &       3.9 &    6.4 &     526 &     0.148 &  0.146  & 3   & 2 &     2.3\\ 
 2255$-$282 &     531 &      13.8 &    23.5 &       4.6 &   11.2 &     525 &     0.139 &  0.152  & 2   & 2 &     2.6\\ 
 2309$+$454 &     752 &       8.3 &    14.1 &       3.9 &    8.3 &    2491 &     0.091 &  0.142  & 2   & 2 &     2.1\\ 
 2355$-$106 &    1078 &       16.1&    27.3 &        6.0&   12.2 &    3257 &     0.150 &  0.295  & 2   & 2 &     2.8\\ 
 3C309.1    &     350 &      63.5 &    98.3 &      17.6 &   27.0 &     388 &     0.481 &  0.756  & 4   & 4 &     4.2\\ 
 3C371      &   27133 &       20.4&    46.6 &       18.0&   27.9 &  136392 &     0.307 &  0.709  & 3   & 3 &     3.0\\ 
 4C39.25    &    9861 &       17.0&    41.9 &       13.0&   33.2 &   39947 &     0.522 &  0.866  & 3   & 2 &     2.8\\ 
\noalign{\smallskip}\hline
\end{tabular}
\end{table*}
\begin{figure*}
\begin{center}
 \resizebox{0.314\textwidth}{!}{\includegraphics{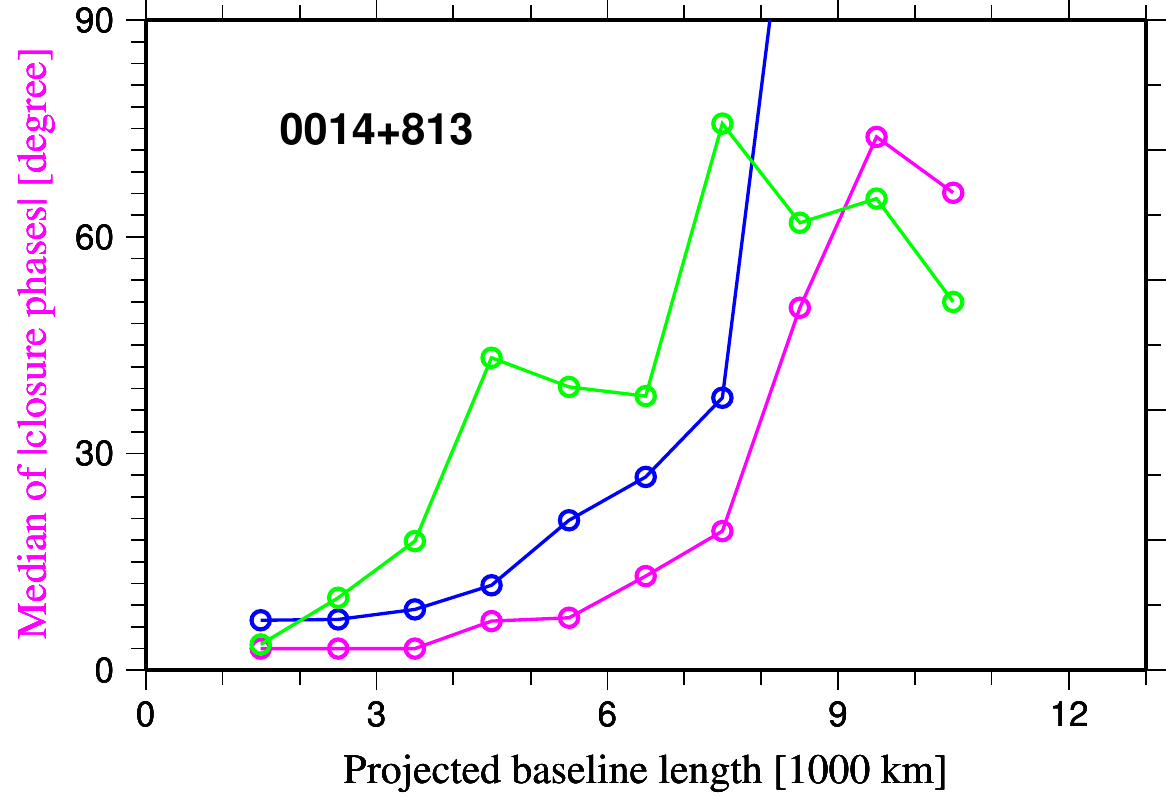}} 
\resizebox{0.28\textwidth}{!}{\includegraphics{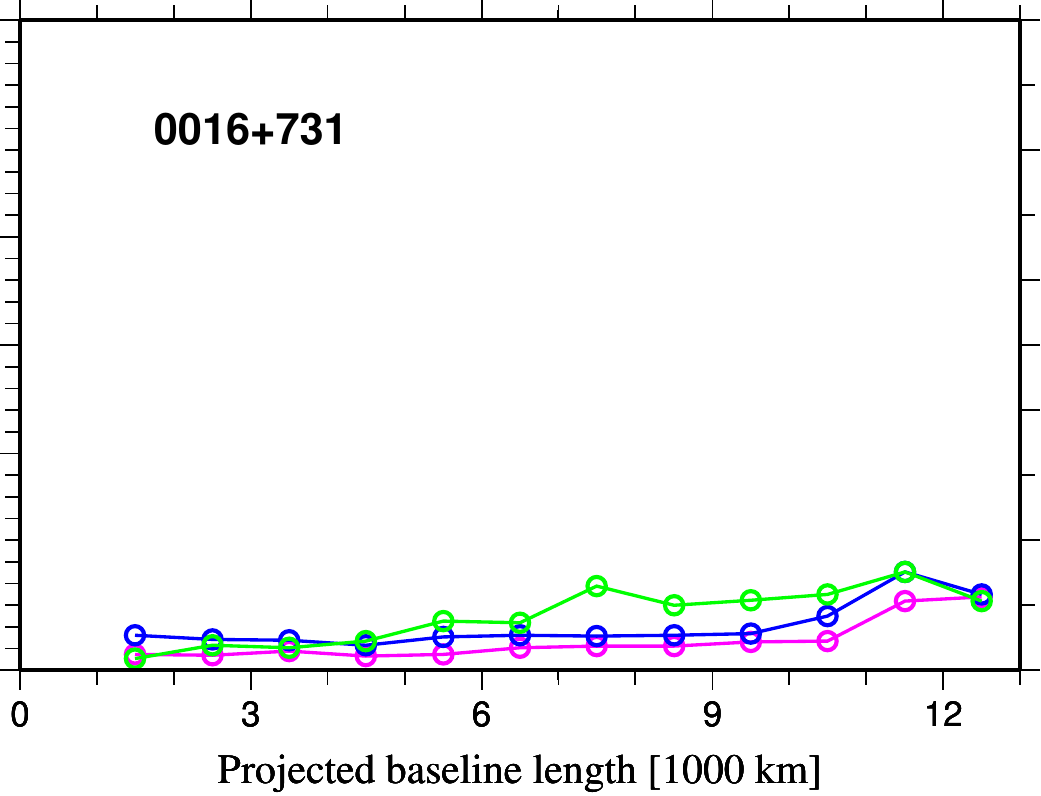}} 
\resizebox{0.368\textwidth}{!}{\includegraphics{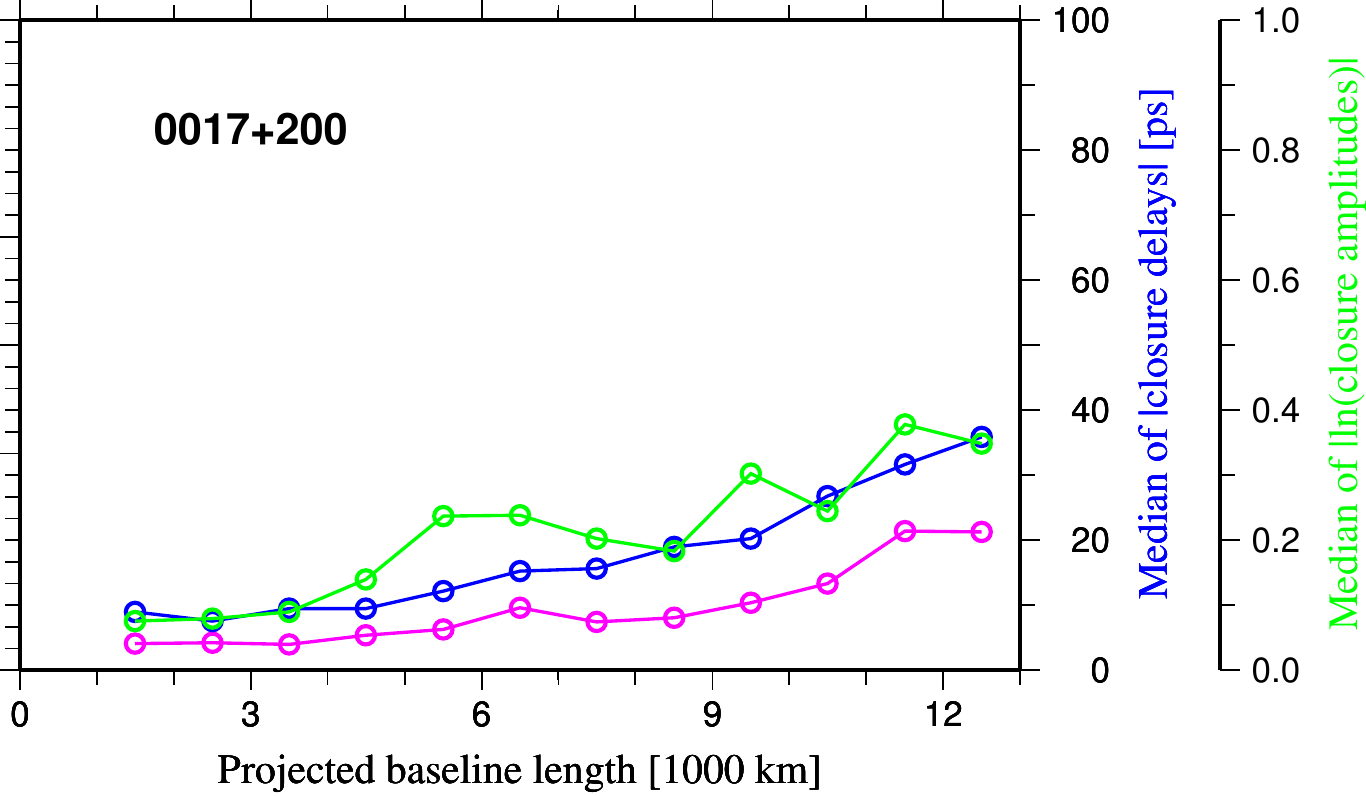}}  
 \resizebox{0.314\textwidth}{!}{\includegraphics{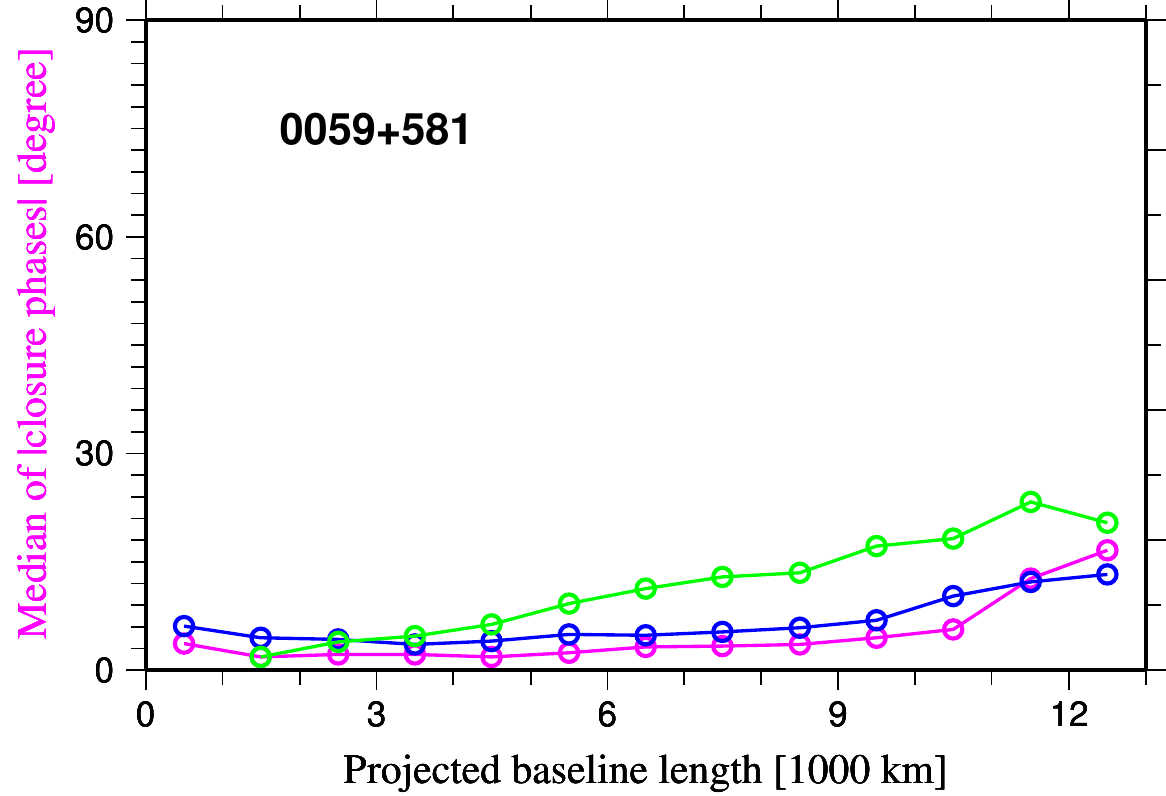}} 
\resizebox{0.28\textwidth}{!}{\includegraphics{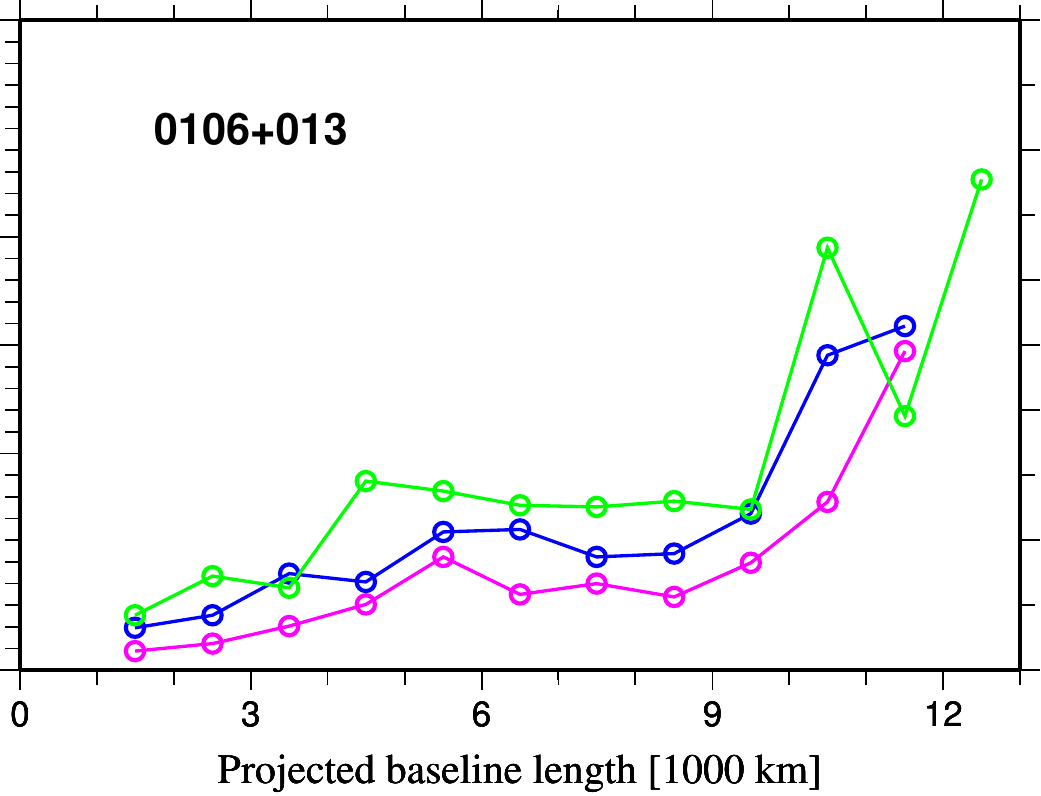}} 
\resizebox{0.368\textwidth}{!}{\includegraphics{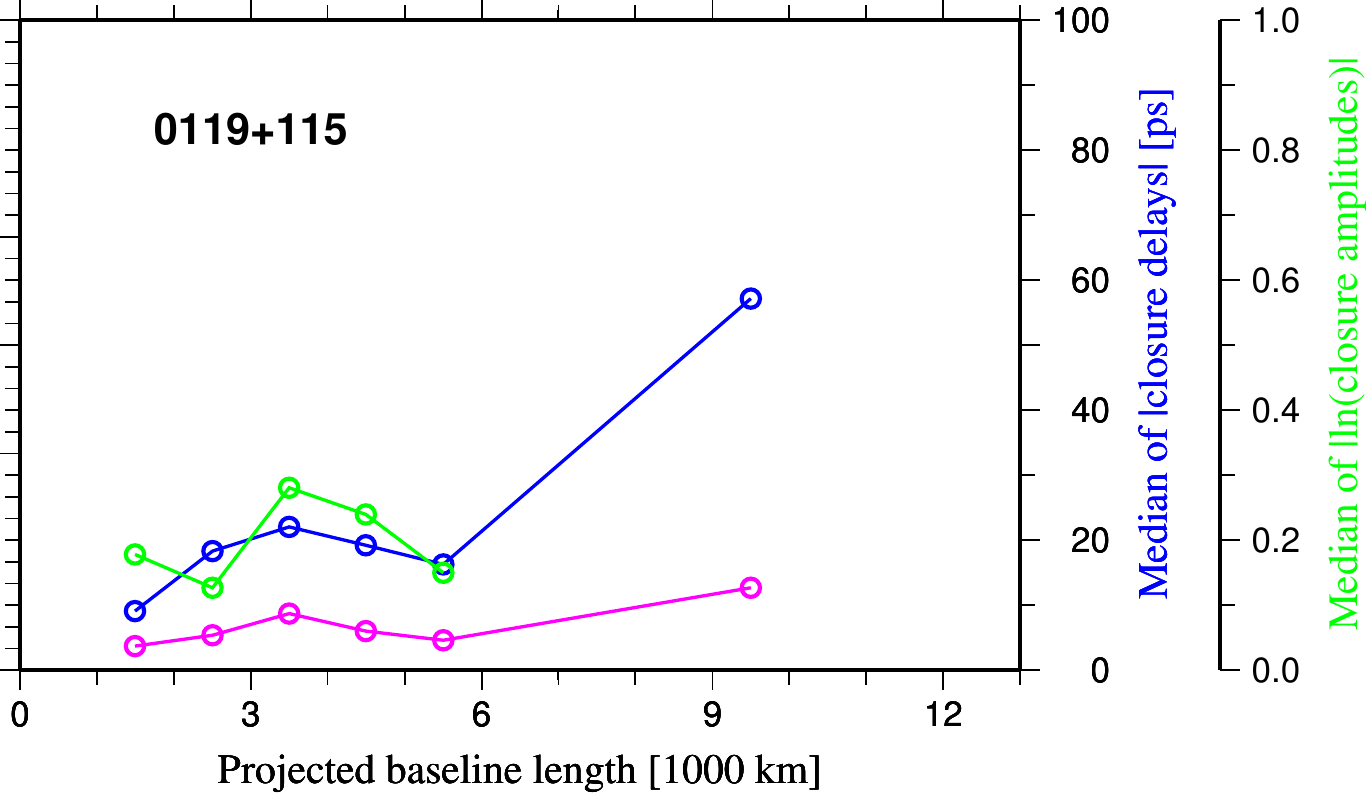}}   
 \resizebox{0.314\textwidth}{!}{\includegraphics{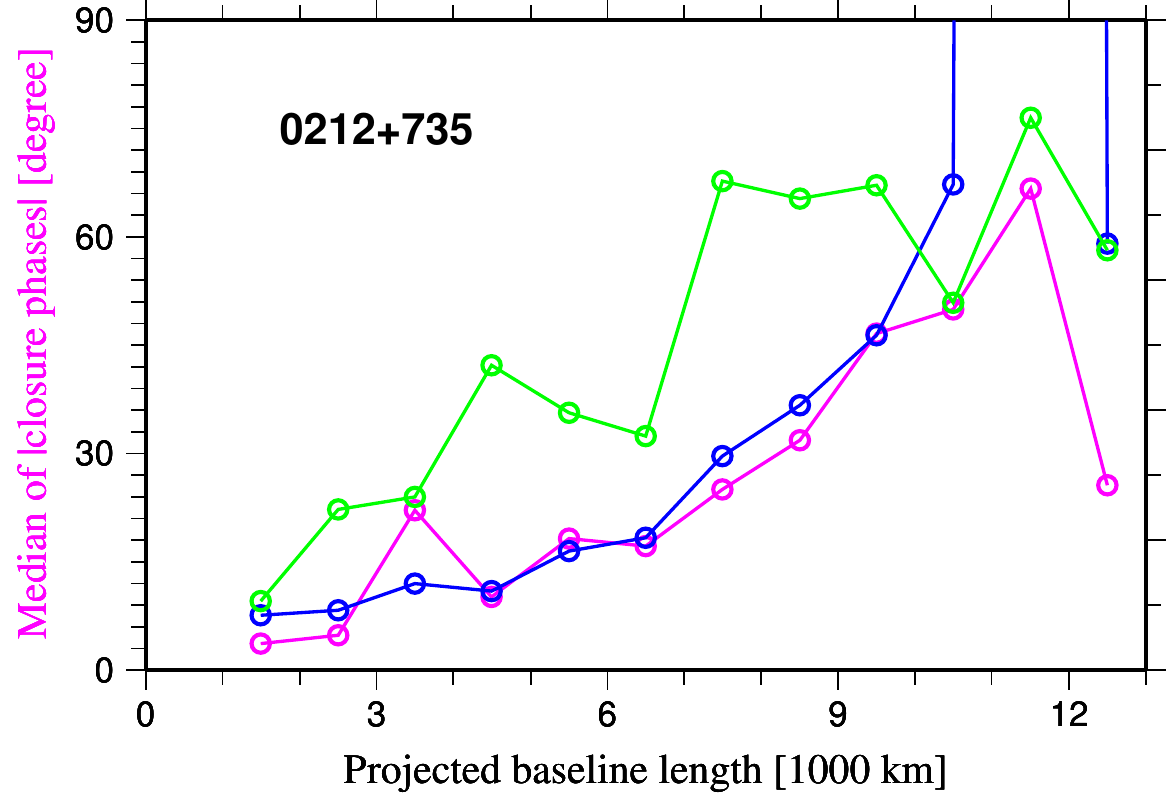}} 
\resizebox{0.28\textwidth}{!}{\includegraphics{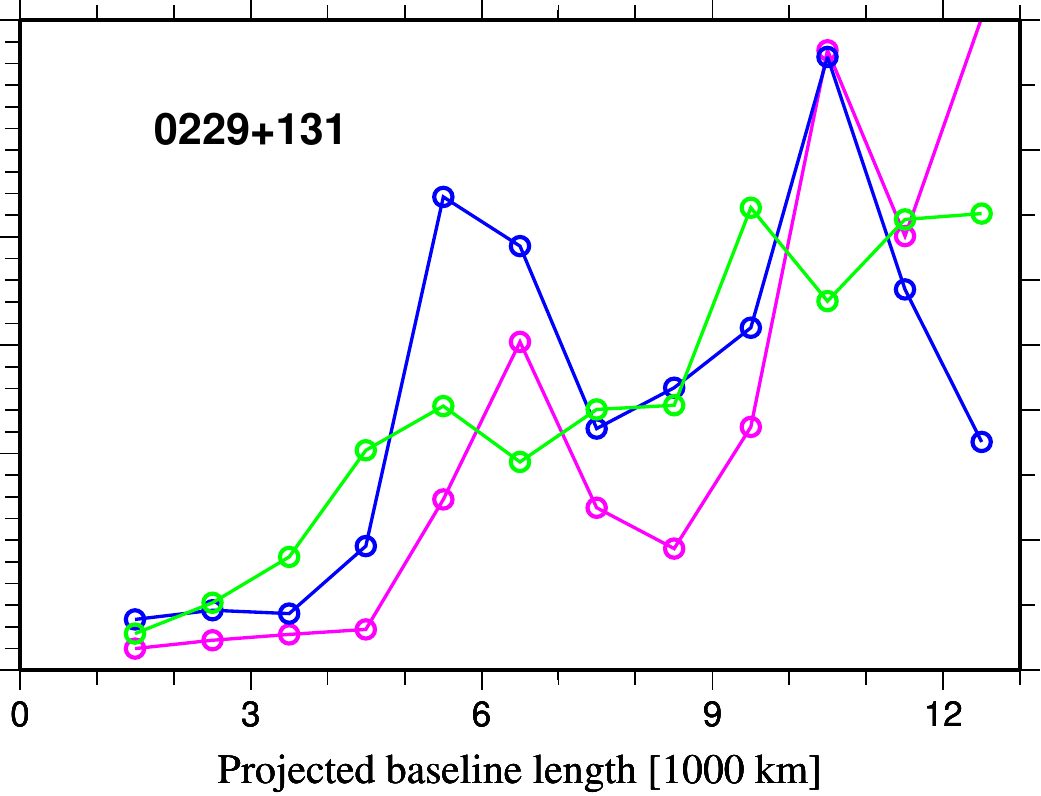}} 
\resizebox{0.368\textwidth}{!}{\includegraphics{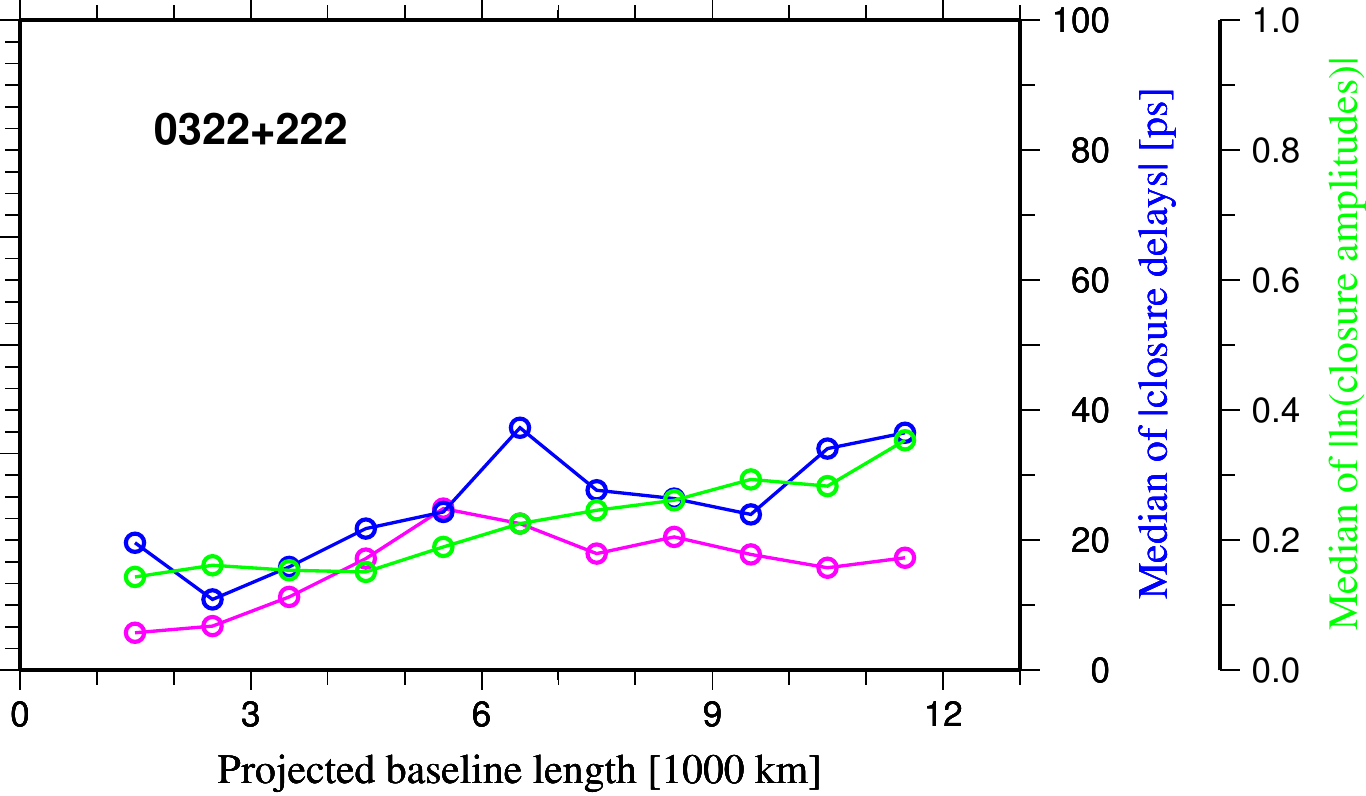}}   
 \resizebox{0.314\textwidth}{!}{\includegraphics{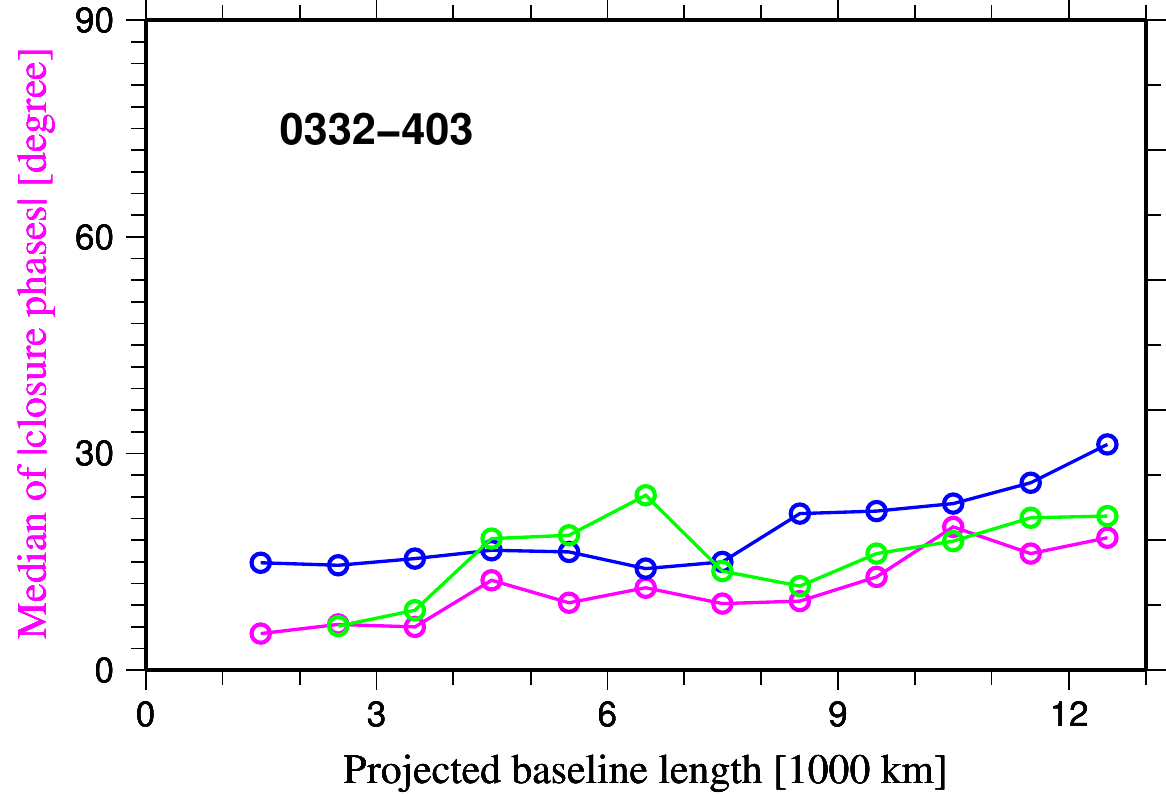}} 
\resizebox{0.28\textwidth}{!}{\includegraphics{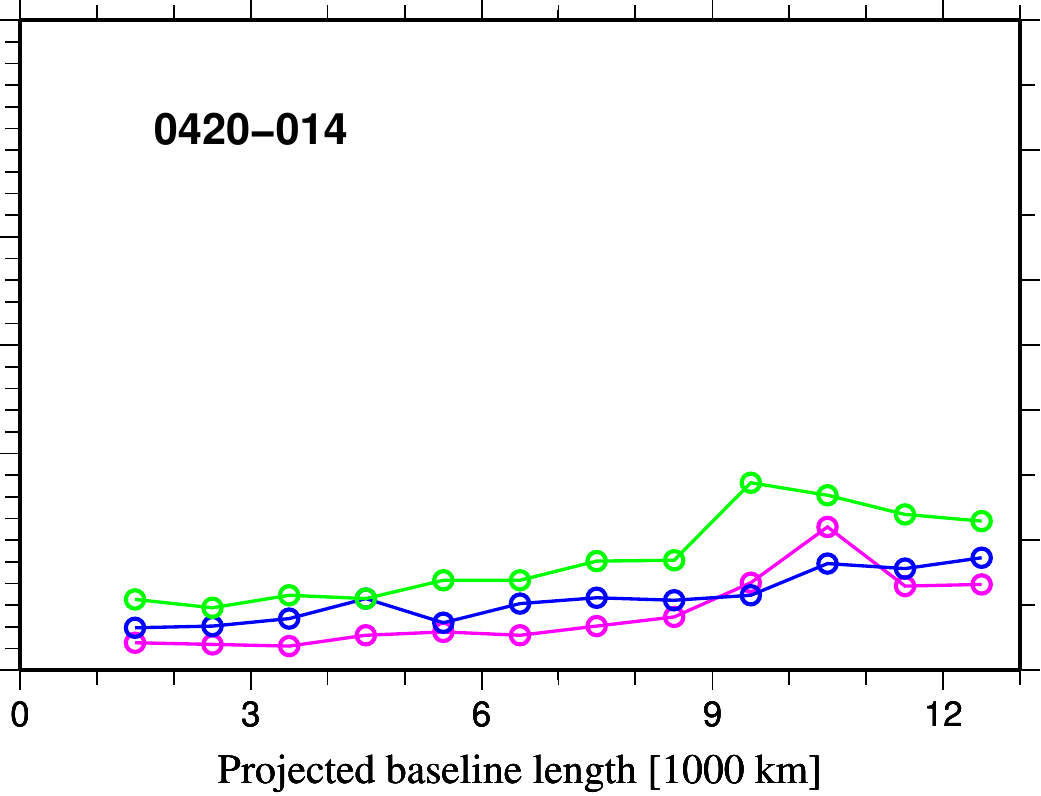}} 
\resizebox{0.368\textwidth}{!}{\includegraphics{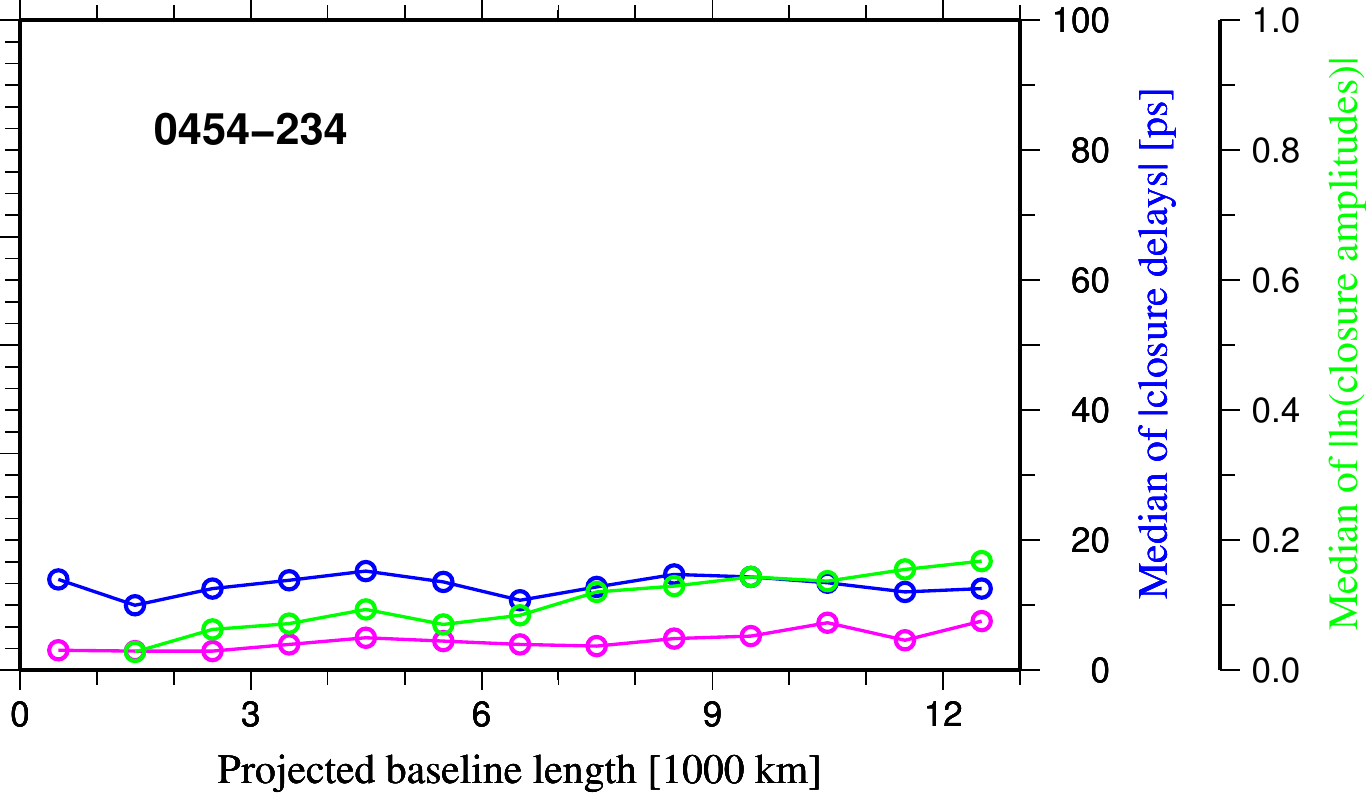}}   
 \resizebox{0.314\textwidth}{!}{\includegraphics{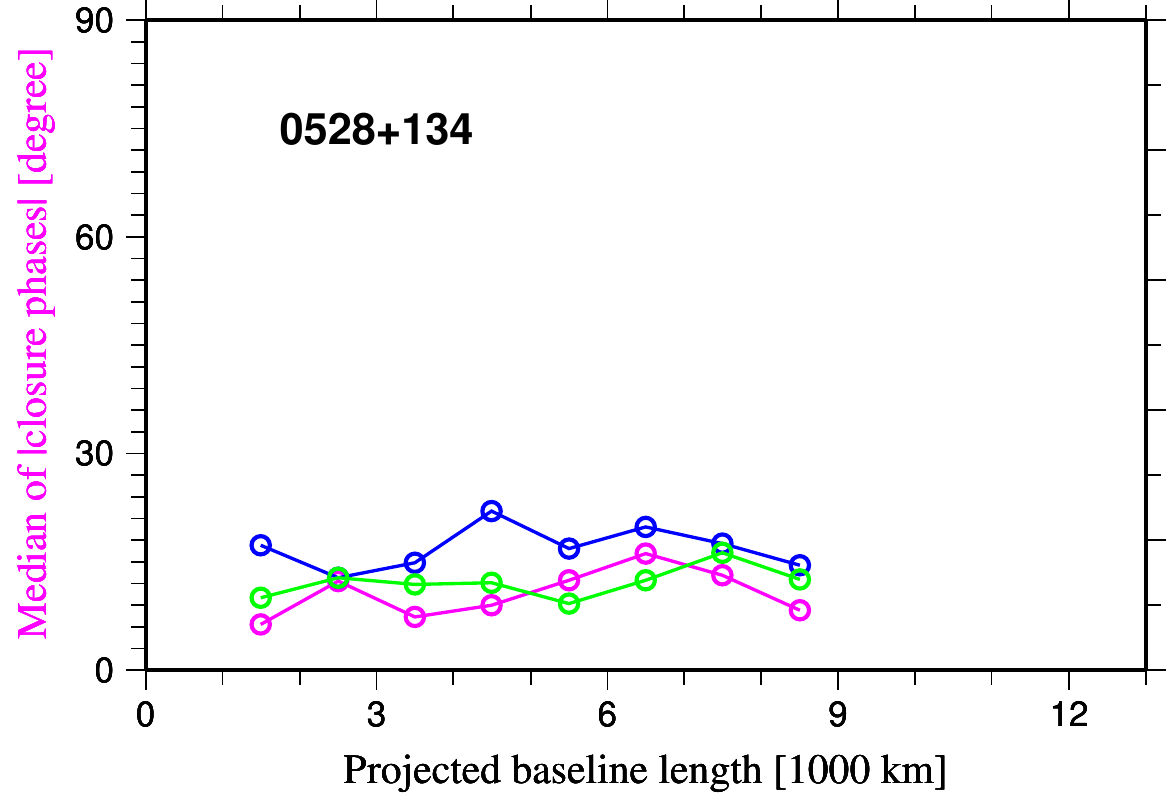}} 
\resizebox{0.28\textwidth}{!}{\includegraphics{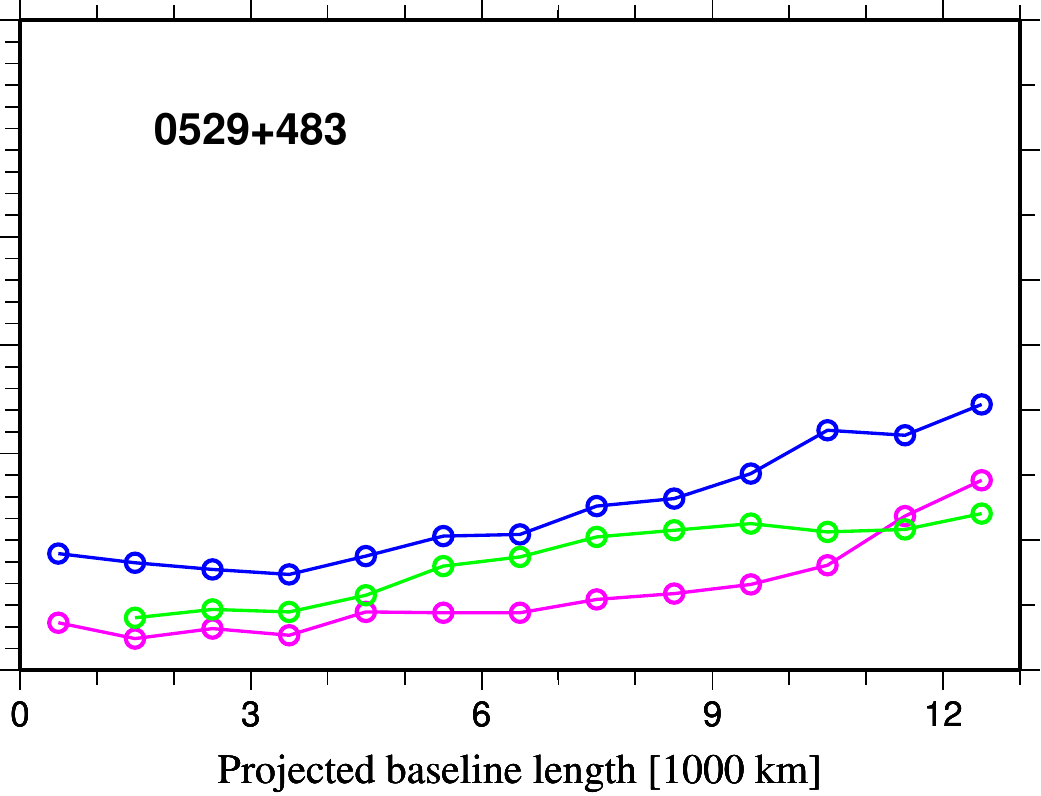}} 
\resizebox{0.368\textwidth}{!}{\includegraphics{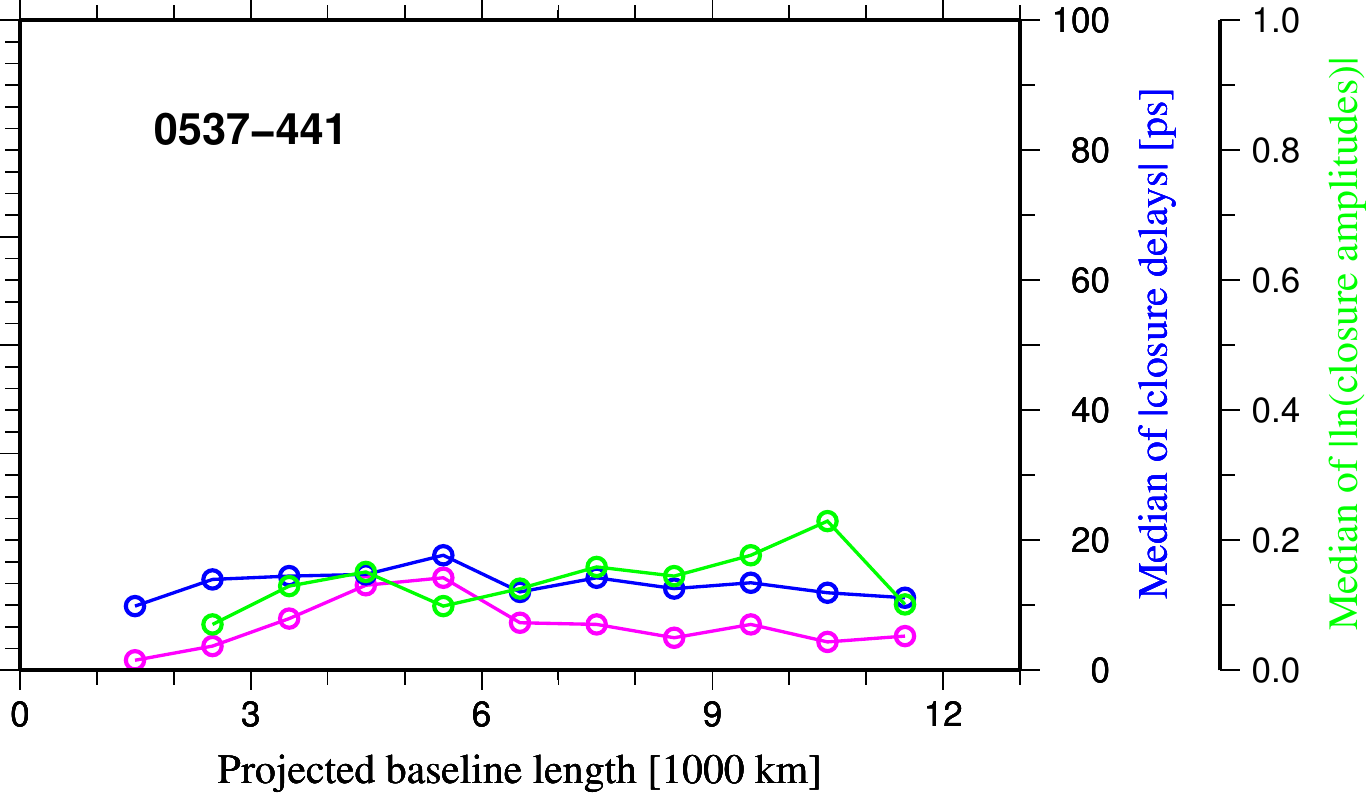}} 
 \resizebox{0.314\textwidth}{!}{\includegraphics{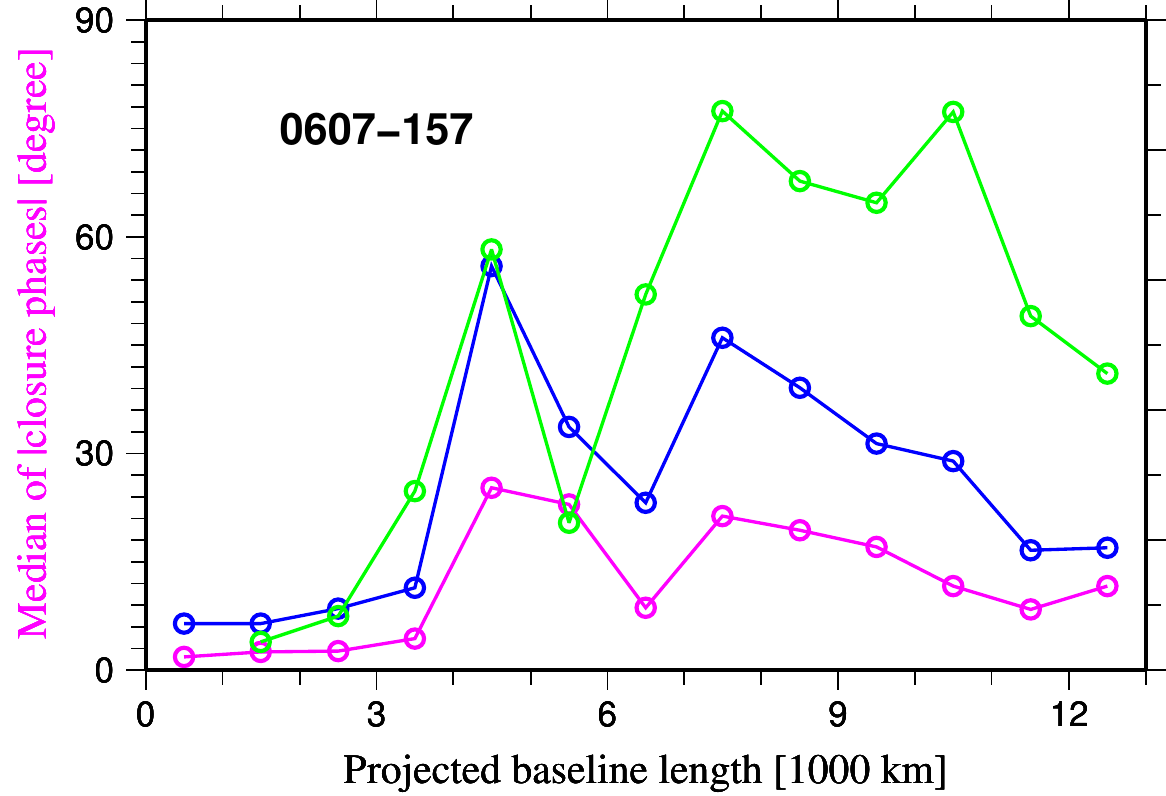}} 
\resizebox{0.28\textwidth}{!}{\includegraphics{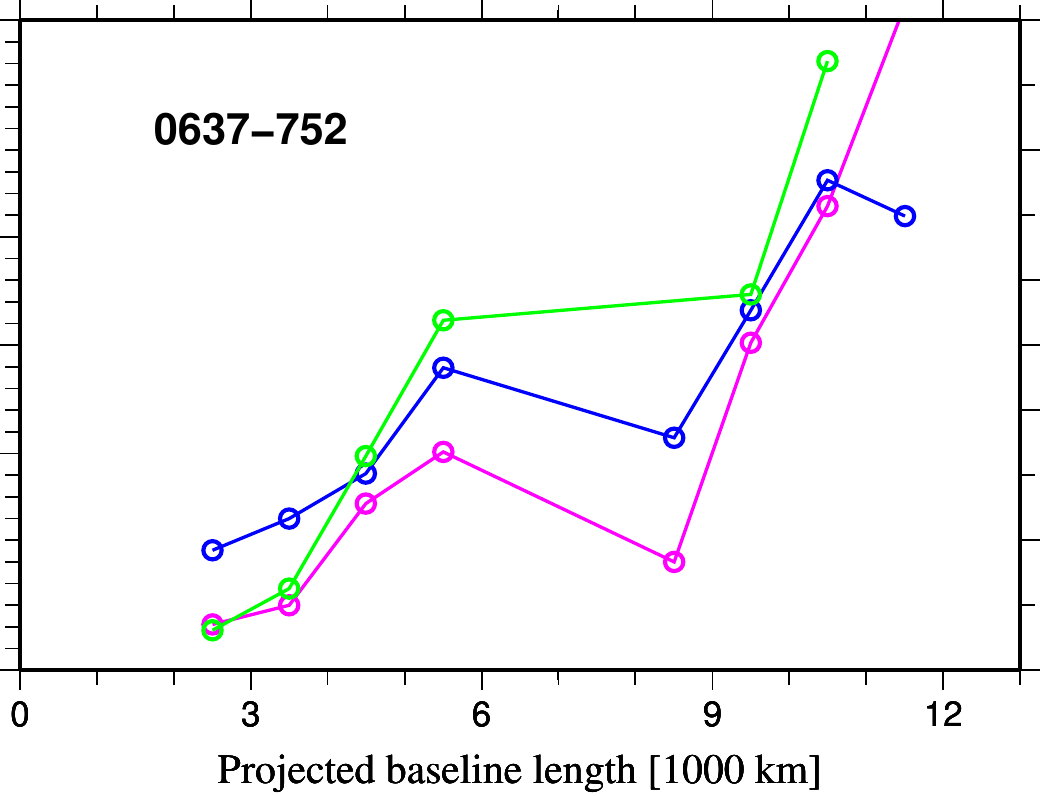}} 
\resizebox{0.368\textwidth}{!}{\includegraphics{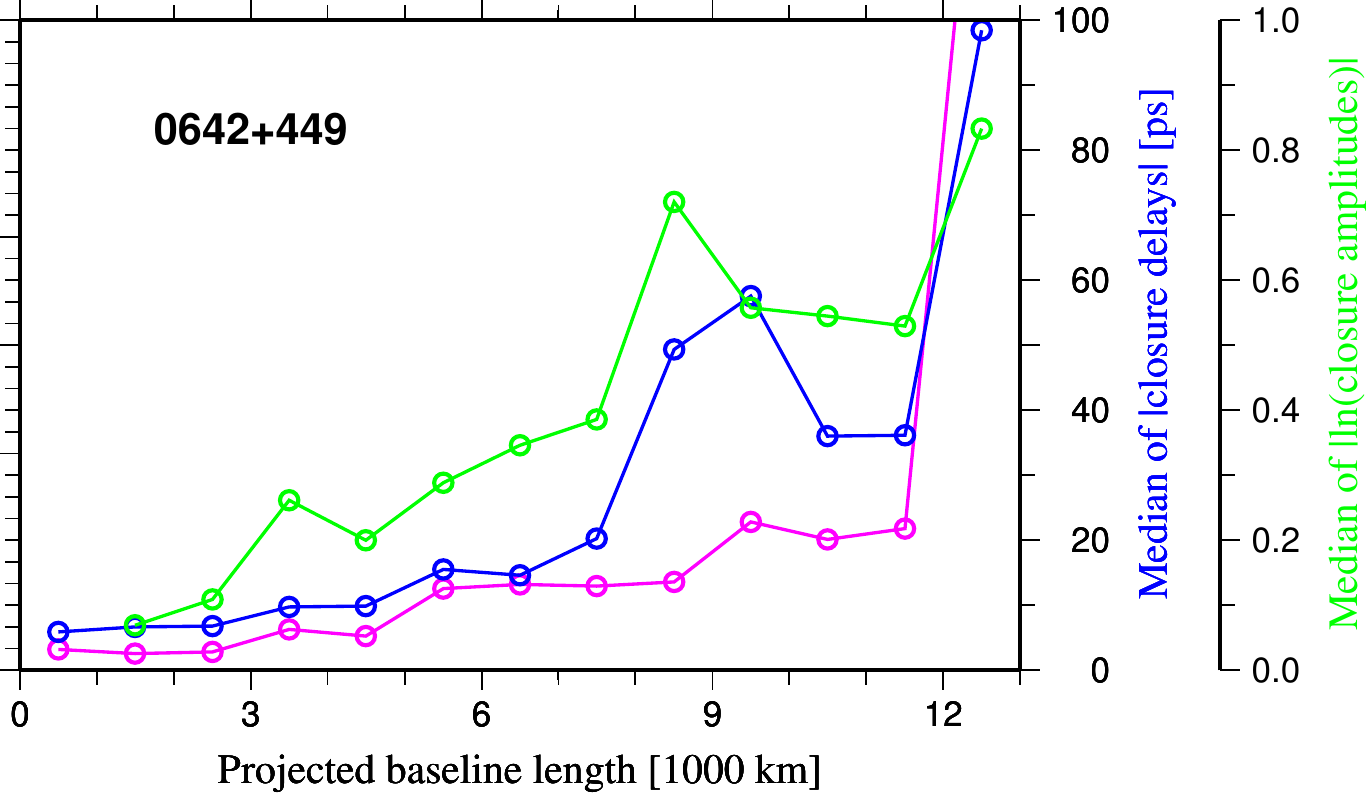}} 
\caption{Variations of the median absolute values of closure delays, closure phases, and natural logarithms of closure amplitudes with respect to the longest baseline length in the triangle or the quadrangle for the 54 most observed radio sources in CONT14.}                                           
\label{closure_rms}                                           
\end{center}                                                   
\end{figure*}                                                  
\begin{figure*}                                                
\begin{center}                                                 
 \resizebox{0.32\textwidth}{!}{\includegraphics{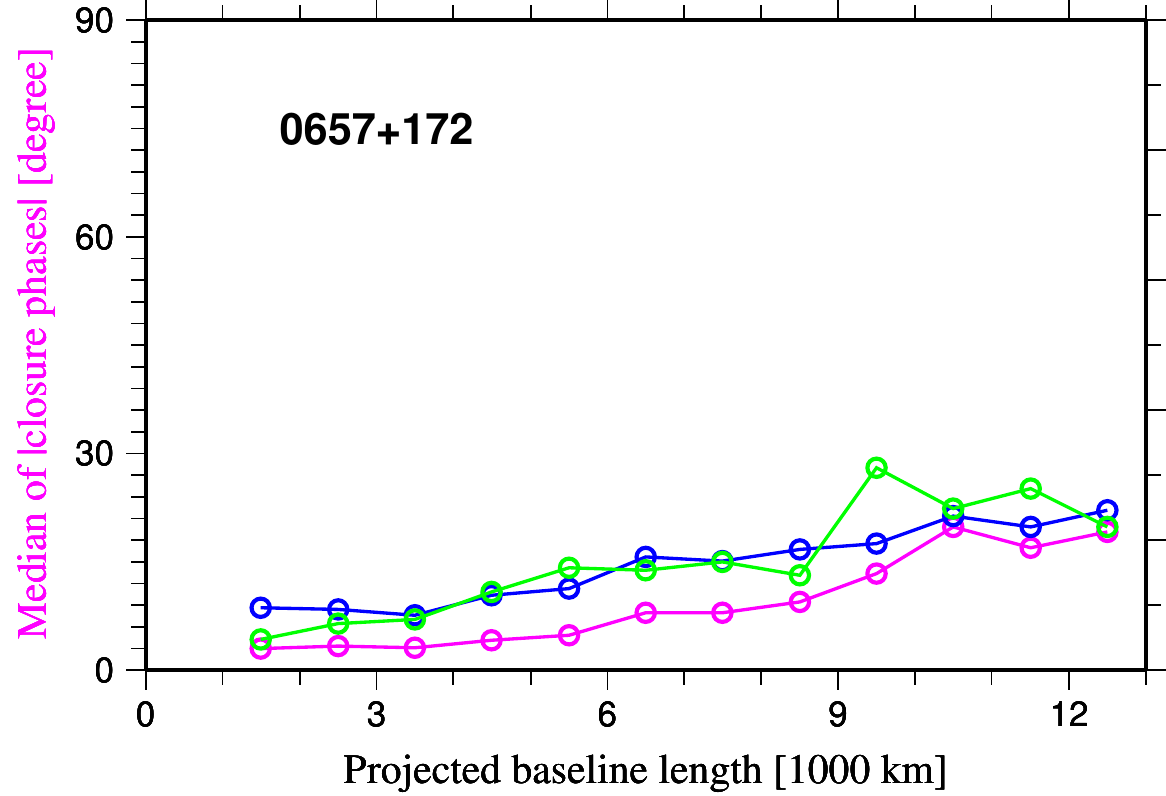}} 
\resizebox{0.285\textwidth}{!}{\includegraphics{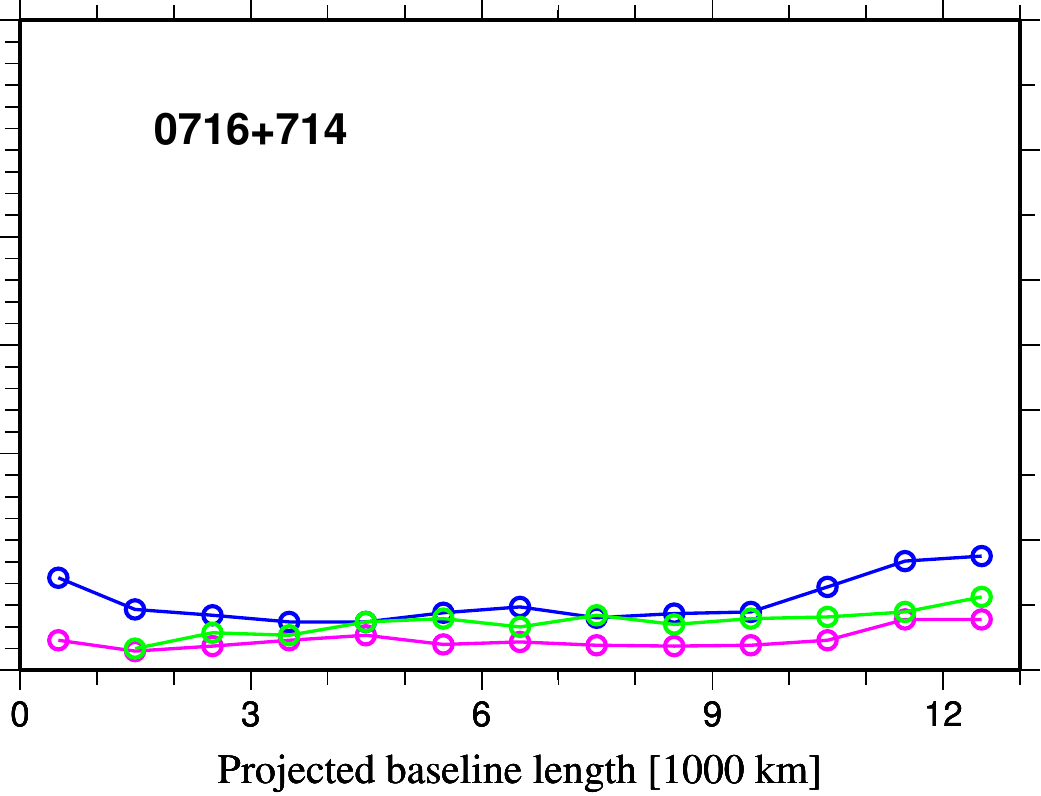}} 
\resizebox{0.376\textwidth}{!}{\includegraphics{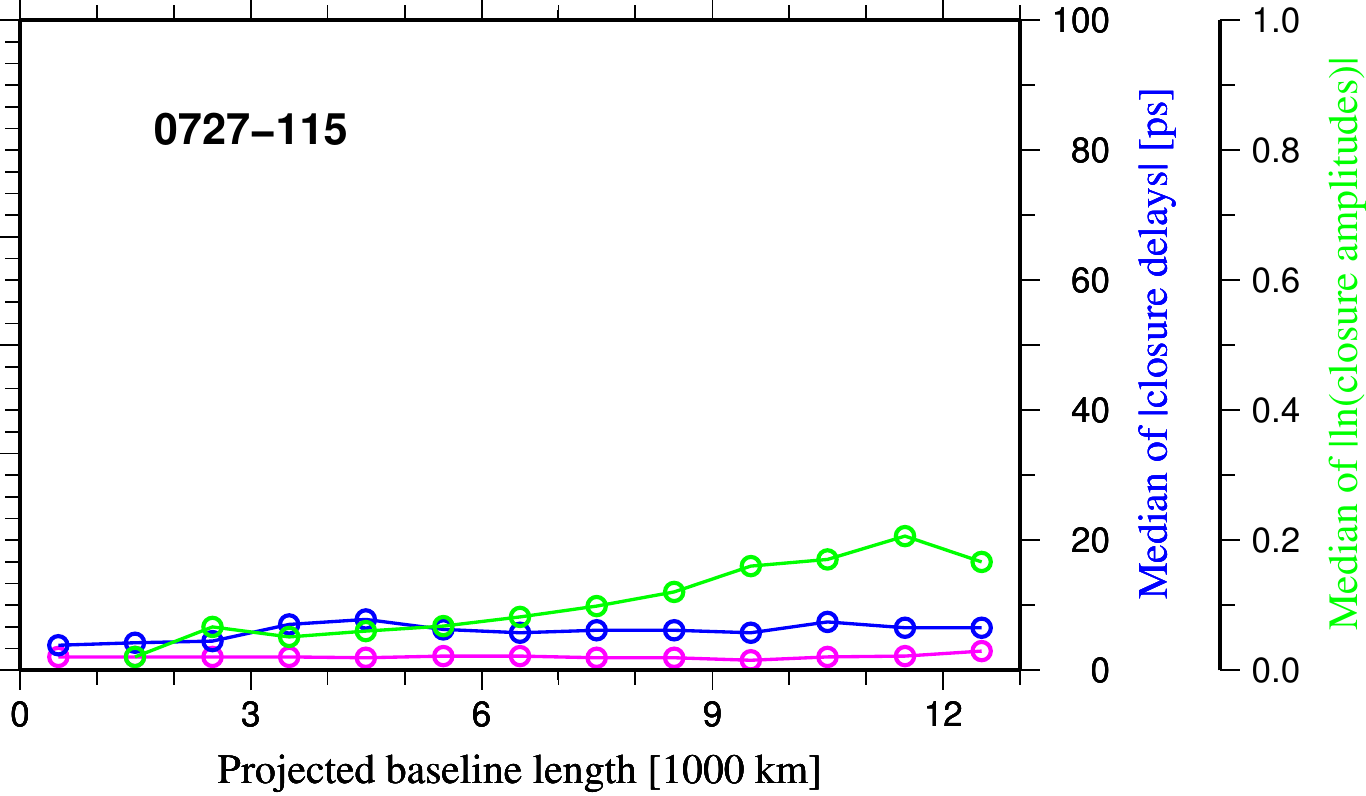}}   
 \resizebox{0.32\textwidth}{!}{\includegraphics{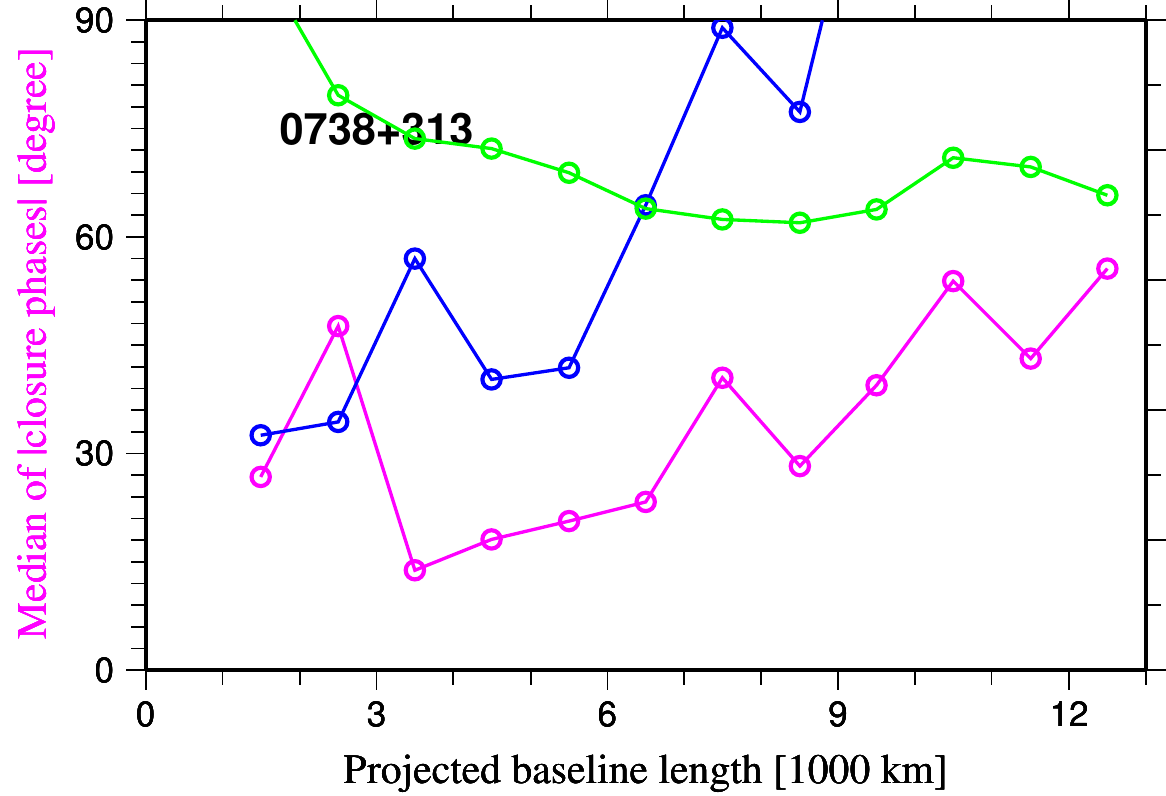}} 
\resizebox{0.285\textwidth}{!}{\includegraphics{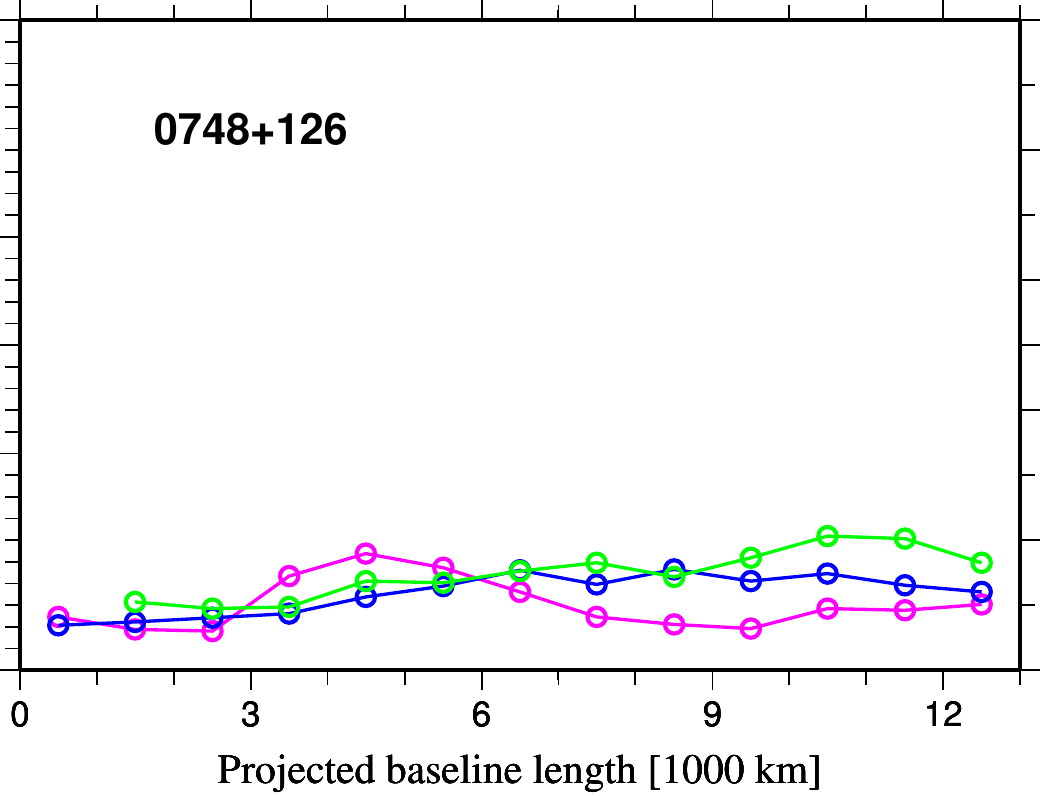}} 
\resizebox{0.376\textwidth}{!}{\includegraphics{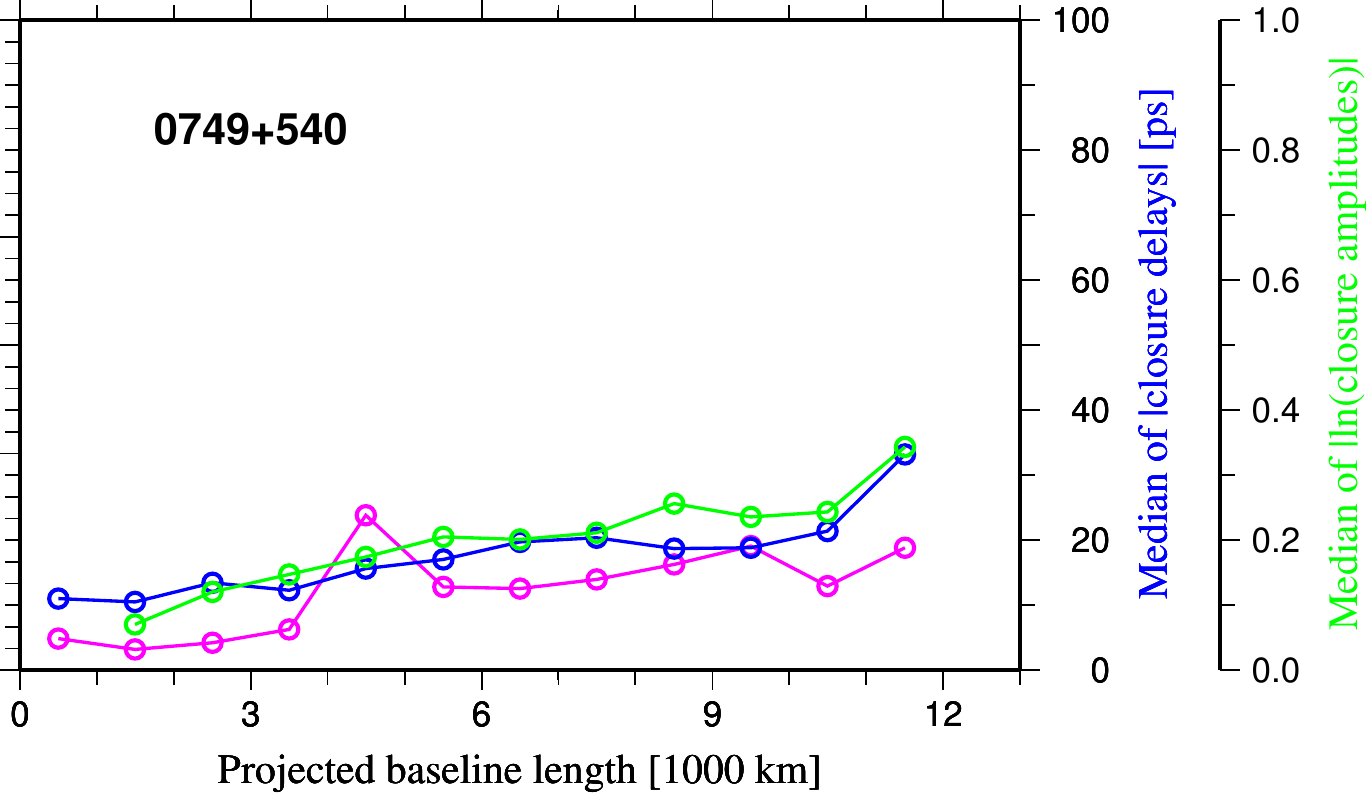}}   
 \resizebox{0.32\textwidth}{!}{\includegraphics{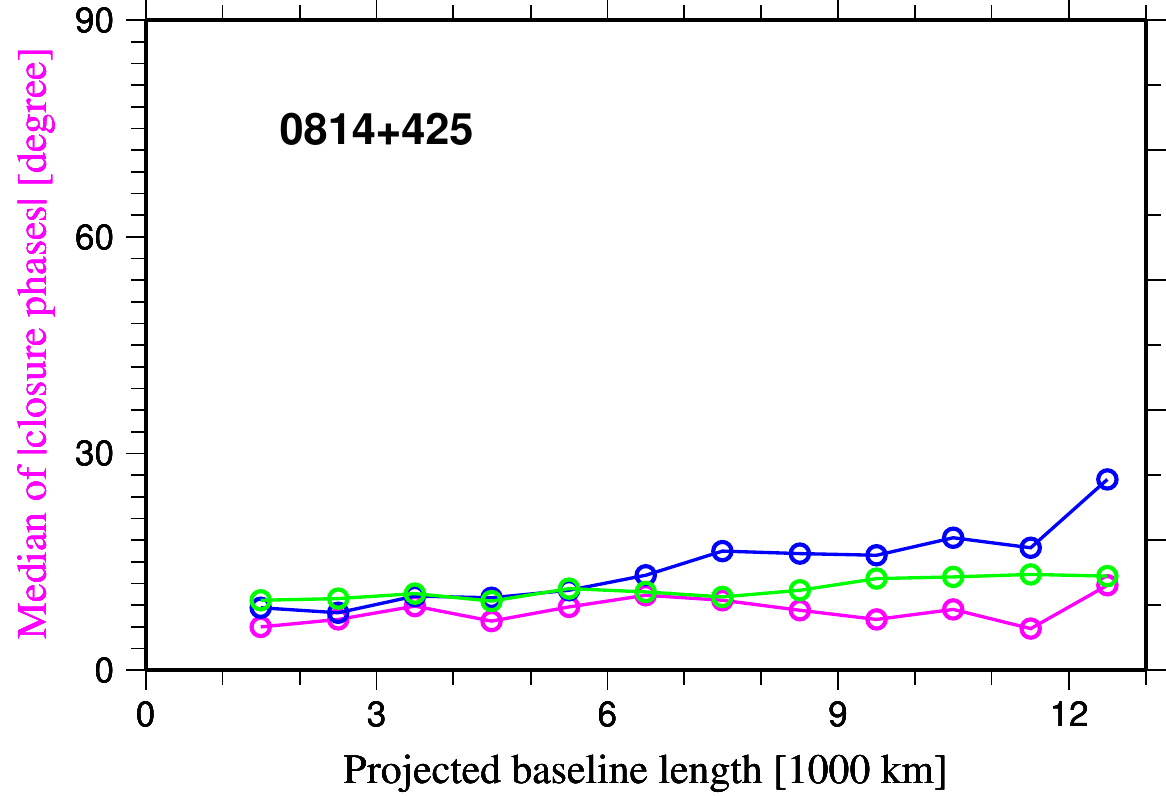}} 
\resizebox{0.285\textwidth}{!}{\includegraphics{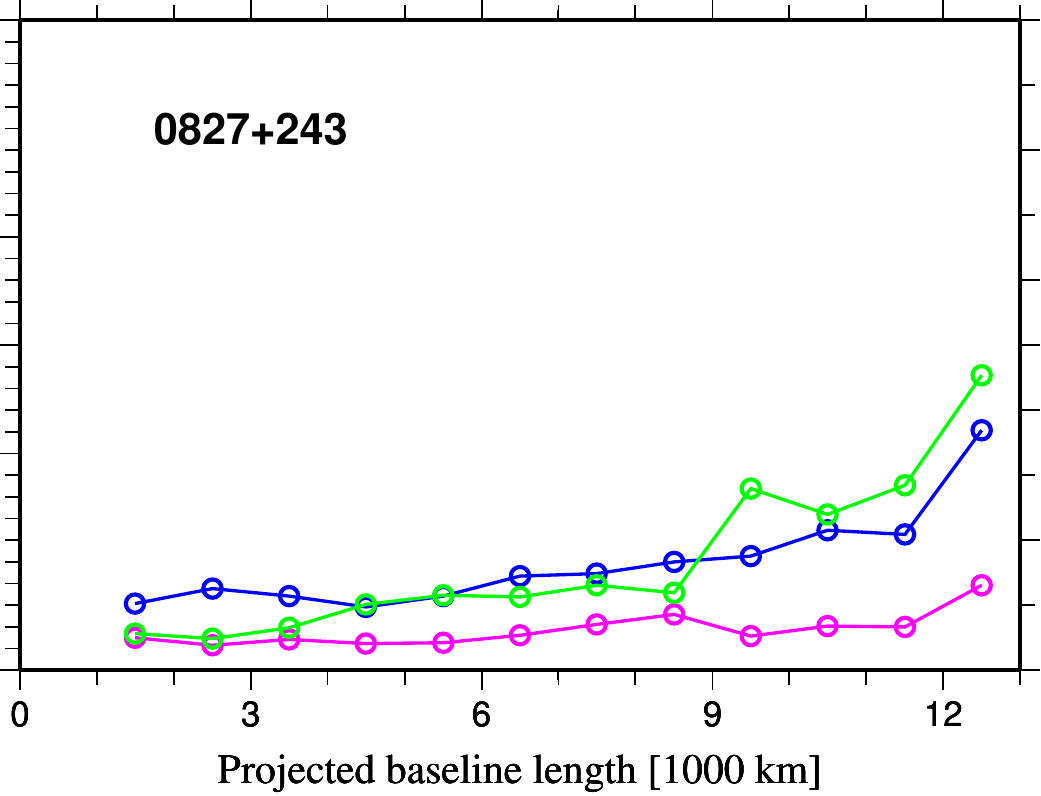}} 
\resizebox{0.376\textwidth}{!}{\includegraphics{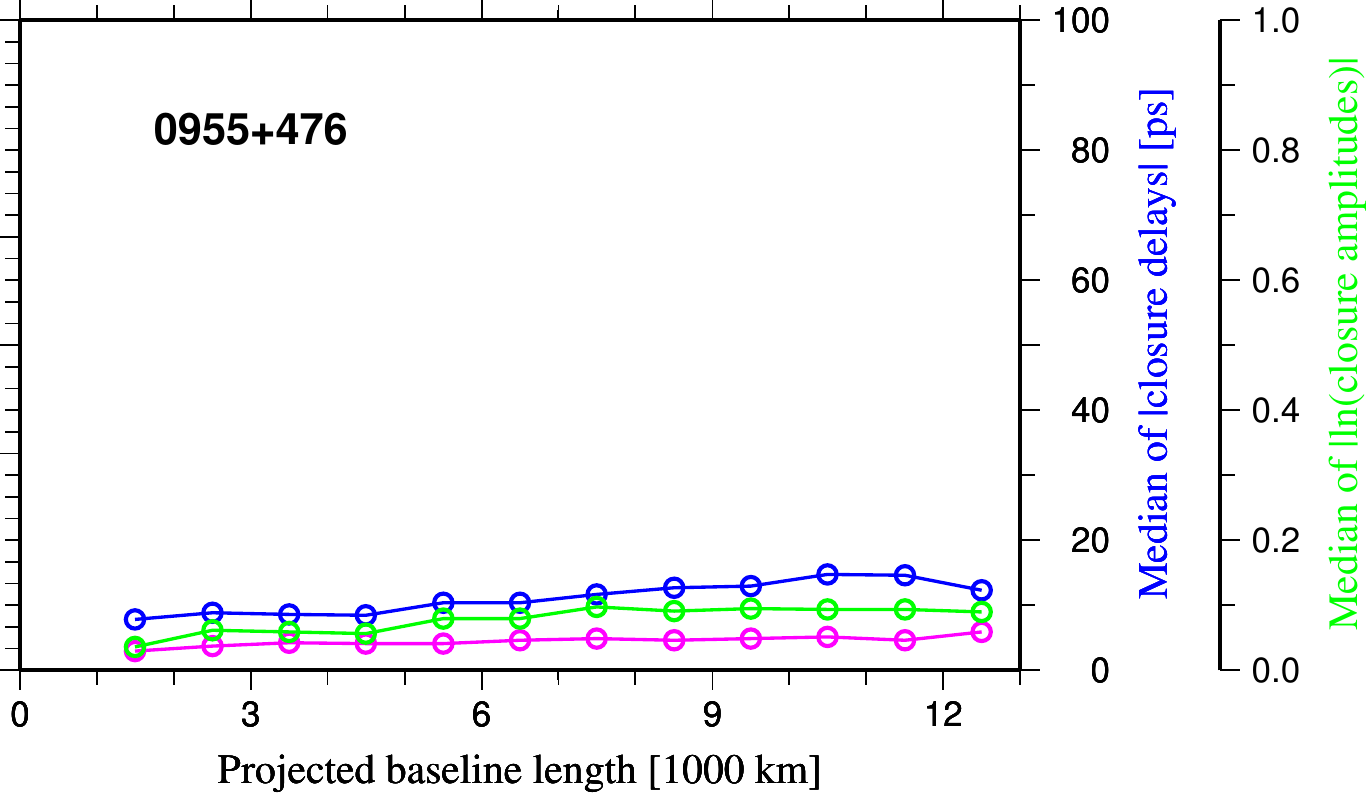}}   
 \resizebox{0.32\textwidth}{!}{\includegraphics{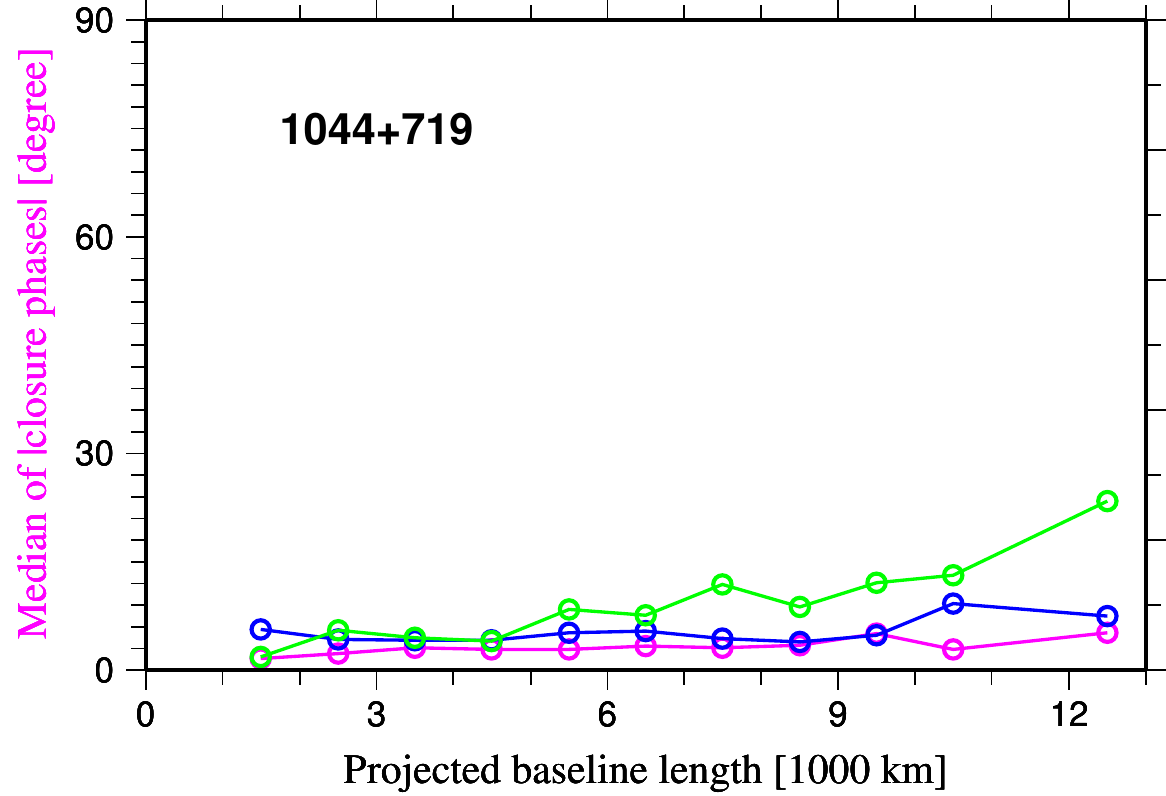}} 
\resizebox{0.285\textwidth}{!}{\includegraphics{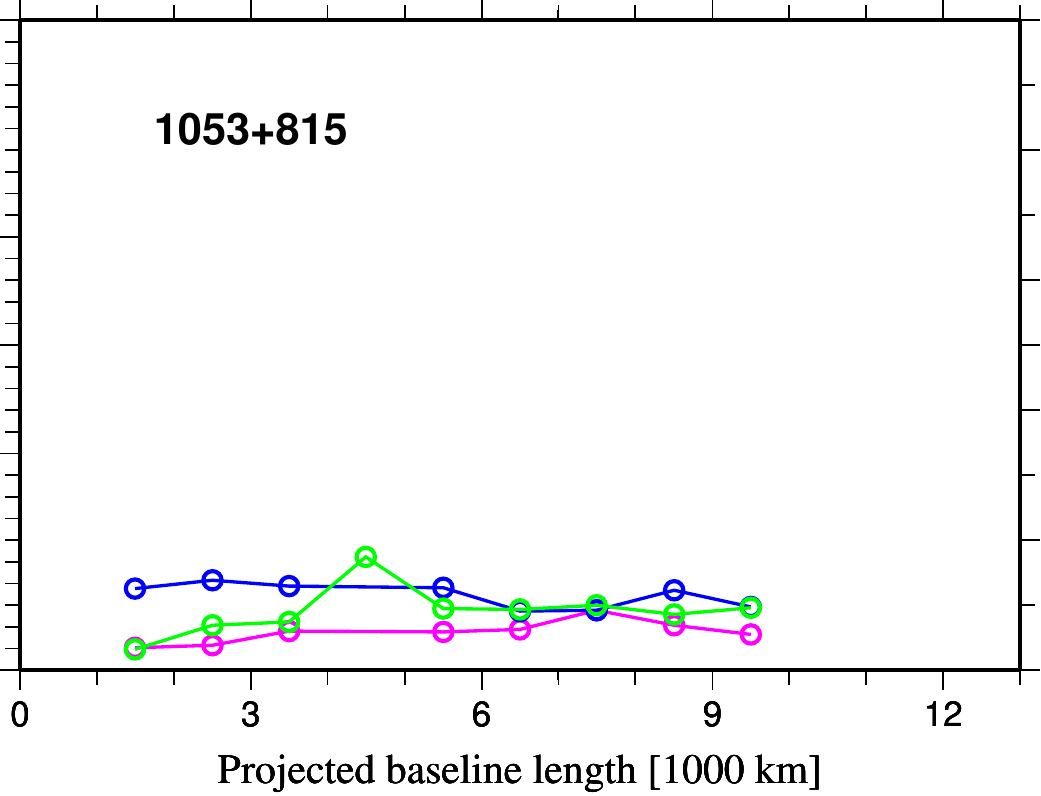}} 
\resizebox{0.376\textwidth}{!}{\includegraphics{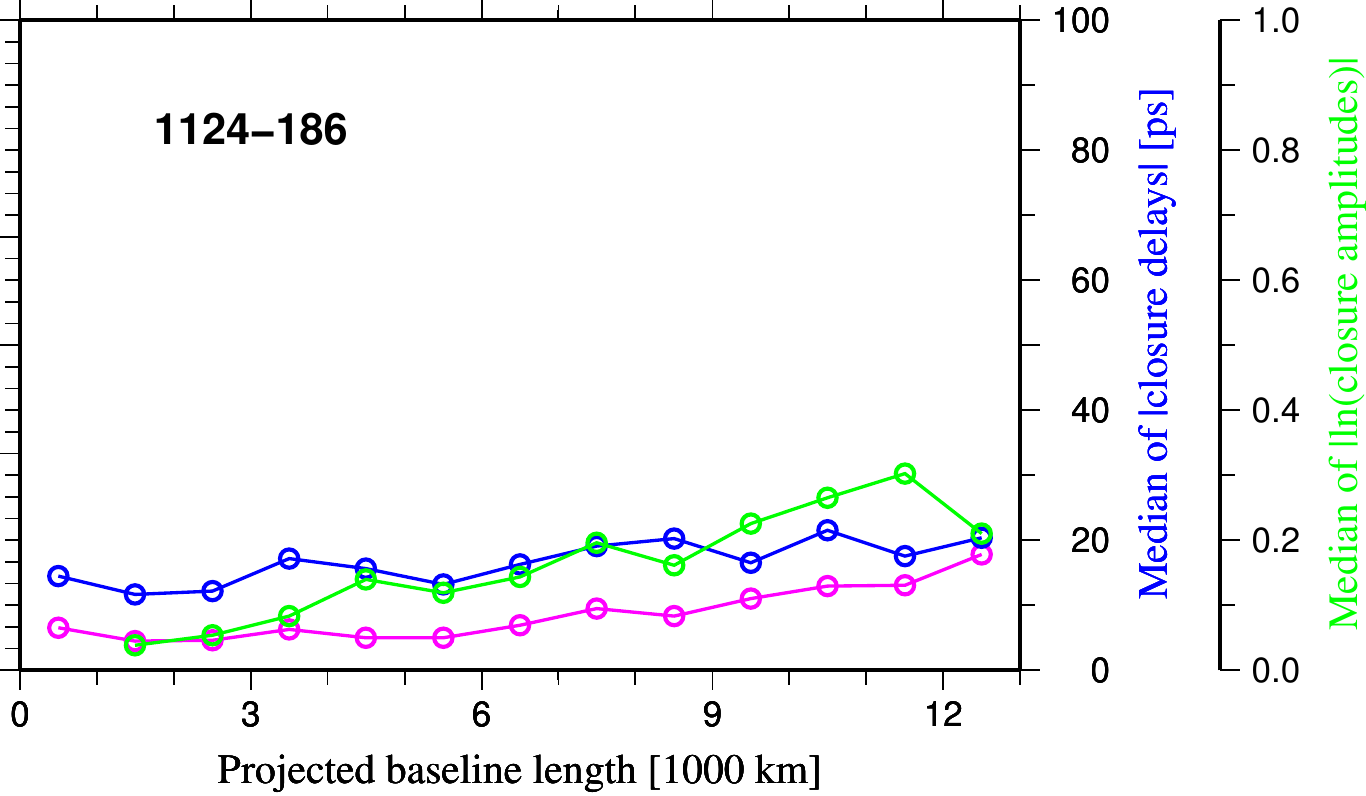}}   
 \resizebox{0.32\textwidth}{!}{\includegraphics{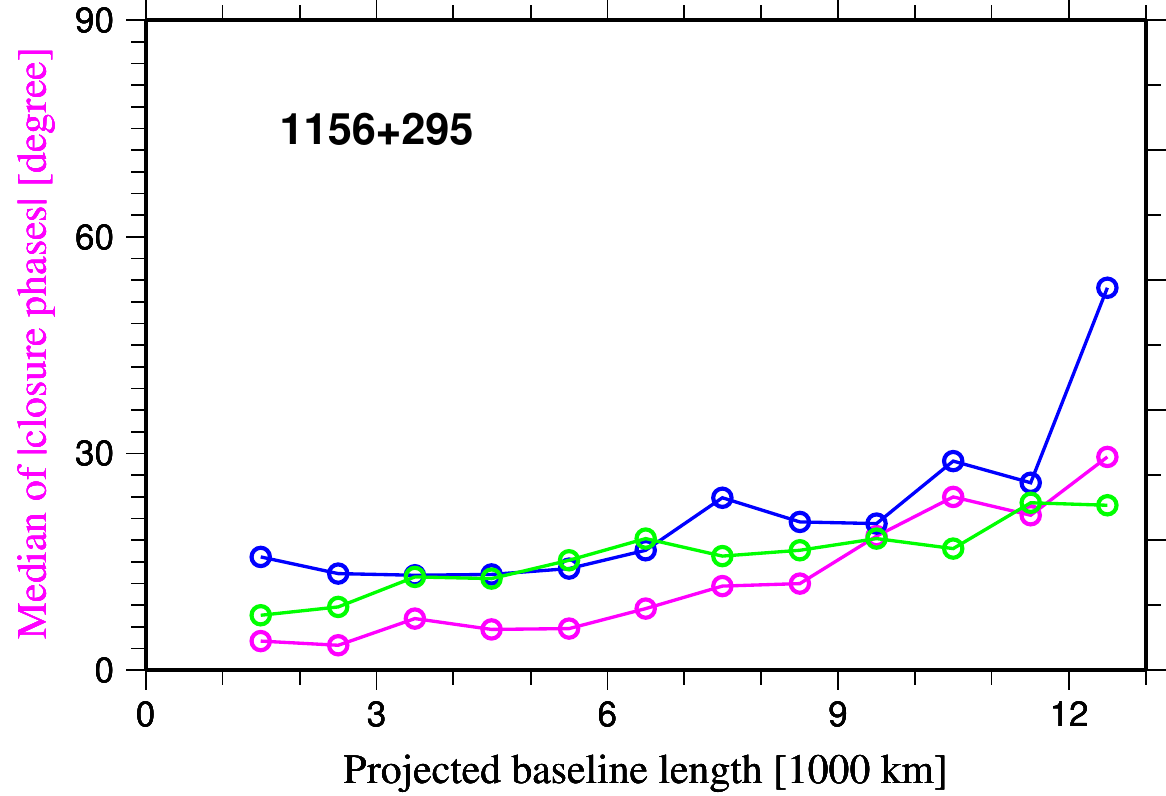}} 
\resizebox{0.285\textwidth}{!}{\includegraphics{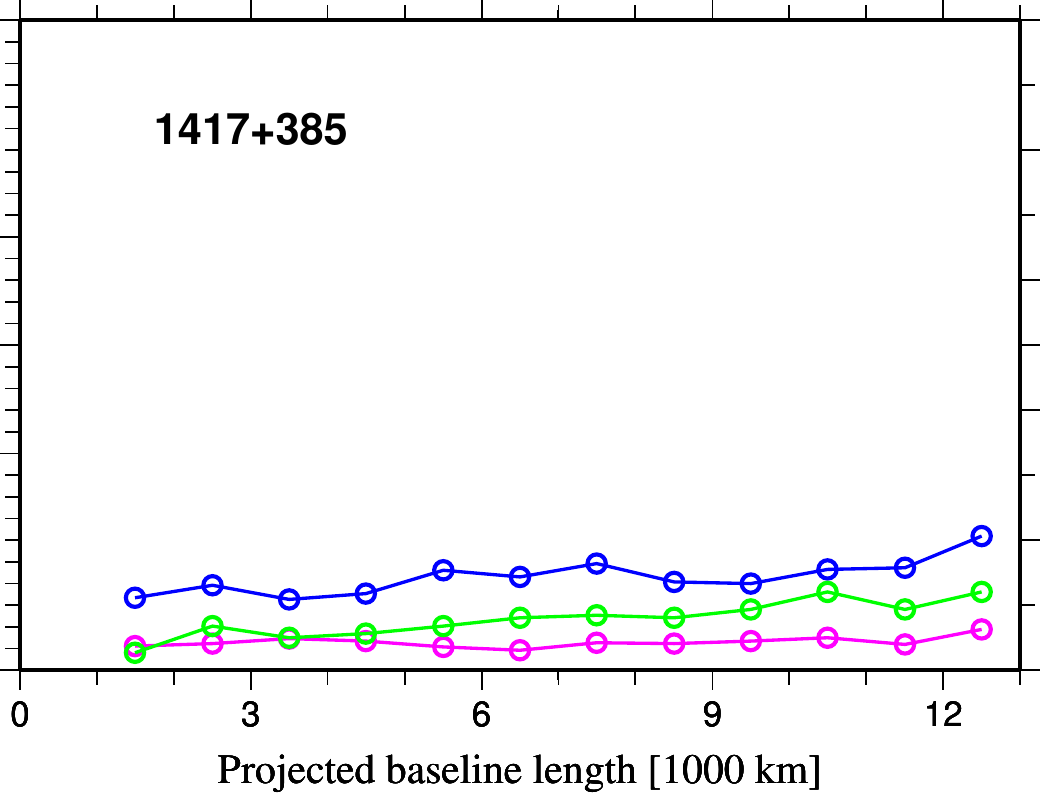}} 
\resizebox{0.376\textwidth}{!}{\includegraphics{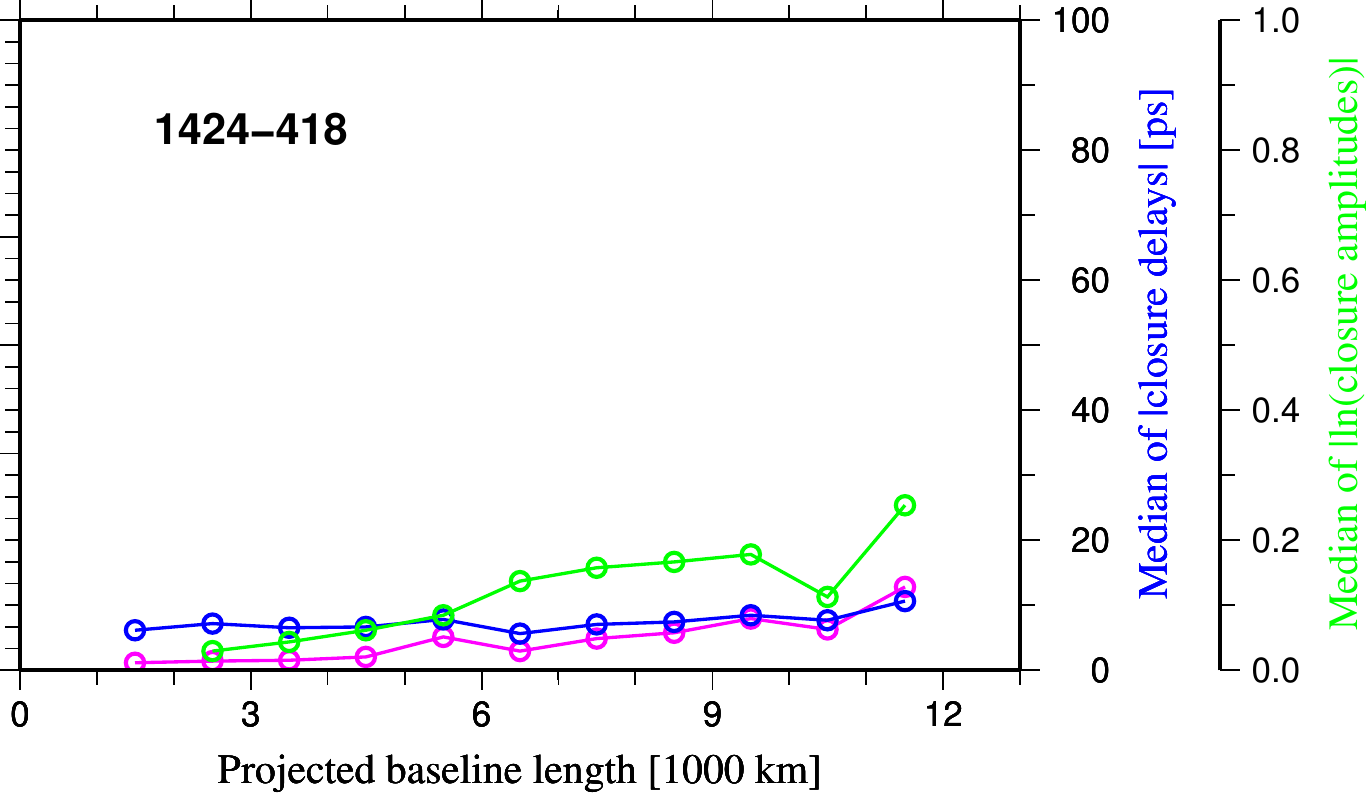}}
 \resizebox{0.32\textwidth}{!}{\includegraphics{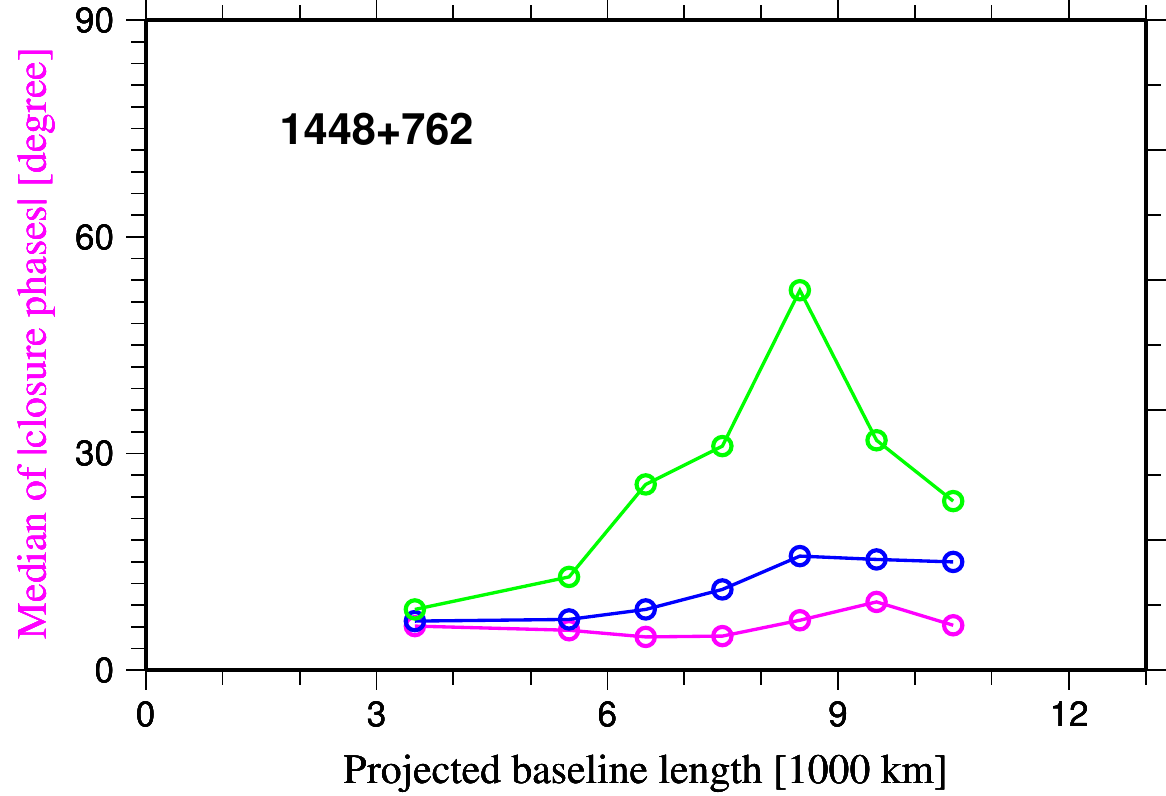}} 
\resizebox{0.285\textwidth}{!}{\includegraphics{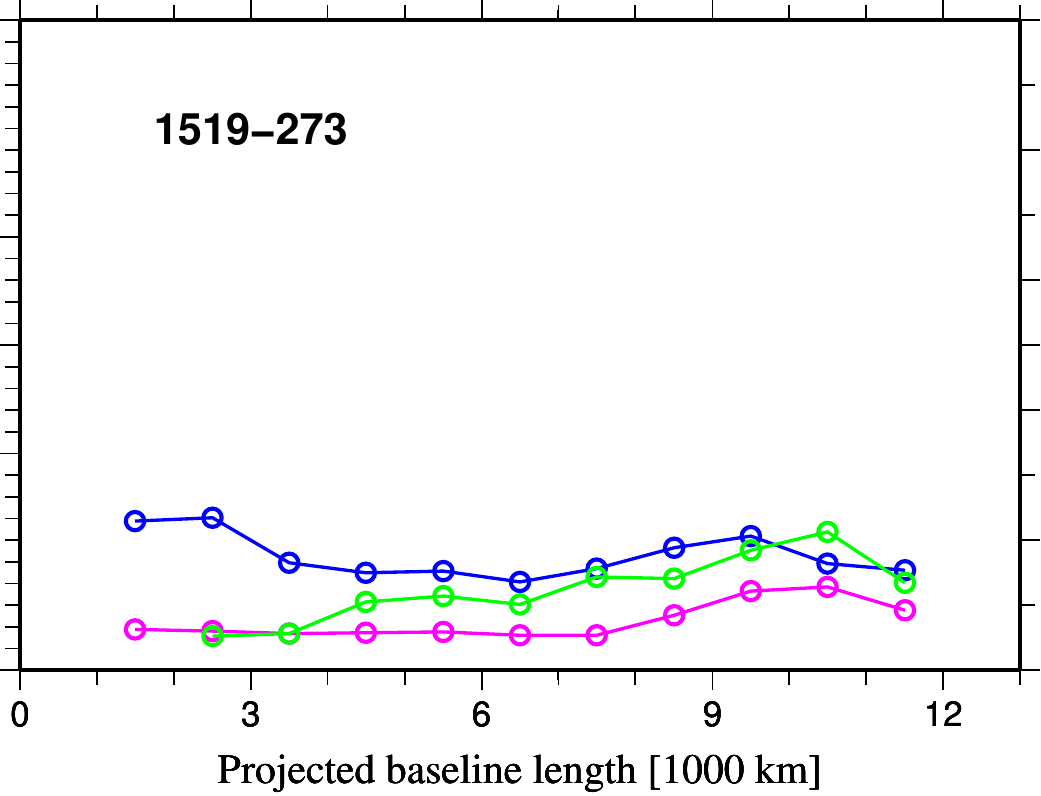}} 
\resizebox{0.376\textwidth}{!}{\includegraphics{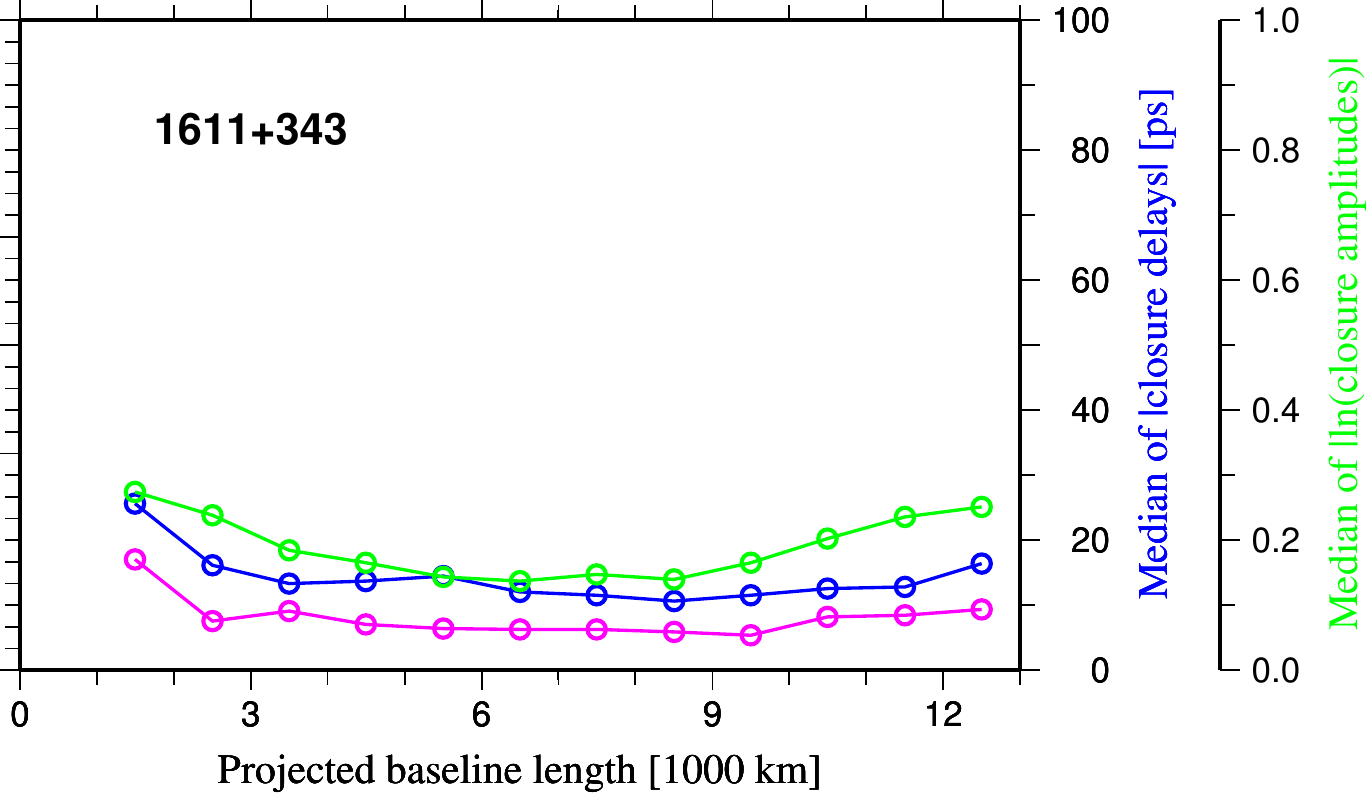}}       \\                 
\label{closure_rms1}      
                                                                    
%
\end{center}                                                   
{\footnotesize{\textbf{Fig. 2.} - Continued}} 
\end{figure*}                                                  
\begin{figure*}                                                
\begin{center}                                                 
 \resizebox{0.32\textwidth}{!}{\includegraphics{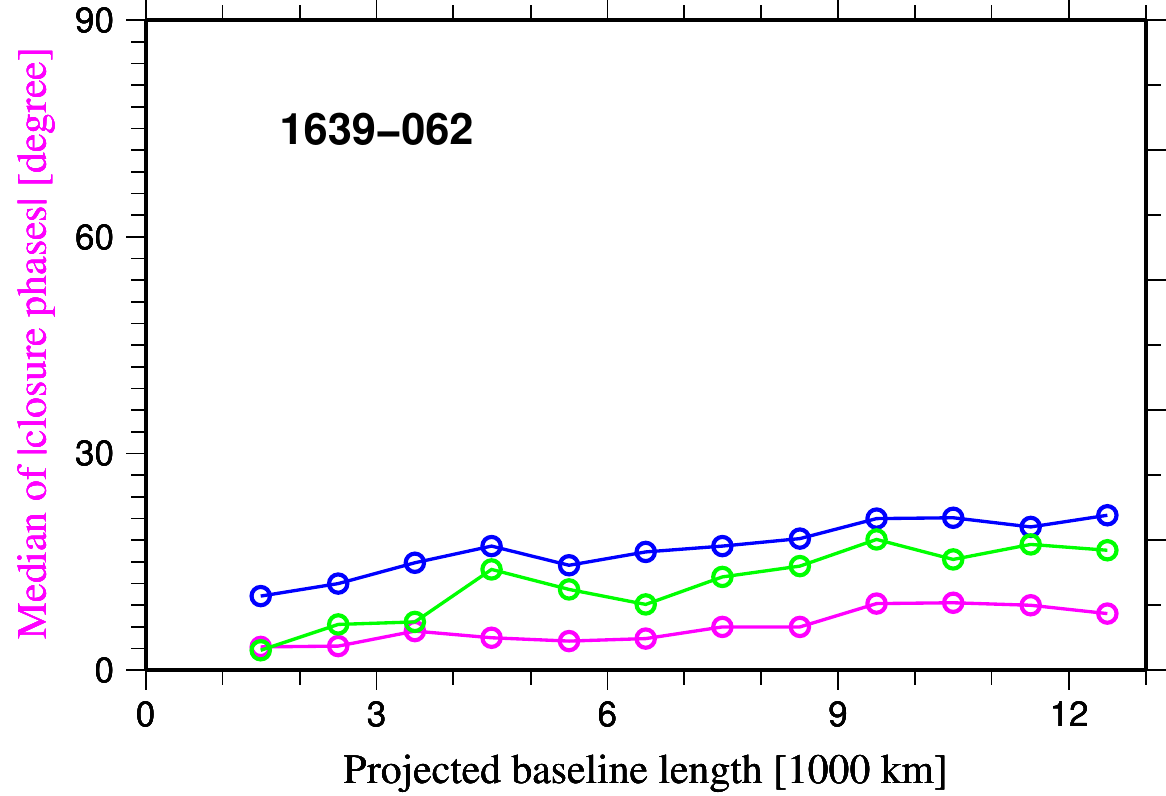}} 
\resizebox{0.285\textwidth}{!}{\includegraphics{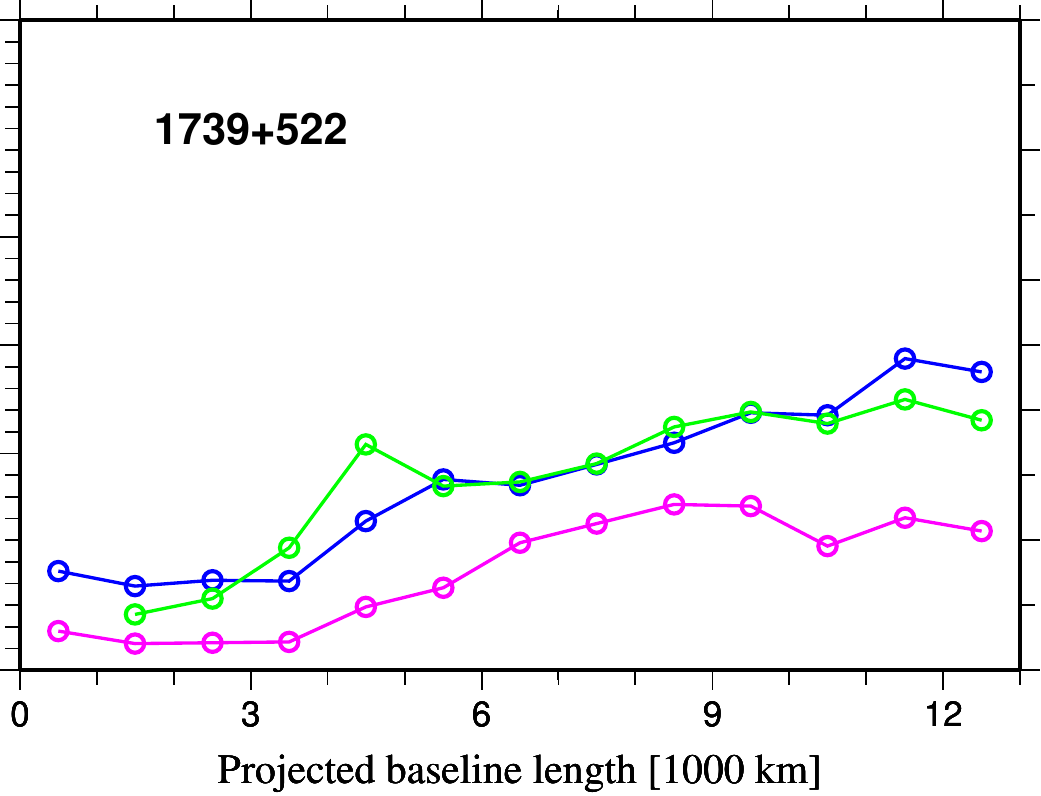}} 
\resizebox{0.376\textwidth}{!}{\includegraphics{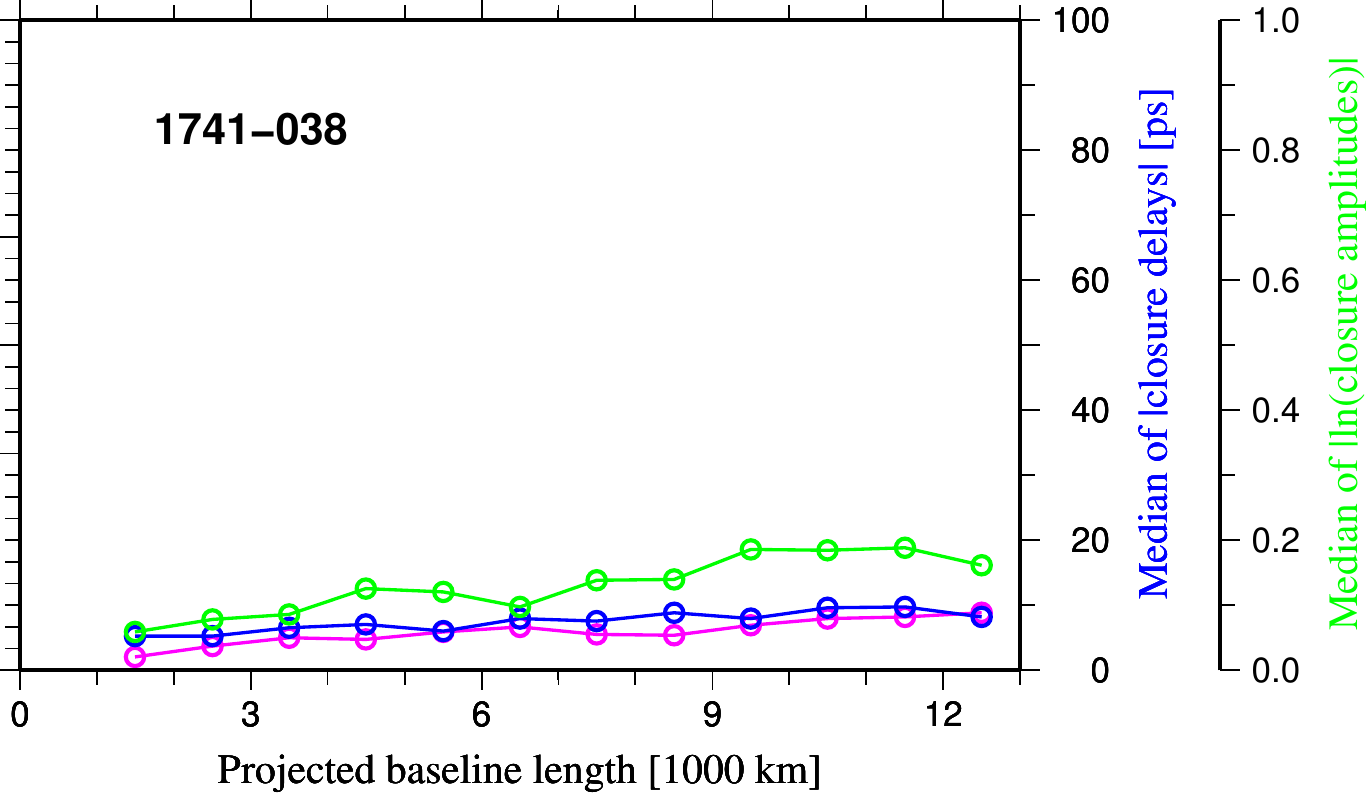}}   
 \resizebox{0.32\textwidth}{!}{\includegraphics{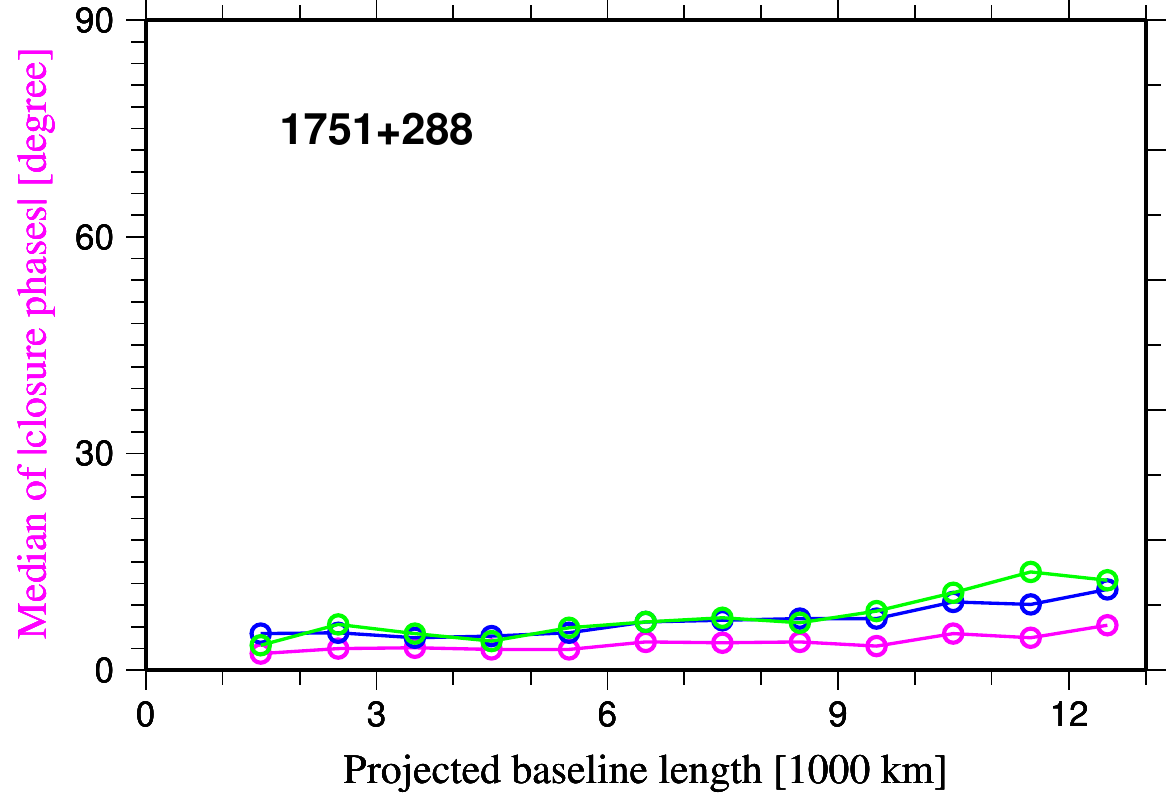}} 
\resizebox{0.285\textwidth}{!}{\includegraphics{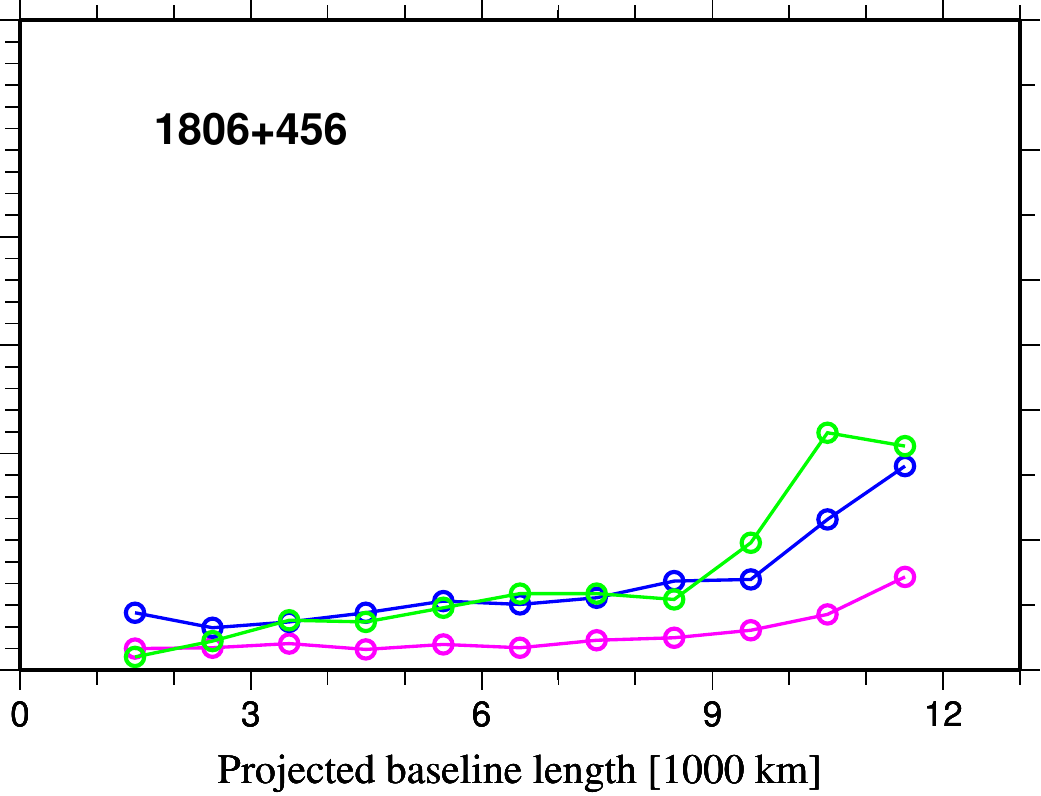}} 
\resizebox{0.376\textwidth}{!}{\includegraphics{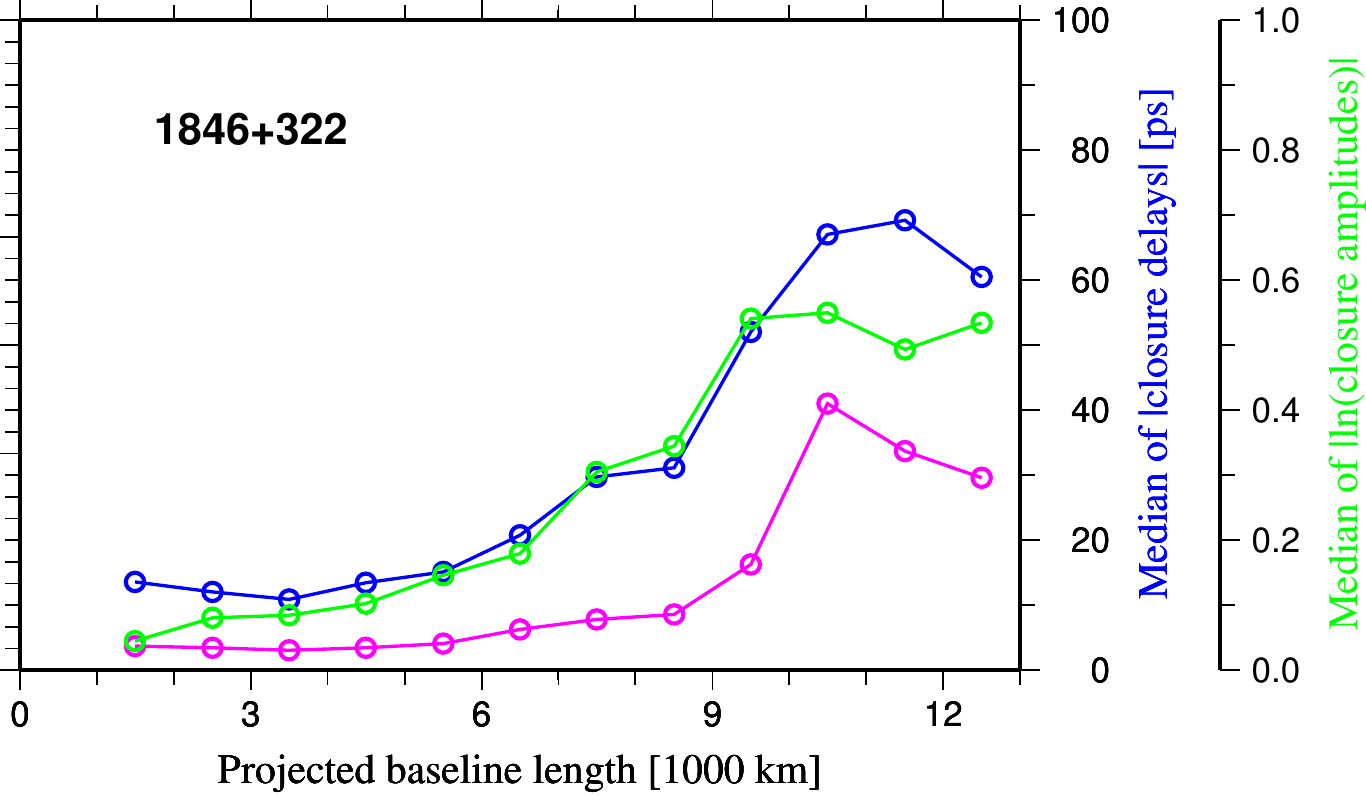}}   
 \resizebox{0.32\textwidth}{!}{\includegraphics{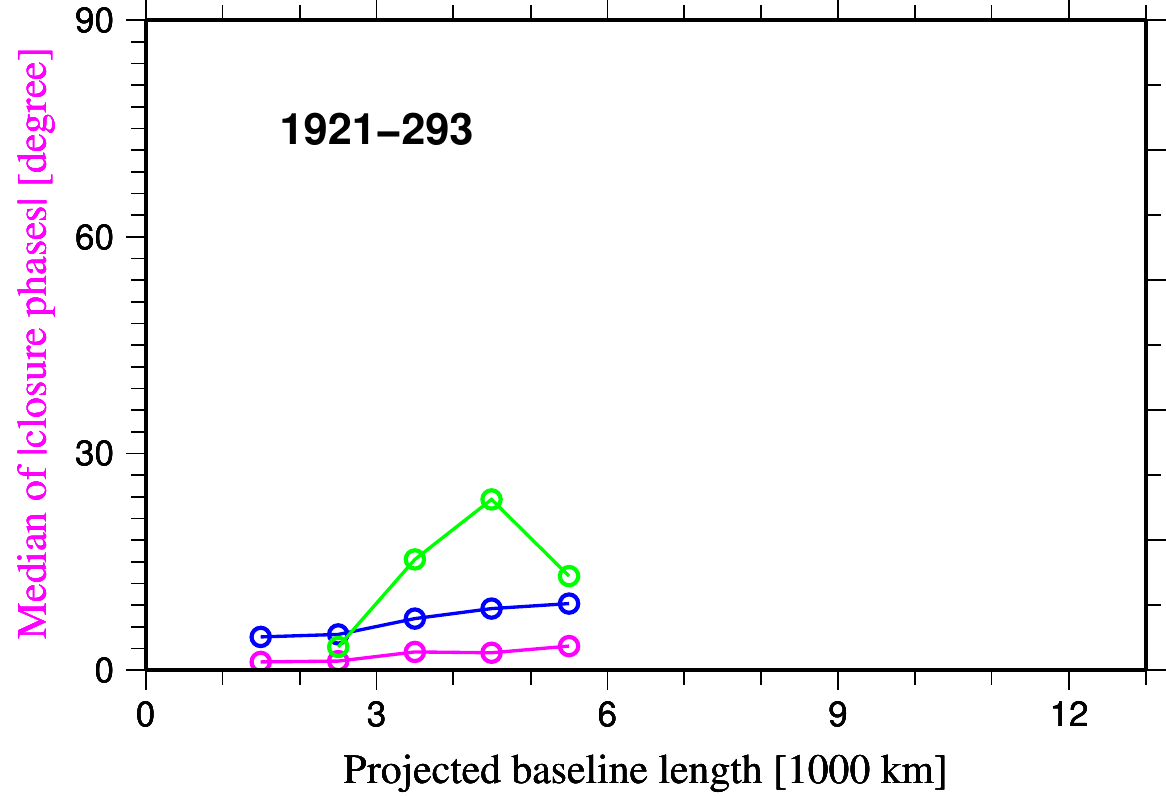}} 
\resizebox{0.285\textwidth}{!}{\includegraphics{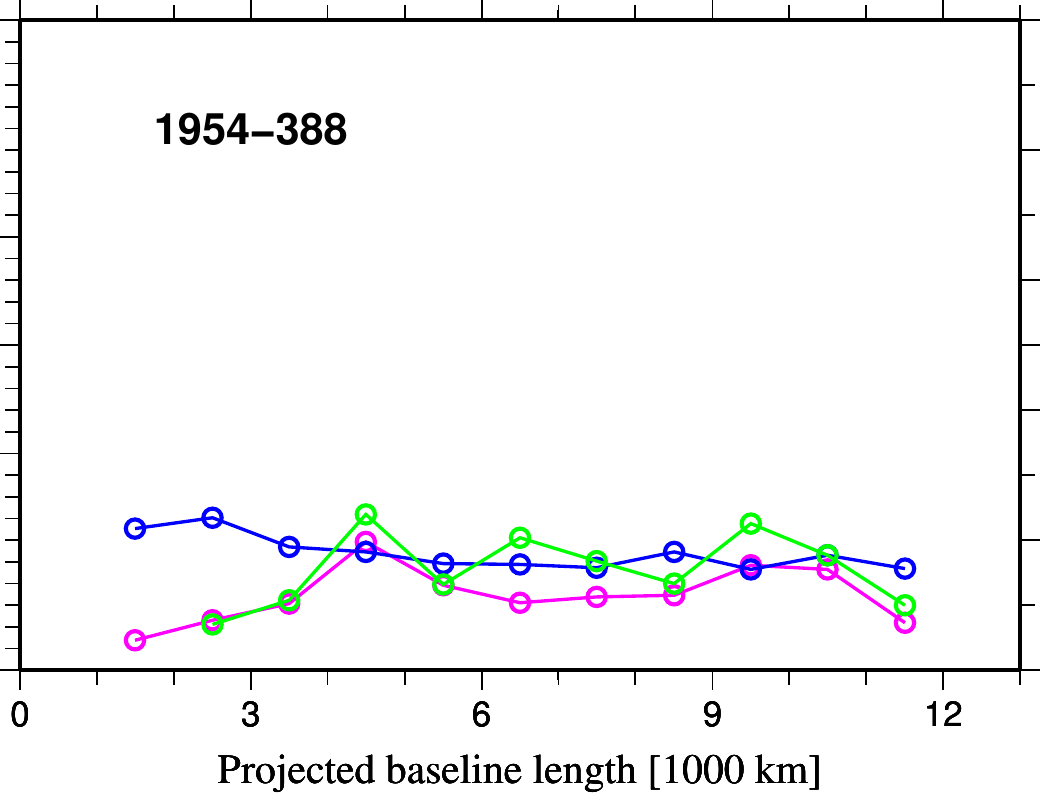}} 
\resizebox{0.376\textwidth}{!}{\includegraphics{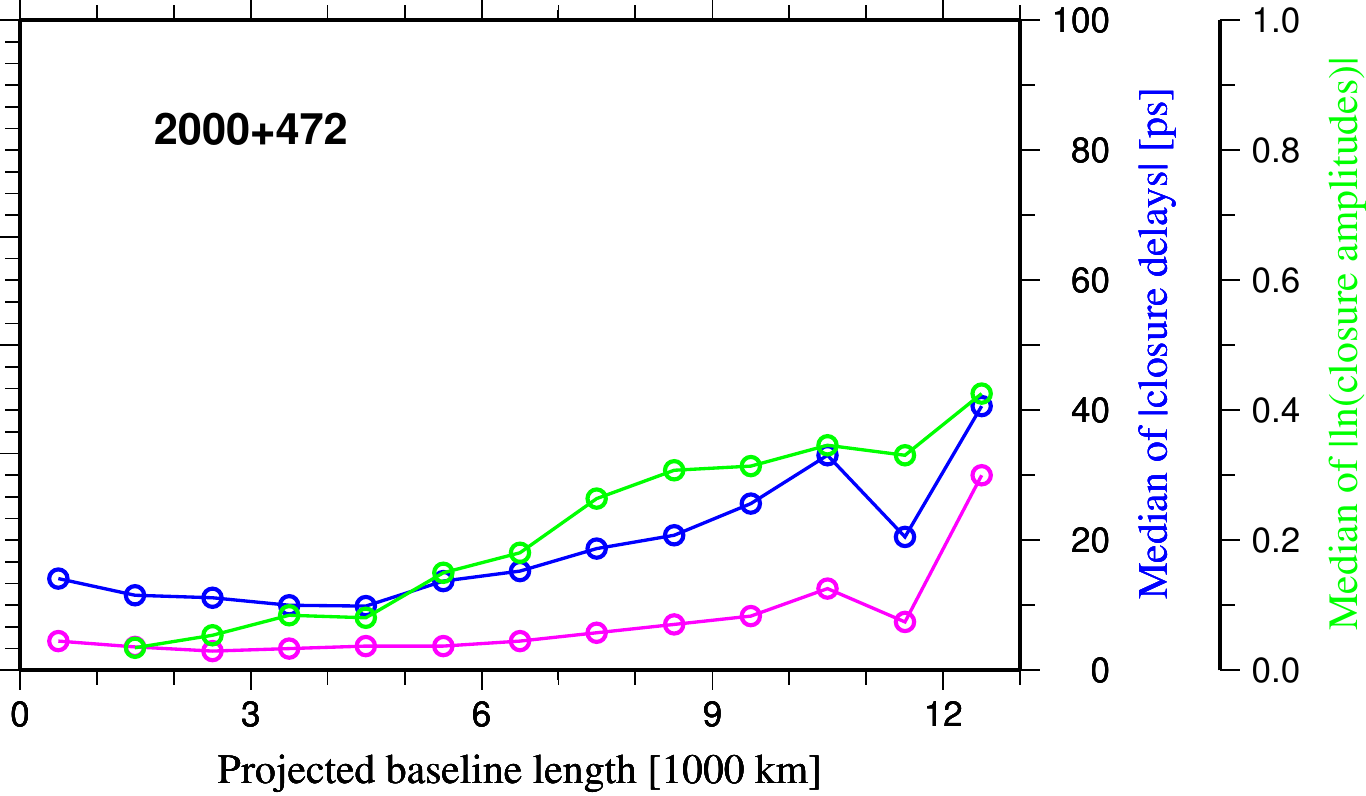}}   
 \resizebox{0.32\textwidth}{!}{\includegraphics{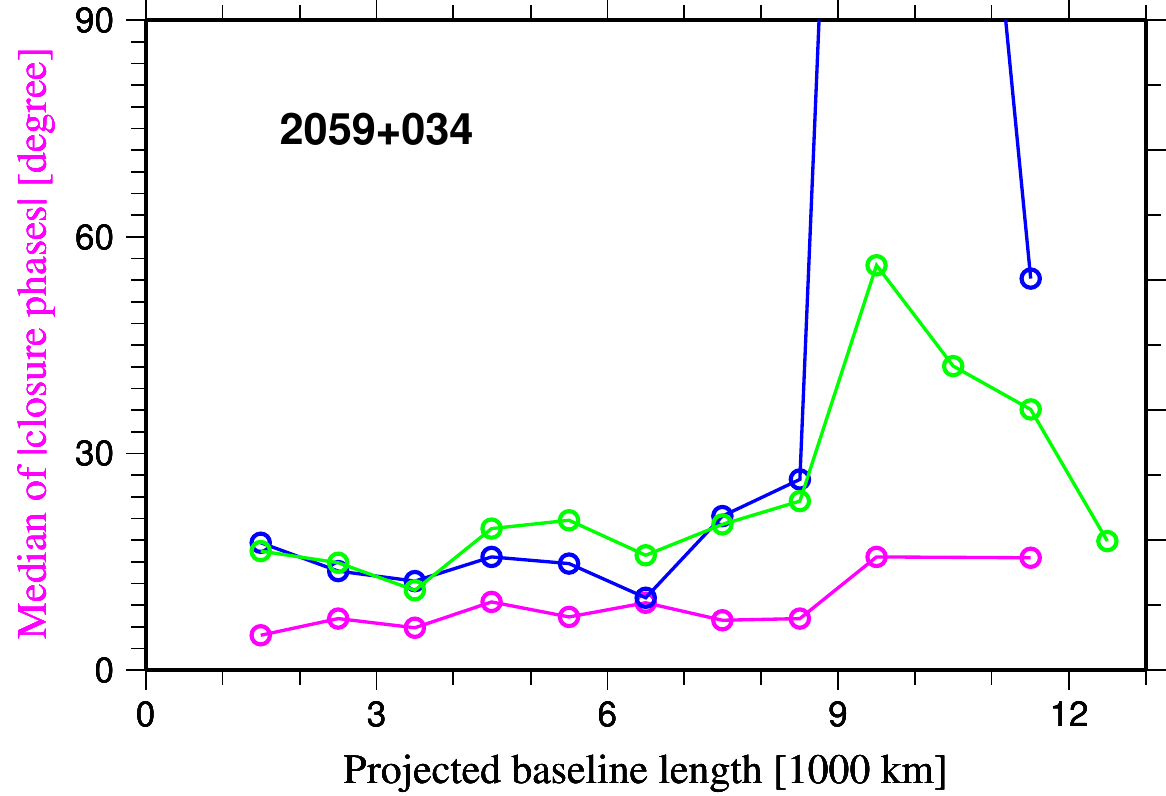}} 
\resizebox{0.285\textwidth}{!}{\includegraphics{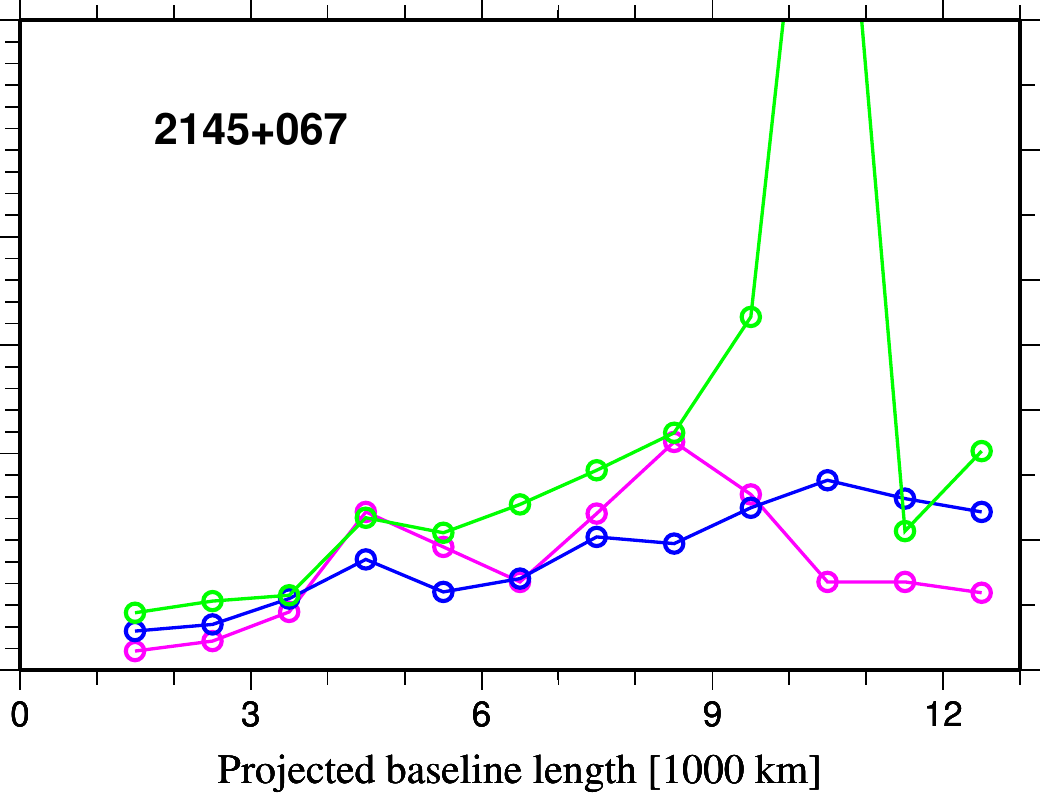}} 
\resizebox{0.376\textwidth}{!}{\includegraphics{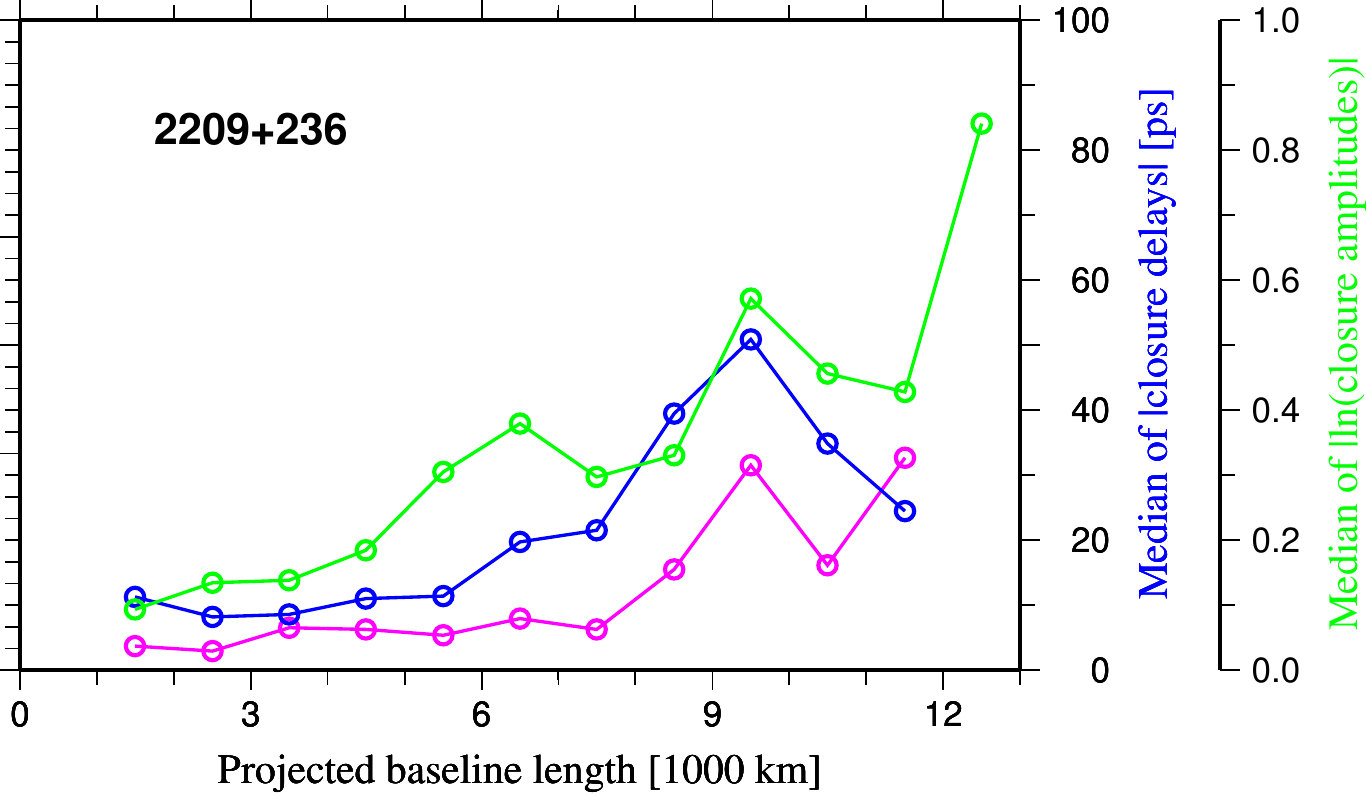}}   
 \resizebox{0.32\textwidth}{!}{\includegraphics{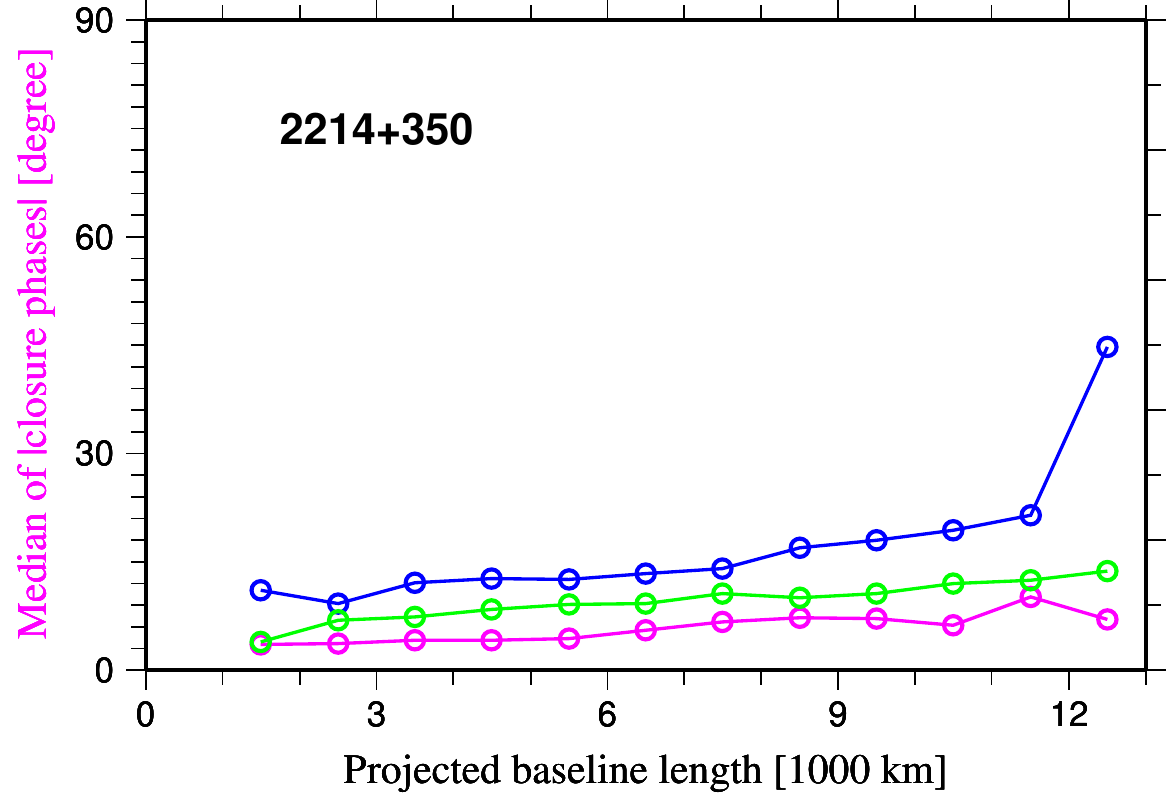}} 
\resizebox{0.285\textwidth}{!}{\includegraphics{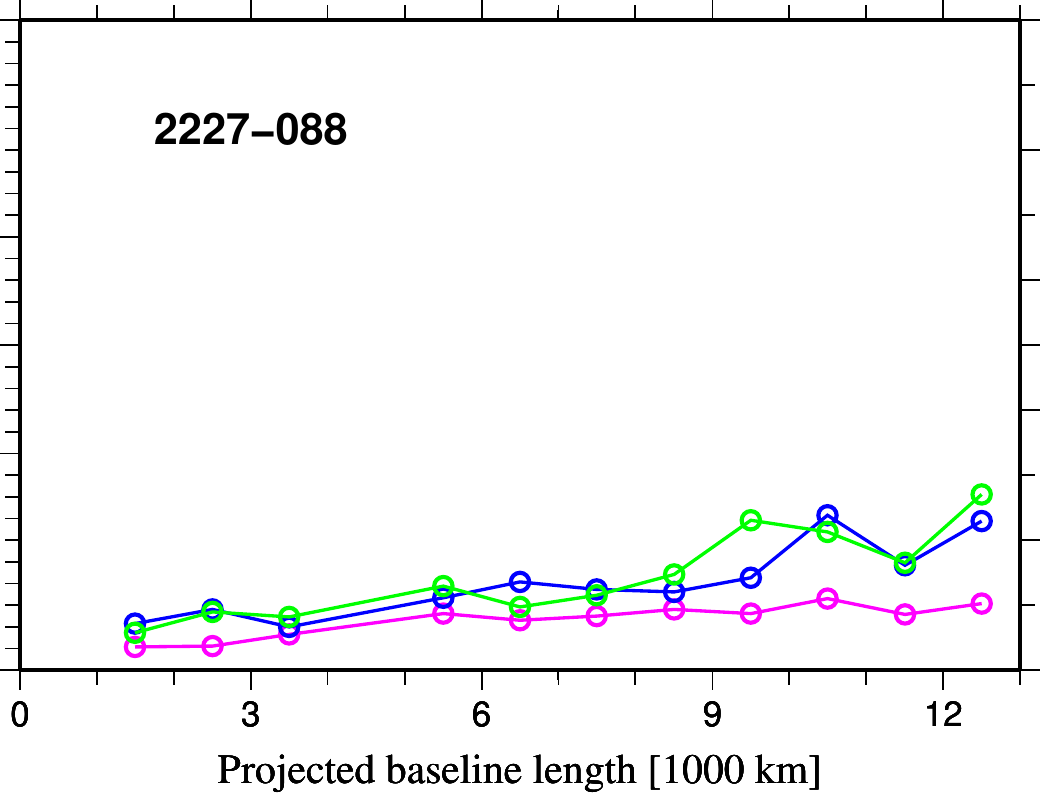}} 
\resizebox{0.376\textwidth}{!}{\includegraphics{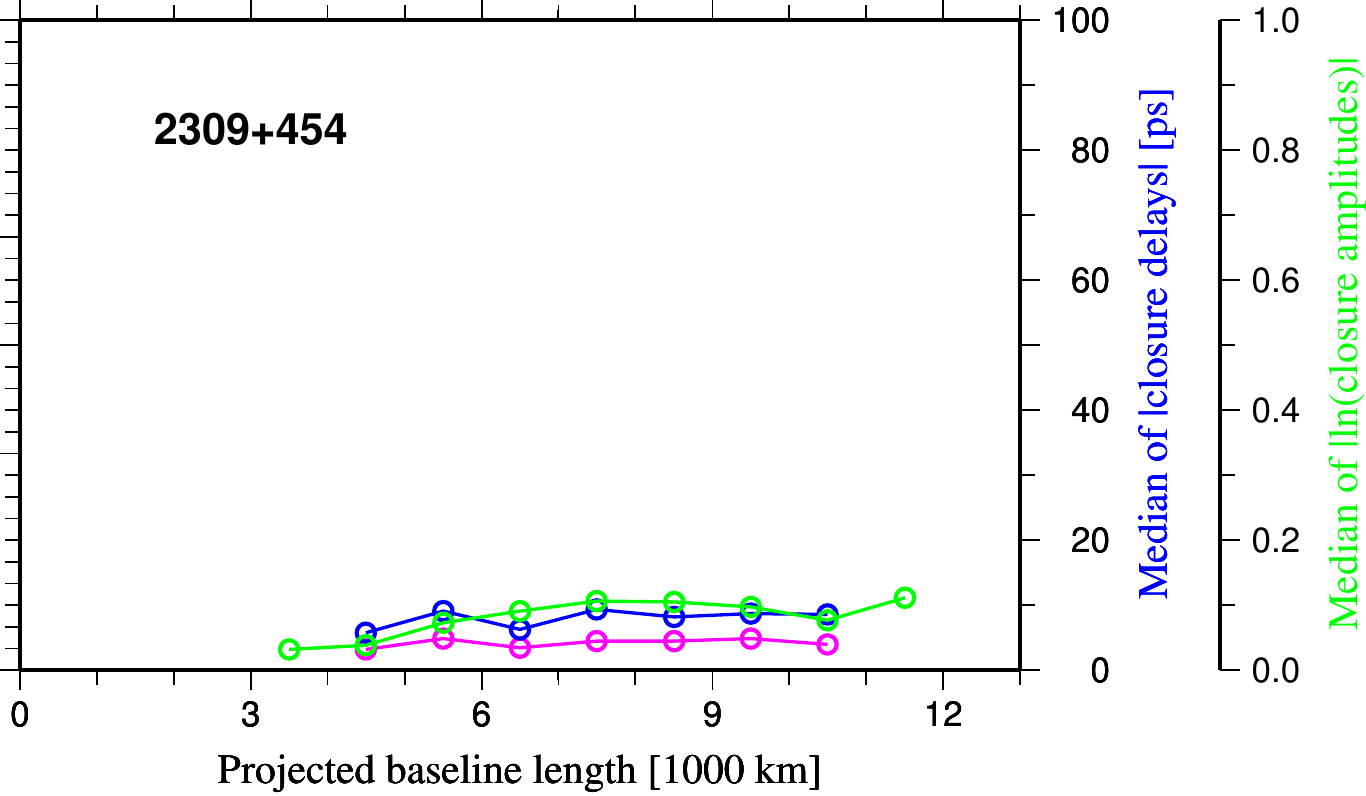}}
 \resizebox{0.32\textwidth}{!}{\includegraphics{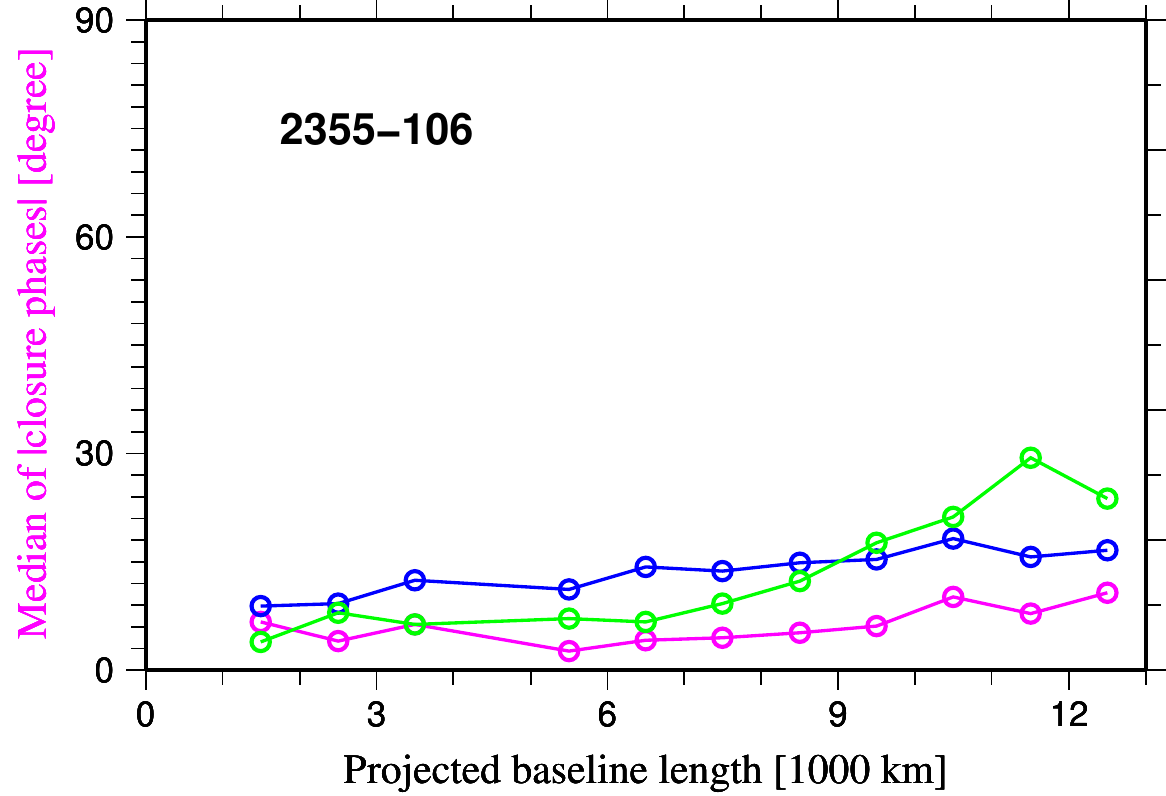}} 
\resizebox{0.285\textwidth}{!}{\includegraphics{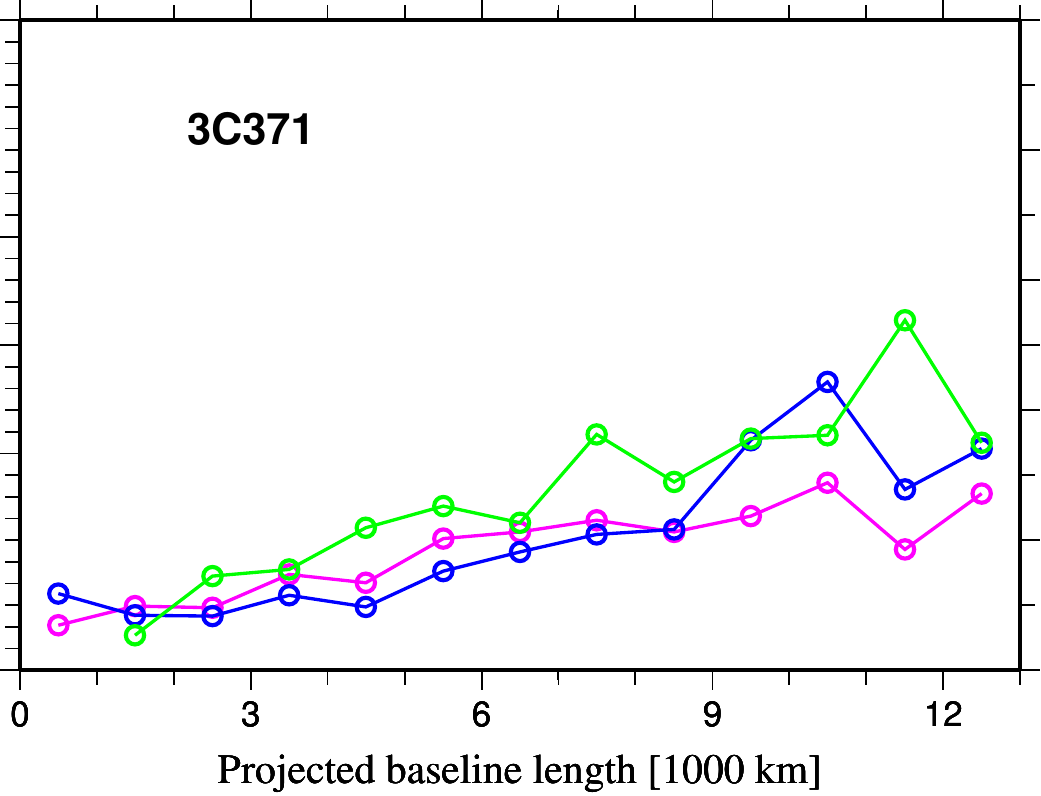}} 
\resizebox{0.376\textwidth}{!}{\includegraphics{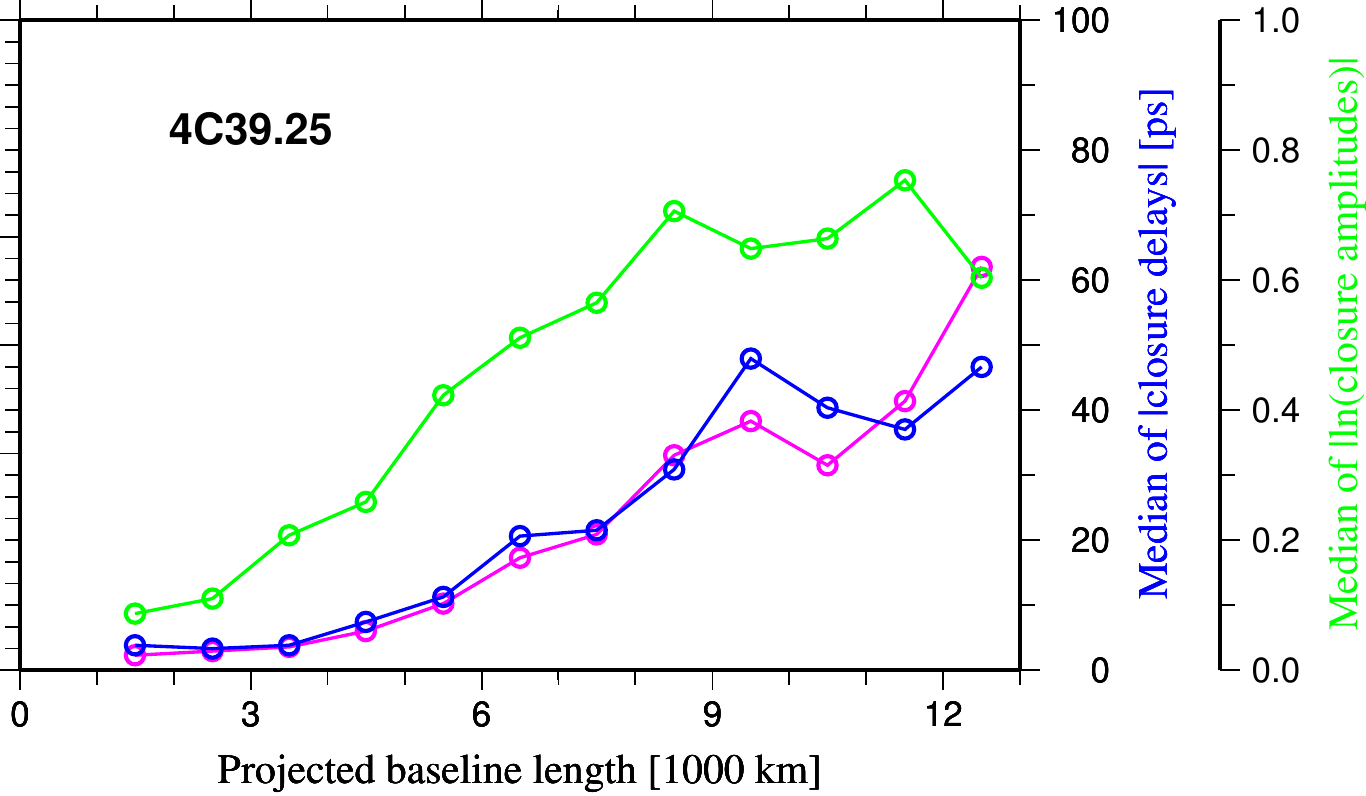}}\\
\label{closure_rms1}                                           
\end{center}       
{\footnotesize{\textbf{Fig. 2.} - Continued}}                                             
\end{figure*}                                                  

Median absolute values of the three closure quantities as a 
function of the longest projected baseline length in the triangle or 
in the quadrangle are shown for 54 radio sources in Fig. \ref{closure_rms}.
The natural logarithms of closure amplitudes for source 0738$+$313 deviate considerably 
from zero even for very small quadrangles, which means that it has a strong structure with
a quite large spatial scale. Almost all other radio sources have a common
pattern in their plots that median absolute values of closure quantities start with small values and 
increase when the projected baseline lengths become larger.
Variations of the three closure quantities agree well with each other.

These plots graphically demonstrate the median absolute closure quantities 
from Table \ref{tab1} that were used to calculate the structure indices, and also 
illustrate that the amount of source structure effects depends on baseline length.
For example, source 0642$+$449 shows
very little structure effect in the three closure quantities for short baselines but has very significant
effects when the projected baselines are larger than about 8000 km. This shows that it is strongly resolved on
small spatial scales, which may be a recent development 
since it was selected as a defining source in ICRF2 and has a structure index
of 2 in BVID. In this case, rms values of closure quantities represent the
magnitude of structure effect much better than median values. 
On the other hand, the median values for source 0016$+$731 are very small and have a flat pattern, and only 
slightly increase when the projected baseline lengths are larger than 11000 km; it is classified 
as having structure index of 1 based on closure delays. It is worth noting that the median values of the three closure quantities in general are much smaller than the rms values.
So called ``good'' sources with a structure index of 2 still tend to have quite 
significant rms closure delays. Also, the structure indices of ``good'' and ``extended'' sources depend strongly
on the date of observation, with 36\% of ``good'' and ``extended'' sources 
changing structure index between the BVID and the CONT14 sessions.

\section{Investigating impacts of source structure effect}
\label{sec3}

The results presented in Tab. \ref{tab1} suggest that the source 3C371 is a good candidate for the preliminary study of
impacts of structure effect because: (1) it has tens of thousand closure quantities in CONT14 and a structure index 
of 3; (2)the  median values of its closure quantities are large even when the projected baseline lengths are small so that
many observables are affected by structure; and (3) compared to the structure effects of some extended sources, such 
as 0014+813 and 0738+313, the structure effects of 3C371 are not so strong such 
that observables with significant structure effects 
would be excluded as outliers during VLBI data analysis.
From Fig. \ref{closure_rms}, we also notice that 3C371 is a good representative of the radio sources 
in CONT14 in the sense of the magnitude of source structure effects.

\subsection{Structure model for 3C371}
The structure of source 3C371 was assumed to have multiple point components and was determined by model-fitting of closure phases 
directly instead of the traditional Fourier imaging.
The method of forward modeling was developed to determine the multi-component structure: (1) closure phases of 
small triangles with longest baseline lengths shorter than a certain value, such as 2000 km, are used to determine 
the relative position and the flux-density ratio of two components based on the model of structure 
phase in \citet{cha90b};
(2) closure phases of triangles with larger baseline lengths are gradually added and 
used to test the obtained multi-component model
by the previous step until a significant mismatch between modeled closure phases and observed closure phases occurs; 
(3) another component is proposed and fitted from closure phases and then the second step is repeated; and (4) fitting continues until the closure phases of triangles with the longest baseline length are exploited. In the whole procedure, the 
identified components are kept and only one new component will be proposed to add in at one time. 

If one does not have any a priori
information about the structure of a source, different a priori values for the two-component model
may need to be tested. In general, however, 
the changing pattern of closure phases of triangles with the same three stations over 24-hours of GMST should give useful insight for that.

Based on this method, a three-component model was determined for the source 3C371. The result is presented in Table \ref{tab2}
and shows that this source is extended in one direction with a position angle of about $260^{\circ}$. 
From publicly available maps of 3C371 in Feb. 2014 and Sep. 2015\footnote{http://astrogeo.org/vlbi\_images/}, we find a good agreement in 
the position offset and direction of extended structure between our modeling and imaging results. 
To make a direct modeling to imaging comparison, we have imaged the CONT14 sessions' visibility data for 3C371, 
shown in Fig. \ref{3C371_image}. Our image shows a core with a one-sided jet extending about 6~mas from the core along a position angle of about $260^{\circ}$. No significant structure is visible farther than 6~mas from the core, in contrast to the VLBA imaging results mentioned directly above. However, the IVS CONT14 observations have longer baselines, more observations with long baselines, and fewer short-VLBI baselines, than observations provided by the VLBA, so it is not surprising that our imaging results show no emission at large separations from the core, where the emission is expected to be more extended and therefore resolved-out. The details of imaging based on geodetic VLBI sessions and a detailed comparison between the images from geodetic sessions and VLBA sessions will be presented in our future publication (Anderson et al., in prep.).

Fig. \ref{fig_3C3711} shows observed closure phases in magenta dots, modeled closure phases from the proposed analysis in blue dots, and closure phases from imaging results in green dots for two triangles as a function of GMST. The rms of closure phases was reduced from $27.9^{\circ}$ to $12.5^{\circ}$ using the three-component model. As we can see, the results from imaging have only a slightly better agreement with the observed closure phases and delays, and the model based on closure quantities does give results close to what full imaging gives.

\begin{figure}
\resizebox{0.45\textwidth}{!}{%
  \includegraphics{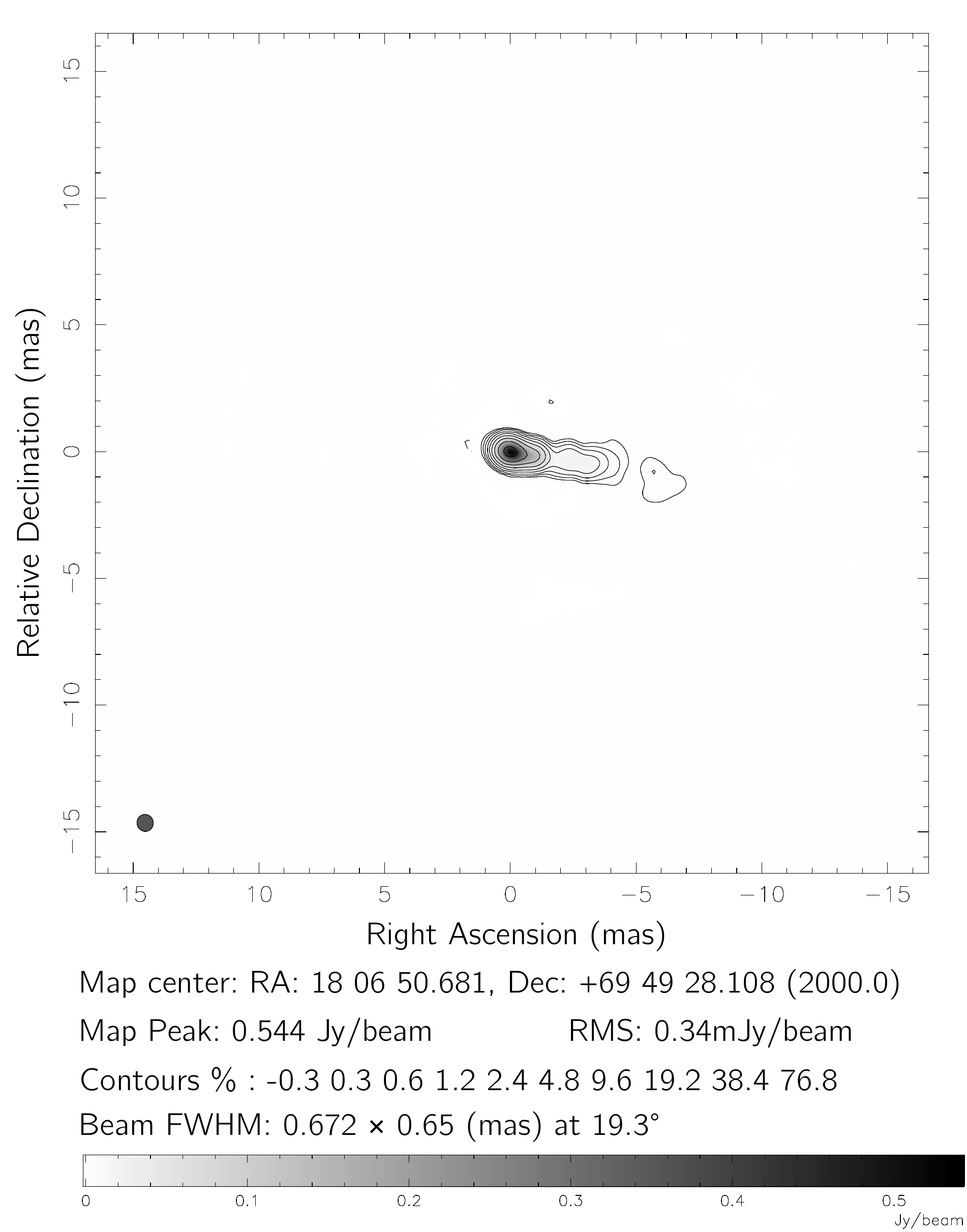}
}
\caption{The image of 3C371 based on the visibility data from CONT14 sessions using natural weighting. 
 The extended direction from our proposed method using closure phases agrees well with the imaging result.}
\label{3C371_image}       
\end{figure}

The three-component model then was used to calculate the structure corrections for delay observables at X band
and the effect at S band was ignored.
By applying this structure model to group delays, the rms of closure delays of 3C371 was reduced from 46.6 ps
to 36.4 ps. Fig. \ref{fig_3C3712} shows observed 
closure delays, modeled closure delays from the three-component model, and closure phases from imaging for the same two triangles in Fig. \ref{fig_3C3711}. 
Modeled closure delays generally have the same pattern as that in observations, however, 
the scatter in the variations of closure delays is much larger than that in closure phases, and 
the structure delays from the structure models cannot exactly follow the variations in observed closure delays. 
The improvement in closure phases after modeling the source structure effects is 55 \%, 
while that in closure delays is only 22 \%. This shows the expected
result that the observed phases are more accurate than the observed
delays and provide better modeling constraints.

The theoretical delay software CALC11 was modified to be capable of correcting the structure effect. 
The theoretical delays for all observations in CONT14 were recalculated to generate new databases. 
Two solutions were then made based on the new databases
and the original databases.

%
%

\begin{table}
\caption{Structure model of 3C371 determined by model-fitting from closure phases. 
The first component with the peak intensity was set to be the reference point. 
The flux-density ratios $k$ and the relative offsets $r$ are with respect to the reference point. $PA$ is
the position angle of the vector of the component and the reference point, measured in the sky counterclockwise with respect to the north. }
\label{tab2}       
\begin{tabular}{cccc}
\hline\noalign{\smallskip}
Component& $k$ & $r$ & $PA$ \\
& & (mas) & (deg.) \\
\noalign{\smallskip}\hline\noalign{\smallskip}
1 & 1 & 0  & 0 \\
2 & 0.302 $\pm$ 0.007 & 0.504 $\pm$ 0.052 & 257.0 $\pm$ 3.2 \\
3 & 0.235 $\pm$ 0.010 & 1.001 $\pm$ 0.058 & 261.3 $\pm$ 4.5 \\
\noalign{\smallskip}\hline
\end{tabular}
\end{table}

\begin{figure}
\begin{center}
\resizebox{0.44\textwidth}{!}{%
  \includegraphics{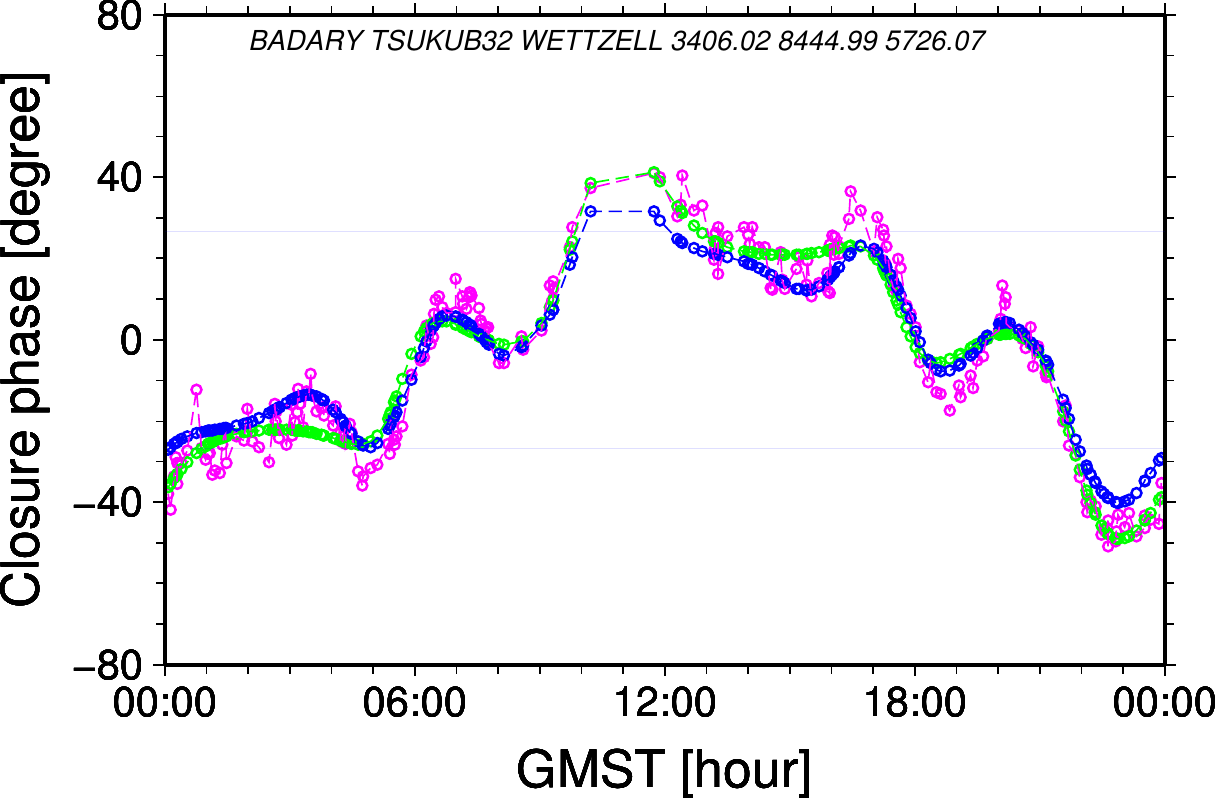}
}
\resizebox{0.44\textwidth}{!}{%
  \includegraphics{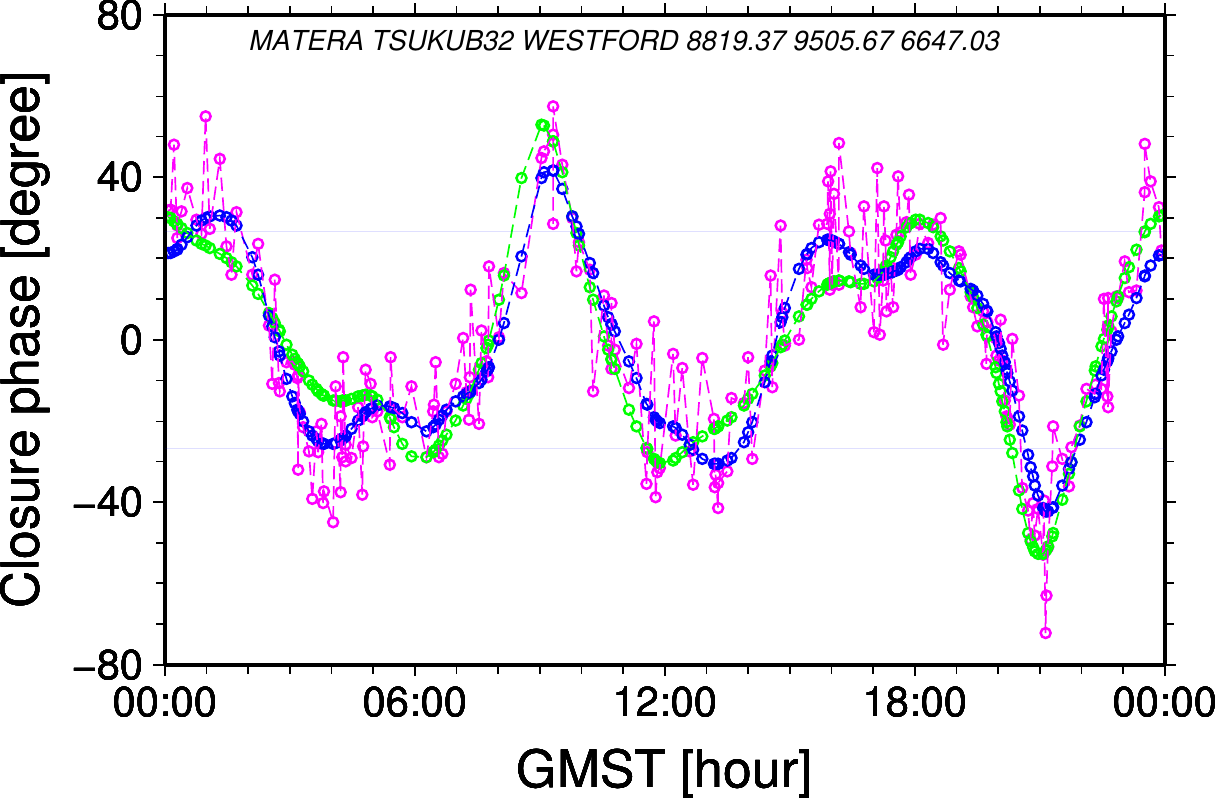}
}
\end{center}
\caption{Comparison between closure phases from the three-component model (blue), the imaging result (green) and the observations (magenta) for two closure triangles. }
\label{fig_3C3711}       
\end{figure}

\begin{figure}
\begin{center}
\resizebox{0.47\textwidth}{!}{%
  \includegraphics{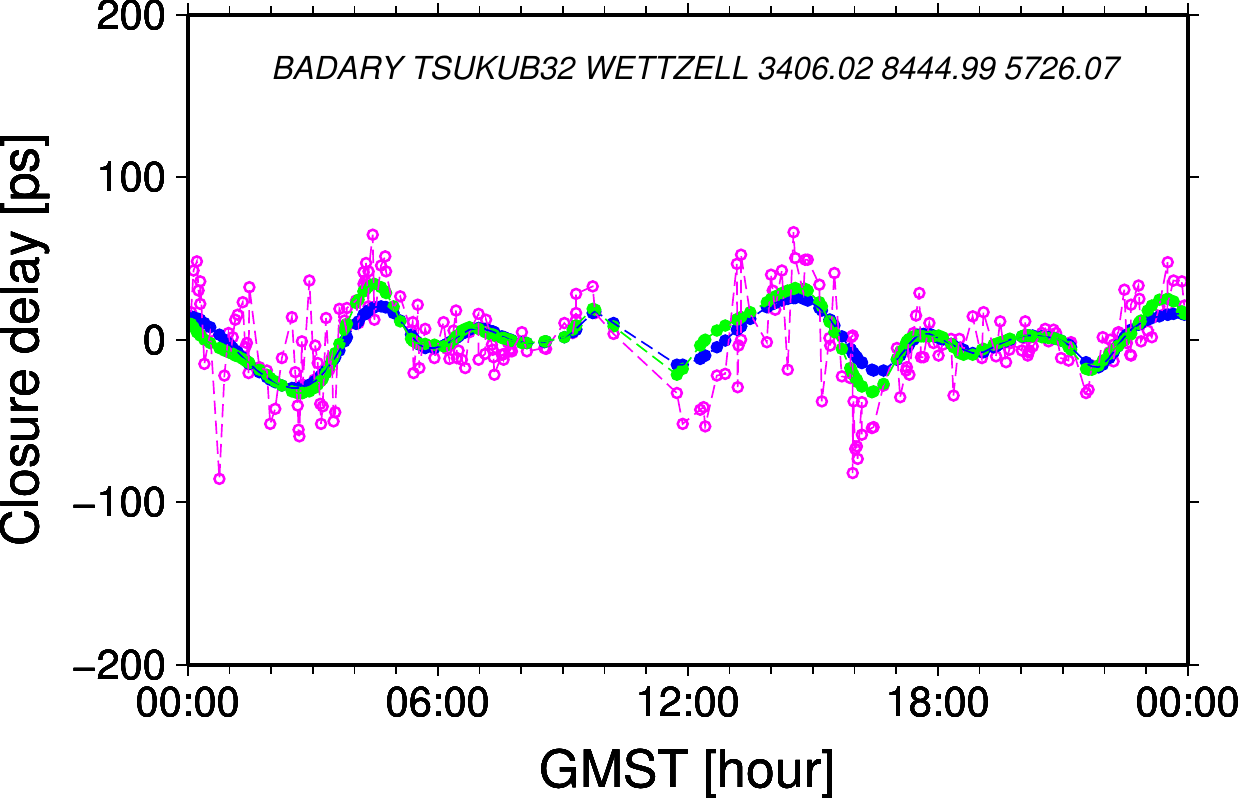}
}
\resizebox{0.47\textwidth}{!}{%
  \includegraphics{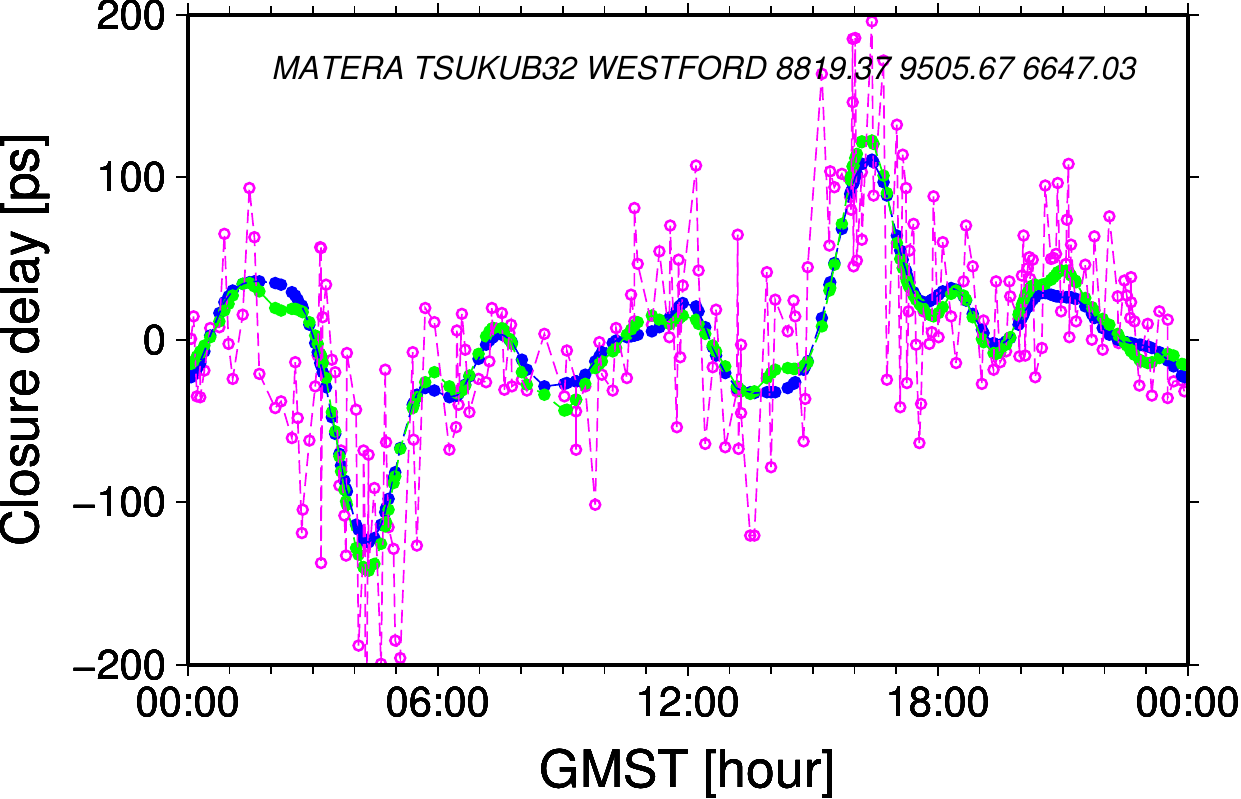}
}
\end{center}
\caption{Comparison between closure delays from the three-component model (blue), the imaging result (green) and the observations (magenta) for two closure triangles. }
\label{fig_3C3712}       
\end{figure}

\subsection{Impacts of source structure effect}

A detailed comparison of results obtained from the two sets of databases 
was made to investigate the impacts of structure effect introduced by one single radio source, 3C371.
In total there are about 254\,000 observations. The overall rms and chi-square for the solution 
of the original IVS databases were 26.65 ps and 0.832, respectively, while those for the new databases are 26.61 ps 
and 0.830. Comparisons of residual rms for 3C371 over 15 sessions are presented in Table \ref{tab3}.
About 12\,000 observables of 3C371 were included in the data analysis.
The overall rms for this source was reduced from 26 ps to 25 ps, and the chi-square was decreased from
0.863 to 0.835. 

\begin{table}
\caption{Comparison of residual delay rms for source 3C371. The second column
shows the numbers of used observables in the data analysis and the numbers of 
usable observables in the databases. Two solutions are based on exactly the same 
ensemble of observables.}
\label{tab3}       
\begin{tabular}{cccc}
\hline\noalign{\smallskip}
& No. of obs. & rms from & rms from \\
SESSION&used/usable & new databases & IVS databases \\
&& (ps) & (ps)\\
\noalign{\smallskip}\hline\noalign{\smallskip}
14MAY06XA &  1141/1161 &    25.4 & 26.1 \\     
14MAY07XA &   868/878 &    24.7 & 25.5 \\      
14MAY08XA &   950/964 &    25.3 & 26.3 \\      
14MAY09XA &   627/650 &    24.8 & 26.1 \\      
14MAY10XA &   812/827 &    26.0 & 26.5 \\      
14MAY11XA &   678/684 &    25.5 & 26.4 \\      
14MAY12XA &   660/681 &    28.2 & 29.0 \\      
14MAY13XA &   704/712 &    26.3 & 27.3 \\      
14MAY14XA &   827/843 &    27.3 & 28.2 \\      
14MAY15XA &   709/720 &    24.8 & 25.5 \\      
14MAY16XA &   787/827 &    24.7 & 25.5 \\      
14MAY17XA &   682/702 &    23.1 & 24.2 \\      
14MAY18XA &   764/796 &    23.2 & 24.1 \\      
14MAY19XA &   549/556 &    24.4 & 25.4 \\      
14MAY20XA &   861/902 &    22.1 & 22.8 \\      

\noalign{\smallskip}\hline
\end{tabular}
\end{table}

The baseline repeatability for most of baselines was improved in a range up to 0.04 mm.
The comparison for the baselines related to YARRA12M is shown in Table \ref{tab4} as an example.

The differences between results of polar motion and nutation/precession parameters from the two datasets over 15 days were 
at the level of microarcseconds with a maximum of 4.4 microarcseconds, and those for UT1
were below 0.1 microsecond. The coordinates of stations WESTFORD and ZELENCHK 
have the biggest differences, about 1.4~mm in the U direction, while the rest of the stations 
have agreement in three coordinates at the level of 0.2~mm.

\begin{table}
\caption{
Comparision of the rms of time-series of baseline lengths over 15 days for all baselines of YARRA12M.
(Unit: mm) }
\label{tab4}       
\begin{tabular}{lcc}
\hline\noalign{\smallskip}
 &  \multicolumn{2}{c}{rms of baseline length} \\
BASELINE & new databases & IVS databases \\
\noalign{\smallskip}\hline\noalign{\smallskip}
KATH12M-YARRA12M &  3.12 & 3.12 \\ 
HOBART12-YARRA12M &  3.55 & 3.55 \\          
HOBART26-YARRA12M &  3.56 & 3.55 \\       
WARK12M-YARRA12M &  3.82 & 3.82 \\  
TSUKUB32-YARRA12M &  5.57 & 5.57 \\  
HART15M-YARRA12M &  8.61 & 8.62 \\  
BADARY-YARRA12M &  3.54 & 3.55 \\              
KOKEE-YARRA12M &  4.19 & 4.20 \\ 
YARRA12M-ZELENCHK &  3.64 & 3.66 \\        
MATERA-YARRA12M &  3.90 & 3.94 \\   
WETTZELL-YARRA12M &  3.90 & 3.91 \\  
NYALES20-YARRA12M &  4.01 & 4.03 \\   
ONSALA60-YARRA12M &  3.94 & 3.96 \\          
YARRA12M-YEBES40M &  4.01 & 4.02 \\   
\noalign{\smallskip}\hline
\end{tabular}
\end{table}

The direction of source 3C371 was estimated as a global parameter to be
(18$^{\mathrm h}$~06$^{\mathrm m}$~50{\fs}680\,675,
$+$69\arcdeg~49\arcmin~28{\farcs}108\,484)
from the IVS databases and to be (18$^{\mathrm h}$~06$^{\mathrm m}$~50{\fs}680\,664,
$+$69\arcdeg\ 49\arcmin\ 28{\farcs}108\,472)
from the new databases. The difference is 165 microarcseconds in right ascension and 12 microarcseconds in declination.
This difference should be due to the reference point used for the calculation of the structure effect, which 
is the peak intensity of the brightness distribution in the study. Another pair of solutions, 
in which the position of source 3C371 
was fit as session-wise parameter, were made to get the time series of source's position. The results are presented 
in Fig. \ref{fig_m3C371}. The main difference is a constant offset in right ascension, which can be removed 
by choosing an appropriate reference point
for calculating the structure delay corrections. The variation in the source's
position remains at the level of a few hundred microarcseconds, which suggests that the modeled structure effect for this
source in bandwidth synthesis and ionospheric effect free delay observables does not perform as well as that for phases.
The differences in the positions of the remaining sources are about 3 microarcseconds or below that for global sources and 
up to 10 microarcseconds for session-wise sources.

\begin{figure}
\resizebox{0.52\textwidth}{!}{%
  \includegraphics{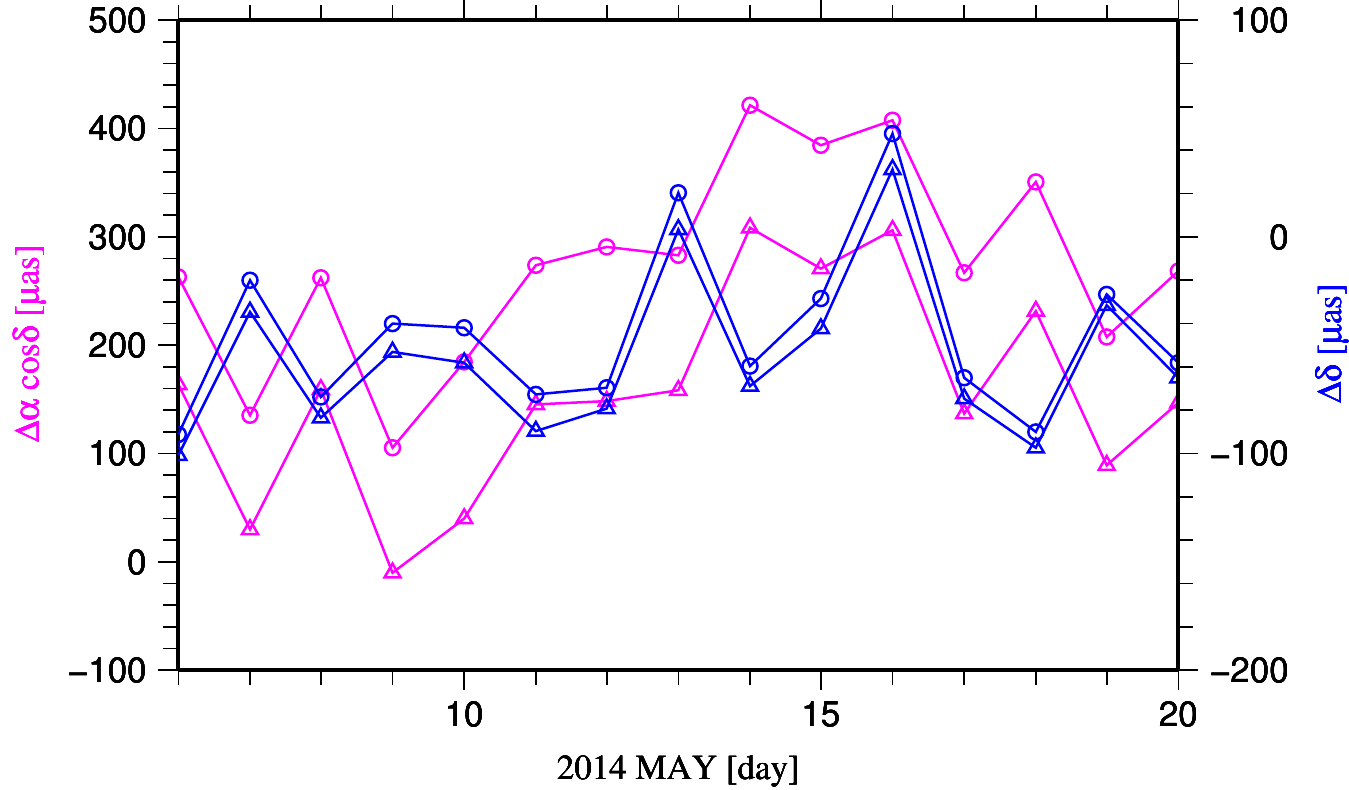}
}
\caption{Comparison of estimated position corrections 
of source 3C371 to its ICRF2 position with its structure effect corrected or not.
The circles represent the time series from the original IVS 
databases, while the triangles represent that from the new
databases with the aim of correcting the structure effect. }
\label{fig_m3C371}       
\end{figure}

\section{Discussion and conclusion}
\label{sec4}
The source structure effects in CONT14 observations were studied in terms of closure delays, 
closure phases, and closure amplitudes. 
A method of deriving structure index based on closure quantities was proposed, and structure 
indices for 65 radio sources with at least 30 closure triangles were obtained
according to this method. This result is comparable to structure indices from 
the BVID and the closure quantities in principle capture important information about source structure.
The equations of closure delay structure indices are derived from exactly the same 
thresholds of structure delays as that were used by \citet{fey1997}. This allows our closure delay 
structure indices to hold the same meaning of structure index as its original definition. There are, 
however, two main differences between the structure indices from these two methods: (1) because 
our closure delay structure indices are derived from actual observables and the BVID structure
 indices are from theoretical predictions of only structure delay based on images, our closure delay structure indices have to 
take the measurement noise into account, while this is not the case for the BVID structure indices; (2) our closure delay structure indices are determined 
from the actual ($u,v$) coordinates sampled in a session, whereas the structure indices of \citet{fey1997} 
are calculated from a grid sampling all possible ground-based VLBI ($u,v$) coordinates. 
Further investigations, based on the same datasets of the structure indices from the BVID,
of our method and of the influence of the different observing networks are definitely needed.
Therefore, the structure indices
for geodetic sources can be regularly updated without making images, for instance,
from all the VLBA and VLBA plus global VLBI stations sessions, IVS terrestrial reference frame sessions, 
and even from IVS R1 and R4 sessions for a fraction of radio sources. 
Structure index is conservatively defined by the median absolute value
of structure effect corrections, although the median values of 
closure quantities are in general significantly smaller than
their rms values. The structure effects in geodetic VLBI may have been underestimated.
According to our study the rms values are better than the median values
in terms of demonstrating the magnitude of structure effect.

Source 3C371 was selected as a starting point for the study of impacts of structure 
effect on geodetic VLBI data analysis. 
A three-component model of the structure of 3C371, derived by model-fitting 
from closure phases, was used to correct its structure delay.
The structure model derived from phases does not fit with the delay observables 
as well as the phase observables. 
The results show the impacts of structure effect by this \emph{individual} source on 
EOPs is up to about 4.4 microarcseconds and on station's position
in some cases are beyond 1 mm. The estimated source position is strongly 
dependent on the reference point of the structure model.

Even though this preliminary study of the structure effects for an
individual source can be limited to summarizing the significance of
correcting for the structure effect, there are at least three
conclusions that can be made. First, although source structure effects
may be averaged out to some extent when solving for geodetic
parameters, in particularly for an individual source in this study, 
they are crucial for determining the position of each
individual source.  Without correcting the source structure effects, a
source position determined from geodetic observations does not have a
clear reference point, neither the location of the peak intensity nor
the center of the brightness, because the estimated source position
strongly depends on the specific baseline geometries of
observations. But with the effects corrected, we can explicitly say
where the determined source position is located with respect to the
source structure. For instance, we can say that the reported source
position in the case of 3C371 after correcting these effects is the
location of the peak intensity identified from our method.  If we use
a map of a source to correct the effects, then we can provide the
location of the reference point for the estimated source position in
the map. Since the difference between the positions with and without
correcting structure effects is at the level of sub-milliarcseconds,
and larger differences can be expected for some sources, this is very
important for high-accuracy relative astrometry. In addition, with the
possibility of identifying the cores of sources, we can realize a more
stable celestial reference frame. Second, the impact
on station position in some cases is already beyond 1~mm. 
Third, the residual rms for source 3C371 was reduced by
1~ps. This is significant, even though there is still room for the
improvement of the structure model. From our study, an improvement in
geodetic VLBI data analysis of, at least, the picosecond level can be
expected after a complete consideration of structure effects. 

This study demonstrates the preliminary results of structure effect. 
Only 10\% of radio sources have a structure index of 1
in CONT14. Sources 
with a structure index of 2, the majority in geodetic VLBI observations,
actually have rms closure delay at the
level of 30 ps, which significantly contribute to the total
residual rms of VLBI data analysis. 
A rigorous and consistent method of handling the source structure is
to correct the structure phases, 
based on the brightness distribution obtained from that epoch, for
the raw phases in all channels 
used in the recording system during the post-processing procedure
and then re-determine multi-band group delays and ionosphere corrections.

\begin{acknowledgement}

We acknowledge the International VLBI Service for
Geodesy and Astrometry (IVS) and all of its components for
their efforts in observing, correlating, and providing the VLBI
data used in this study. We thank the three anonymous referees for thoughtful readings and valuable comments,
and Leonid Petrov (Astrogeo Center, USA) for the helpful discussions about this work.
This research has made use of material from the Bordeaux VLBI Image Database
(BVID). This work was done while MH
worked as a guest scientist at GFZ, Germany, and was supported by the National
Natural Science Foundation of China (grant No. 11473057 and 11303077).
\end{acknowledgement}

%

%
%

\begin{appendix}

\section{Specification for the calculations of closure quantities}
\label{sec6}
To the accuracy of the
second order in delay, the closure delay $\tau_{\mbox{\scriptsize abc}}(t)$ at reference epoch $t$ 
for three stations $a$, $b$, and $c$, is 
calculated from geodetic VLBI observations\footnote{For astronomical observations 
that reference all observables for a scan to the same wavefront,
the model of closure delay has the simpler form 
$\tau_{\mbox{\scriptsize abc}}(t)=\tau_{\mbox{\scriptsize ab}}(t)+\tau_{\mbox{\scriptsize bc}}(t)-\tau_{\mbox{\scriptsize ac}}(t)$.} by
    \begin{equation}
     \label{eq_closure1}
\begin{aligned}
\tau_{\mbox{\scriptsize abc}}(t)=&\tau_{\mbox{\scriptsize ab}}(t)+\tau_{\mbox{\scriptsize bc}}(t)-\tau_{\mbox{\scriptsize ac}}(t)\\
&+[\dot{\tau}_{\mbox{\scriptsize bc}}(t)\cdotp{\tau^{\prime}_{\mbox{\scriptsize ab}}(t)}+\frac{1}{2}\ddot{\tau}_{\mbox{\scriptsize bc}}(t)\cdotp{{\tau^{\prime}_{\mbox{\scriptsize ab}}(t)}^2}],
\end{aligned}
    \end{equation}
where, for instance, $\tau_{\mbox{\scriptsize ab}}$ is the group delay observable from station $a$ to
station $b$, and $\tau_{\mbox{\scriptsize bc}}$ is the group delay observable from station $b$ to
station $c$, for the same wavefront received by three stations. A prime on a delay symbol indicates that
the term refers only to the geometric delay without dependence on station clock offset, and
a superposed dot and double superposed dots denote differentiation
with respect to time once and twice, respectively.
The definition and model of closure delay was discussed in detail by \citet{xu16}.
By convention in geodetic VLBI measurements, the time tag of a VLBI observable is
referred to the epoch when the wavefront passes the first station in the
baseline. In order to have the three delay observables in the closure refer to the the same wavefront,
there are the corrections in the brackets of equation \ref{eq_closure1} to the group delay for the 
second baseline in the triangle.

Similar to closure delay, closure phase can be calculated from geodetic VLBI observations\footnote{For astronomical data, 
$\phi_{\mbox{\scriptsize abc}}(t)=\phi_{\mbox{\scriptsize ab}}(t)+\phi_{\mbox{\scriptsize bc}}(t)-\phi_{\mbox{\scriptsize ac}}(t)$.} by using
    \begin{equation}
     \label{eq_closure2}
\begin{aligned}
\phi_{\mbox{\scriptsize abc}}(t)=&\phi_{\mbox{\scriptsize ab}}(t)+\phi_{\mbox{\scriptsize bc}}(t)-\phi_{\mbox{\scriptsize ac}}(t)\\
   &+[\dot{\tau}_{\mbox{\scriptsize bc}}(t)\cdotp{\tau^{\prime}_{\mbox{\scriptsize ab}}(t)}+\frac{1}{2}\ddot{\tau}_{\mbox{\scriptsize bc}}(t)\cdotp{{\tau^{\prime}_{\mbox{\scriptsize ab}}(t)}^2}] \cdot 2\pi{\nu},
\end{aligned}
    \end{equation}
where, for instance, $\phi_{\mbox{\scriptsize ab}}$ is fringe phase observable on baseline $ab$, and $\nu$ is the reference frequency.
We should be aware that $\tau$, $\dot{\tau}$, and $\ddot{\tau}$ in equation \ref{eq_closure2} are phase delay and 
the derivatives of phase delay, and are different from the group delay terms in equation \ref{eq_closure1}.

For closure amplitude, in order to have closure amplitude quantities 
be zero for point-like sources, like closure delay and closure phase, 
we calculate the absolute value of natural logarithm of the closure amplitude by

    \begin{equation}
     \label{eq_closure4}
     C_{\mbox{\scriptsize amp}}=| \ln(A_{\mbox{\scriptsize abcd}}) |.
    \end{equation}

\section{Equations for determing structure index from closure quantities}

Continuous structure index from closure quantities is defined as follows:

    \begin{equation}
     \label{eq_ind6}
 \mbox{\emph{SI}}_{\mbox{\scriptsize clo-dela}} \equiv  \ln \frac{\left|\tau_{\mbox{\scriptsize closure}}\right|^{\mbox{\scriptsize med}}}{1~\mbox{ps}},
    \end{equation}

    \begin{equation}
     \label{eq_ind3}
 \mbox{\emph{SI}}_{\mbox{\scriptsize clo-phas}} \equiv  0.77 \ln \frac{\left|\phi_{\mbox{\scriptsize closure}}\right|^{\mbox{\scriptsize med}}}{1~\mbox{deg}} + 1.26 ,~\mbox{and}
    \end{equation}

    \begin{equation}
     \label{eq_ind4}
  \mbox{\emph{SI}}_{\mbox{\scriptsize clo-amp}} \equiv  2.67 \left| \ln A_{\mbox{\scriptsize closure}} \right|^{\mbox{\scriptsize med}} +  2.14.
    \end{equation}
The form of equation \ref{eq_ind6} was selected to match the integer steps of equation 3, not to match the continuous structure index of \citet{fey1997}. A least square fit was performed based on the CONT14 observations to determine the coefficiences in equations \ref{eq_ind3} and \ref{eq_ind4}, by matching the closure phase and closure amplitude structure indices as well as possible to the closure
delay structure indices. Generalized, exact forms of these equations for their applications to other observations need to
be investigated carefully.

\end{appendix}

\end{document}